\documentclass[11pt]{article}

\usepackage{geometry} % set the page margins
\usepackage{graphicx} % insert graph use command "includegraphics"
\usepackage{subfigure}
\usepackage{caption}
\usepackage{amsmath}
\usepackage{amsthm}
\usepackage{mathrsfs}
\usepackage{amsfonts}
\usepackage{amssymb}
\usepackage{color}
\usepackage{array}
\usepackage{booktabs}
\usepackage{tabu}
\usepackage{tabularx}
\usepackage{arydshln} % dashline
\usepackage{algorithm}
\usepackage{algorithmic}
\usepackage{ulem}
\usepackage{cancel}
\usepackage{bm}
\usepackage{multirow}
\usepackage{txfonts}

\newcommand{\bzeta}{\boldsymbol{\zeta}}

\newtheorem{theorem}{Theorem}
\newtheorem{lemma}{Lemma}
\newtheorem{coro}{Corollary}
\newtheorem{defi}{Definition}

\newtheorem{remark}{Remark}

\newcommand{\ba}{\begin{array}}
\newcommand{\ea}{\end{array}}
\newcommand{\bt}{\begin{tabular}}
\newcommand{\et}{\end{tabular}}
\newcommand{\btb}{\begin{table}}
\newcommand{\etb}{\end{table}}
\newcommand{\C}{\begin{center}}
\newcommand{\ec}{\end{center}}
\newcommand{\bea}{\begin{eqnarray}}
\newcommand{\eea}{\end{eqnarray}}
\newcommand{\Bea}{\begin{eqnarray*}}
\newcommand{\Eea}{\end{eqnarray*}}
\newcommand{\beq}{\begin{equation}}
\newcommand{\eeq}{\end{equation}}
\newcommand{\thickhline}{%
    \noalign {\ifnum 0=`}\fi \hrule height 1pt
    \futurelet \reserved@a \@xhline
}

\def \bfm#1{\mbox{\boldmath$#1$}}

  \def \C {{\bfm C}}

  \def \0 {{\bfm 0}}

\def \O {{\bfm O}}

 \def \u {{\bfm u}} 
  \def \v {{\bfm v}}
\def \V {{\bfm V}}
\def \x {{\bfm x}} \def \X {{\bfm X}}
 
 \def \z {{\bfm z}}

\def \cd {{\mathcal D}} \def \ce {{\mathcal E}} 
  
\def \cl {{\mathcal L}}  
 \def \cn {{\mathcal N}}
 
\def \cp {{\mathcal P}} 
\def \cq {{\mathcal Q}}
\def \cs {{\mathcal S}}
\def \ct {{\mathcal T}}
\def \cx {{\mathcal X}} \def \cy {{\mathcal Y}} \def \cz {{\mathcal Z}}

\def \bk {{\mathbb K}}
\def \br {{\mathbb R}}

\begin{document}
\title{Model-free Subsampling Method Based on Uniform Designs}

\author{Mei Zhang$^{a,b}$,\ \ Yongdao Zhou$^b$,\ \ Zheng Zhou$^b$,\ \ Aijun Zhang$^{c,}$\thanks{Corresponding author. Email: ajzhang@umich.edu}\\
{\small \it $^a$College of Mathematics, Sichuan University,
Chengdu 610064, China}\\
{\small \it $^b$School of Statistics and Data Science, LPMC $\&$ KLMDASR,
Nankai University, Tianjin 300071, China}\\
{\small \it $^c$Department of Statistics and Actuarial Science,
The University of Hong Kong, Hong Kong SAR, China} \date{} }
\maketitle

\noindent {\bf Abstract:}
Subsampling or subdata selection is a useful approach in large-scale statistical learning.
Most existing studies focus on model-based subsampling methods which significantly depend on the model assumption.
In this paper, we consider the model-free subsampling strategy for generating subdata from the original full data.
In order to measure the goodness of representation of a subdata with respect to the original data, we propose a criterion, generalized empirical $F$-discrepancy (GEFD),  and study its theoretical properties in connection with the classical generalized $\ell_2$-discrepancy in the theory of uniform designs.
These properties allow us to develop a kind of low-GEFD data-driven subsampling method based on the existing uniform designs.
{By simulation examples and a real case study, we show that the proposed subsampling method is superior to the random sampling method. Moreover, our method keeps robust under diverse model specifications while other popular subsampling methods are under-performing.} In practice, such a model-free property is more appealing than the model-based subsampling methods, where the latter may have poor performance when the model is misspecified, as demonstrated in our simulation studies.


\section{Introduction} \label{Sect:Intro}
Technological advances have enabled an extraordinary speed in data generation and collection in many scientific fields and practices, such as in astronomy, economics, and industrial problems.
However the growth rate of the storage memory and the
computational power is still far from sufficiently handling the explosion of modern data sets.
Therefore, the demand for extracting a small sample from a large amount of data arises routinely.

Assume $\cx = \{\x_i, i = 1, \dots, N\}$ is an observed large-scale data in $\br^s$.
This paper considers the problem of selecting $n$ points from $\cx$ to form a subdata $\cp$ that captures the main information on the distribution of $\cx$.
In another word, $\cp$ is expected to be a good representation with respect to $\cx$.
The recent literatures have paid attention to the subsampling problem, but most of these literatures studied the randomized algorithms based on weighted random sampling, see \cite{B18,M11,W18,W14}. %Usually, these algorithms belong to model-based approaches,
%which performed in order to pursue specific goals under pre-assumed models.
Some of these algorithms are model-based, such as the optimal subsampling methods for logistic regression proposed by Wang et al. \cite{W18}, where the optimal subsampling probabilities are determined in order to minimize the asymptotic mean squared error of the target subsample-estimator (mMSE or mVc) given the full data.
The mMSE and mVc are respectively  based on A- and L-optimality criteria in the theory of optimal experimental design.
For the deterministic subsampling approach, Wang et al. \cite{W19} propsoed the information-based optimal subdata selection (IBOSS) for the linear regression of the big data.
The basic idea of IBOSS is to select the most informative data points deterministically based on  the D-optimality criterion in  optimal experimental design. % without relaying on random subsampling.
Nevertheless, IBOSS is based on the simplest regression model.  If the underlying model is more complex than a linear regression model, IBOSS would not {keep} the optimal performance.
In statistical learning, we usually do not have prior knowledge about the underlying model, which can be either linear or nonlinear. Thus,  it is meaningful to develop a subsampling method which is robust to the model specification.

Our research aim is a model-free subsampling method that is competitive no matter whether the underlying model is correctly specified.
% which is not more than the model-based subsampling method much worse when the supposed model is correct, and performs better than the model-based subsampling method when the supposed model is incorrect.
For statistical learning from large-scale dataset, the working model usually takes the complex form, either parametric or nonparametric. Such working model is usually misspecified, when comparing to the ground truth.  In this case, a successful model-free subsampling method would be particularly useful.
%While the more accurate a model is, the higher complexity the model will be, in which situation, the modeling procedure on the big data will be seriously difficult and cost a great deal of time.
%For saving time, we need a subdata to setup the modeling procedure and hope the model established according to such subdata has good property in predicting.
%In other words,  model-free subsampling methods are needed.
%As a model-free subsampling method,
Among randomized subsampling techniques, the uniform random sampling (URS) %which randomly sampling of the indices of the points with uniform distribution,
is the simplest strategy and it is often regarded as the baseline method for developing other model-free subsampling methods. It is known that the subdata selected by URS can preserve the distribution of the full data.
%used as a building block for more complex sampling methods.
%%URS is a model-free subsampling method since the subsample is more and more similar to the full data as the subsample's size increasing and it does not need any assumption on the established model.
%However, because of the randomness, the representation of the subdata by URS may be not good in some local area of  the original data especially when the size of subdata is not large.
%%in practice, the distribution of the resulting set of points may diverge quite largely from the original one,
%Then, other model-free subsampling methods which have better performances than URS may be provided.
%
%The sample by URS can be regarded as the random sampling points whose target distribution is the uniform distribution.
Compared to the URS method, some quasi-Monte Carlo methods have better performances for representing uniform or general distribution. The uniform design \cite{F18, FW94} is a popular method that possesses the model-robust property, as is widely used in numerical integration, computer experiments, statistical simulations and other statistical areas.
To measure the uniformity of a point set,  there exist many forms of discrepancy criteria, such as the star discrepancy and the generalized $\ell_2$-discrepancies including
centered $\ell_2$-discrepancy, wrap-around $\ell_2$-discrepancy and
mixture discrepancy, see \cite{hick1, hick2, W16,Z13}.
%There are many methods for sampling on a continuous space based on a given cumulative distribution function (CDF) $F(\x)$. For the most common situation when $F(\x)$ is the uniform distribution on $C^s=[0,1]^s$, to obtain a sample well representing with respect to $F(\x)$ is to construct a uniform design under some specified uniformity criterion.
%So far, many discrepancies have been proposed as criteria
%to assess the uniformity of designs on $C^s$,
%see \cite{hick1, hick2, W72, W16, Z13}.
The latest monograph by Fang et al. \cite{F18} {contained} a comprehensive introduction of the theory of uniform designs. %The target distribution of a uniform design is also the uniform distribution.
Moreover, for a general distribution $F(\x)$ in $\mathbb{R}^s$, Fang and Wang \cite{FW94} proposed the concept of $F$-discrepancy for measuring the goodness of representation of a point set with respect to $F(\x)$.  The smaller the value of $F$-discrepancy, the better the point set represents the $F$-distribution.
%Lower the value of the discrepancy, better the representation performance of the point set. Then, the quasi-Monte Carlo point set pursues a low discrepancy.

Motivated by the model-robust property of uniform designs,  we propose a data-driven subsampling method based on a generalized empirical $F$-discrepancy (GEFD). The main idea is to utilize the uniform design on the unit hypercube and transform it to the observational data space. The proposed GEFD criterion is defined as the $\ell_2$-norm of the difference between empirical distributions of the small data and the big data in the observational space. Under the joint independence assumption, the GEFD criterion can be translated to the unit hypercube upon the suitable transformations. We study the asymptotic equivalence of such transformation, and then develop the subsampling method based on the existing uniform designs in the literature. {Such a uniform design-based subsampling method only depends on the data, but not the model. Therefore, we call it data-driven subsampling (DDS). DDS is demonstrated through several numerical examples to enjoy the model-robust property even when the working model is misspecified. To illustrate the superiority of DDS, we compare the performance of DDS, URS, IBOSS and some popular model-free subsampling method, such as kernel herding\cite{C12} and support points\cite{M18}. As expected, DDS keeps effcient and robust under diverse model sepcification even when the working model is misspecified, while other popular subsampling method are under-performing.}
\par The remainder of this paper is organized as follows. Section~\ref{Sect:GEFD}
proposes the new GEFD criterion in order to
measure the goodness of presentation for a small data with respect to the full data. Section~\ref{Subsect:property} establishes the asymptotic equivalence between the GEFD on the observational space  and the generalized $\ell_2$-discrepancy on the unit hypercube. The corresponding empirical version of Koksma-Hlawka inequality is also derived.
%in which the  generalized empirical $F$-discrepancy gives the upper bound of the difference of averaged function values between different data sets.
Then in Section~\ref{Sect:DDS} we develop the data-driven subsampling method under the proposed GEFD criterion. Section~\ref{Sect:ADDS} analyzes the complexity of the proposed
algorithm and provides an accelerated approach for practical implementation.
Section~\ref{eg} presents numerical examples of the proposed subsampling method for both classification and regression tasks. The last section concludes the paper. All the proofs are defered to the Appendix A. Detailed data and added figures are given in Appendix B.

\section{Generalized Empirical $F$-Discrepancy}
\label{Sect:GEFD}

%Similar  to the representation points of a distribution,

For a given data $\cx =\{\x_i, i = 1, \dots, N\} \subset \br^s$ with large $N$, denote its empirical cumulative distribution function (ECDF) as $F_{\cx}$. To find a small data to represent
% captures the main information of
$\cx$, a natural idea is to find a subdata $\cp\subseteq\cx$ that has {\it low discrepancy} with respect to
the ECDF  $F_{\cx}$.
%%\revA{Zhang et al. \cite{Z19} proposed the empirical $F$-discrepancy
%%to measure the representation of the small data $\cp$ with respect to the big data $\cx$,}
%It is known that the $F$-discrepancy does not have an analytic expression. For easy computation, we may need a discrepancy which has an analytic expression to measure the representation of $\cp$ with regard to $\cx$.
Following the $F$-discrepancy by Fang and Wang \cite{FW94}  and the generalized $\ell_2$-discrepancy by Hickernell \cite{hick1}, we define the generalized empirical $F$-discrepancy (GEFD) in this section. It will be shown that such a discrepancy has an analytic expression.

%The generalized $L_2$-discrepancy which used the tool of reproducing kernel Hilbert spaces.
Let $C^s = [0,1]^s$ be the $s$-dimensional unit hypercube,
and $\bk (\cdot,\cdot)$ be a real-valued kernel function defined on
$C^s \times C^s$,
%and $K(\cdot, \cdot)$ be a bivariate kernel function,
%\bea \bk(\u,\v) = \prod_{j = 1} ^ s K(u_j,v_j), ~~\u,\v \in C ^ s, \eea
satisfying
(i)  symmetric: $\bk(\u, \v) = \bk(\v, \u)$, for any $ \u,\v \in C^s$;
(ii)  non-negative definite:
$\sum_{i,j = 1}^n a_i a_j \bk(\u_i,\u_j) \geq 0$ for any
$n > 0, a_i \in \br$ and $\u_i \in C^s$.
%\begin{itemize}\item [(i)]  symmetric: $\bk(\u, \v) = \bk(\v, \u)$, for any $ \u,\v \in C^s$;
%\item [(ii)]  non-negative definite:
%$\sum_{i,j = 1}^n a_i a_j \bk(\u_i,\u_j) \geq 0$ for any
%$n > 0, a_i \in \br$ and $\u_i \in C^s$.\end{itemize}
% Then $\bk$ is called a kernel function.
Denote the space of real-valued functions on $C^s$ with kernel function $\bk$ by %$\mathcal{W}$ which
$\mathcal{W}_{\bk} = \left\{F: \int_{C^{2s}}\bk (\u, \v) \mbox{d} F(\u)\mbox{d}F(\v) < \infty \right\}$. Then the space $(\mathcal{W}_{\bk},\langle\cdot,\cdot\rangle_{\bk})$ is a Hilbert space, where $\langle\cdot,\cdot\rangle_{\bk}$ denotes the inner product with formula $\langle F,G\rangle_{\bk} = \int_{C^{2s}}\bk (\u, \v) \mbox{d}F(\u)\mbox{d}G(\v)$. Such $\bk$ is a reproducing kernel satisfying that  $\bk(\cdot,\u) \in \mathcal{W}_{\bk}$ and $F(\u) = \langle F, \bk(\cdot,\u) \rangle_{\bk}$ for any $\u \in C^s$ and $F \in \mathcal{W}_{\bk}$. For a point set $\cd = \{\bm{\zeta}_1,\dots, \bm{\zeta}_n\}$ in $C^s$, the reproducing kernel $\bk$ induces the squared generalized $\ell_2$-discrepancy of $\cd$ with respect to the uniform distribution $F_{\mathrm{u}}$ on $C ^ s$ as follows,
\begin{align} \label{uniform_disc} \nonumber
D^2 (\cd; F_{\mathrm{u}}, \bk) &= \int_{C^{2s}}\bk (\u, \v) \mbox{d}
(F_{\cd} - F_{\mathrm{u}})(\u)\mbox{d}(F_{\cd} - F_{\mathrm{u}})(\v) \\
&= \int_{C^{2s}} \bk (\u, \v) \mbox{d} \u\mbox{d}\v
- \frac 2n \sum_{i=1}^n \int_{C^s} \bk (\u ,\bm{\zeta}_k)\mbox {d}\u
+\frac {1}{n^2} \sum_{i=1}^n \sum_{k=1}^n \bk(\bm{\zeta}_i,\bm{\zeta}_k).
\end{align}
By taking different kernel functions $\bk$, we can obtain different kinds of
generalized discrepancy such as the widely used centered ${\ell}_2$-discrepancy, wrap-around ${\ell}_2$-discrepancy and mixture discrepancy,   whose kernel functions  are defined as
\begin{align}
\bk^\mathrm{C}(\u,\v)  & =  \prod_{j = 1}^s \left[ 1 + \frac 12 \left| u_j - \frac 12 \right| +
\frac 12 \left| v_j - \frac12 \right| - \frac 12 \left| u_j - v_j \right| \right], \nonumber\\
\bk^\mathrm{W}(\u,\v) & =
\prod_{j = 1}^s \left[ \frac 32 - \left| u_j - v_j \right| +
\left| u_j - v_j \right| ^ 2 \right], \label{M}\\
\bk^\mathrm{M}(\u,\v) & =
\prod_ {j = 1} ^ { s } \left[ \frac { 15 } { 8 } -
\frac 14 \left| u_j - \frac 12 \right| - \frac 14 \left| v_j - \frac 12 \right|
- \frac 34 \left| u_j - v_j \right|
+ \frac 12 \left| u_j - v_j \right| ^ { 2 }\right ],  \nonumber
\end{align}
respectively. Zhou et al. \cite{Z13} showed that the mixture discrepancy is a better choice for measuring the uniformity of point sets in $C^s$.

%\subsection{Definition of the GEFD}\label{Subsect:def}

%To cope with subsampling problem, consider the space $\br^s$.
Let $\widetilde{\bk}: \br^s \times \br^s \rightarrow \br $ be a kernel function defined on $\br^s \times \br^s$.
%then $\widetilde{\bk}$ also satisfy the properties of symmetric and non-negative definite.
We are concerned with the representation of a small data $\cp$ with respect to the full data $\cx$, which are associated with the ECDFs  $F_{\cp}$ and $F_{\cx}$, respectively.  Similar to the generalized $\ell_2$-discrepancy with respect to the uniform distribution $F_{\mathrm{u}}$ on $C^s$, we consider a norm of the difference between $F_{\cp}$ and $F_{\cx}$,
%Define the norm induced by $\widetilde{\bk}$ as follows
$$
\parallel F_{\cp} - F_{\cx} \parallel _{\widetilde{\bk}} =
\left [ \int_{\br^{2s}}\widetilde{\bk}(\x,\z)\mbox{d}
(F_{\cx}(\x)-F_{\cp}(\x))\mbox{d}(F_{\cx}(\z)-F_{\cp}(\z)) \right ]^{1/2}.
$$
% Then an appropriate kernel function $\widetilde{\bk}$ may give rise to a well-defined induced norm.
%To make full use of the established theory and methods, we adopt a transformation approach such that the object of study falls into the existing framework.
Moreover, we consider the transformation $T_{\cx}: \br^s \rightarrow C^s$ of the form
\begin{equation}\label{F_tra}
T_{\cx}(\x)=(F_{\cx_{(1)}}(x_1), \dots, F_{\cx_{(s)}}
(x_s))^T, \end{equation} where $F_{\cx_{(j)}}$ is the marginal ECDF of the $j$th component of $\cx$, $j = 1, \dots, s$. Clearly  $T_{\cx}$ in (\ref{F_tra}) can translate the data $\cx$ in $\br^s$ to the unit hypercube $C^s$ component by component.
%let us consider a kernel function
Upon such transformation, % nature reference for the selection of $\widetilde{\bk}$ may be
we consider the kernel function
$\widetilde{\bk} (\x, \z) = \bk ( T_{\cx}(\x) , T_{\cx}(\z) )$,
where $\bk (\cdot , \cdot)$ is a reproducing kernel function on $C^s \times C^s$, and
define the GEFD of $\cp$ with respect to $\cx$ as follows.
\begin{defi} \label{GEFD}
Given a  data $\cx=\{\x_1,  \dots, \x_N\} \subset \br^s$,
the squared generalized empirical $F$-discrepancy
for a point set $\cp=\{\bm{\xi}_1,\cdots,\bm{\xi}_n\} \subset \br^s$ is defined by
\begin{equation} \label{dpk2} D^2(\cp; \cx,\bk) = \int_{\br^{2s}}\bk(T_{\cx}(\x), T_{\cx}(\z))
\mathrm{d}(F_{\cx}(\x)-F_{\cp}(\x))\mathrm{d}(F_{\cx}(\z)-F_{\cp}(\z)),\end{equation}
where %$\widetilde{\bk}(\x, \z) = \bk(T_{\cx}(\x), T_{\cx}(\z))$ with
$\bk(\cdot, \cdot)$ is a reproducing kernel function on
$C^s\times C^s$ and $T_{\cx}(\cdot)$ is defined in (\ref{F_tra}).
\end{defi}

%The points in $\cp$ can be different with that in $\cx$, and it can also be chosen from the points in $\cx$.

By the definition of GEFD,  the smaller the GEFD of the point set $\cp$ with respect to
$\cx$, the better it represents $\cx$.
%Since the ECDFs $F_{\cx}$ and $F_{\cp}$ are used as the reference distribution,
From (\ref{dpk2}), the GEFD can be equivalently expressed as
\begin{equation} \label{simple_GEFD}  D^2(\cp;\cx,\bk)
=  \frac{1}{N^2} \sum_{i,k=1}^N \widetilde{\bk}(\x_i,\x_k)
- \frac{2}{Nn} \sum_{i=1}^N \sum_{k=1}^n \widetilde{\bk}(\x_i,\bm{\xi}_k)
+ \frac{1}{n^2} \sum_{i,k=1}^n \widetilde{\bk}(\bm{\xi}_i,\bm{\xi}_k),
\end{equation}
in which $\widetilde{\bk} (\x, \z) = \bk ( T_{\cx}(\x) , T_{\cx}(\z) )$. Therefore, one can
evaluate the GEFD criterion easily.
%Meanwhile, the data set $\cp$ with low GEFD is called
%a good data-driven space-filling design with respect to $\cx$,
%no matter $\cp$ belongs to $\cx$ or not,
%and a good data-driven subdata of $\cx$ if $\cp$ belongs to $\cx$.

\begin{figure}[ht!]
\centering
%\vspace{3mm}
\includegraphics[width=0.8\textwidth]{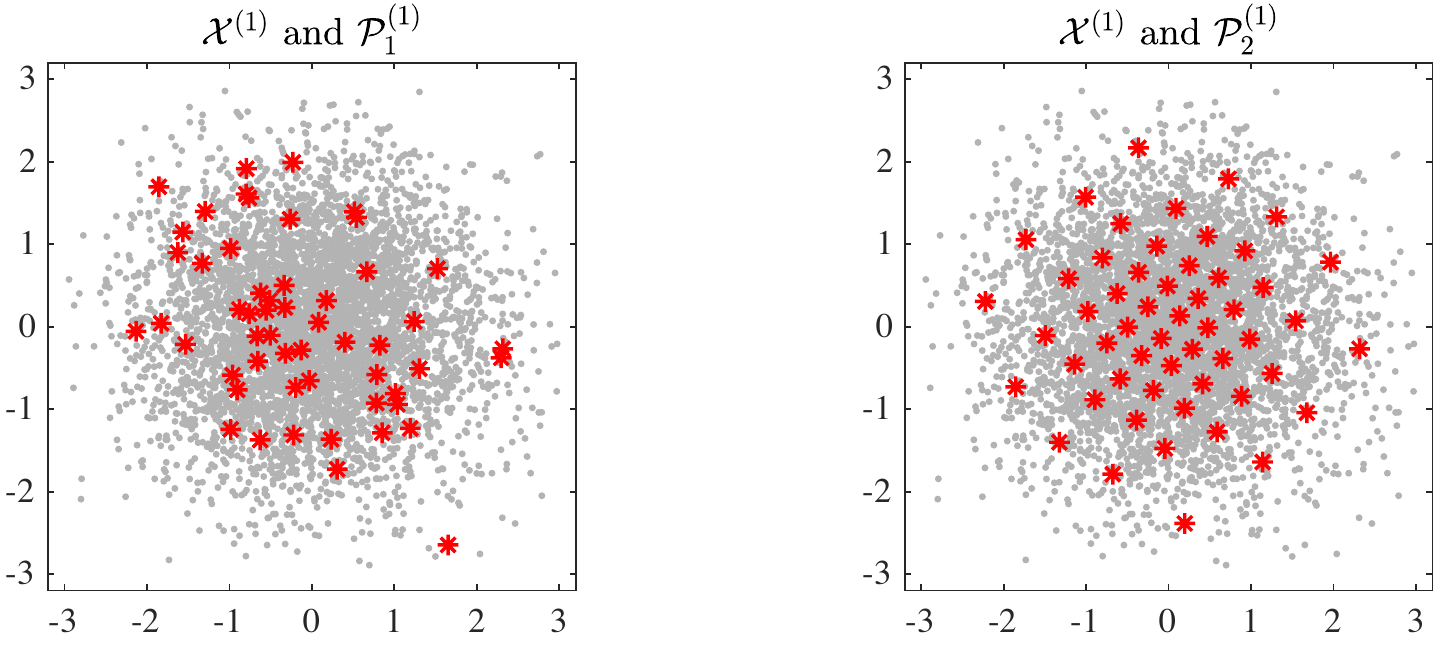}
\includegraphics[width=0.8\textwidth]{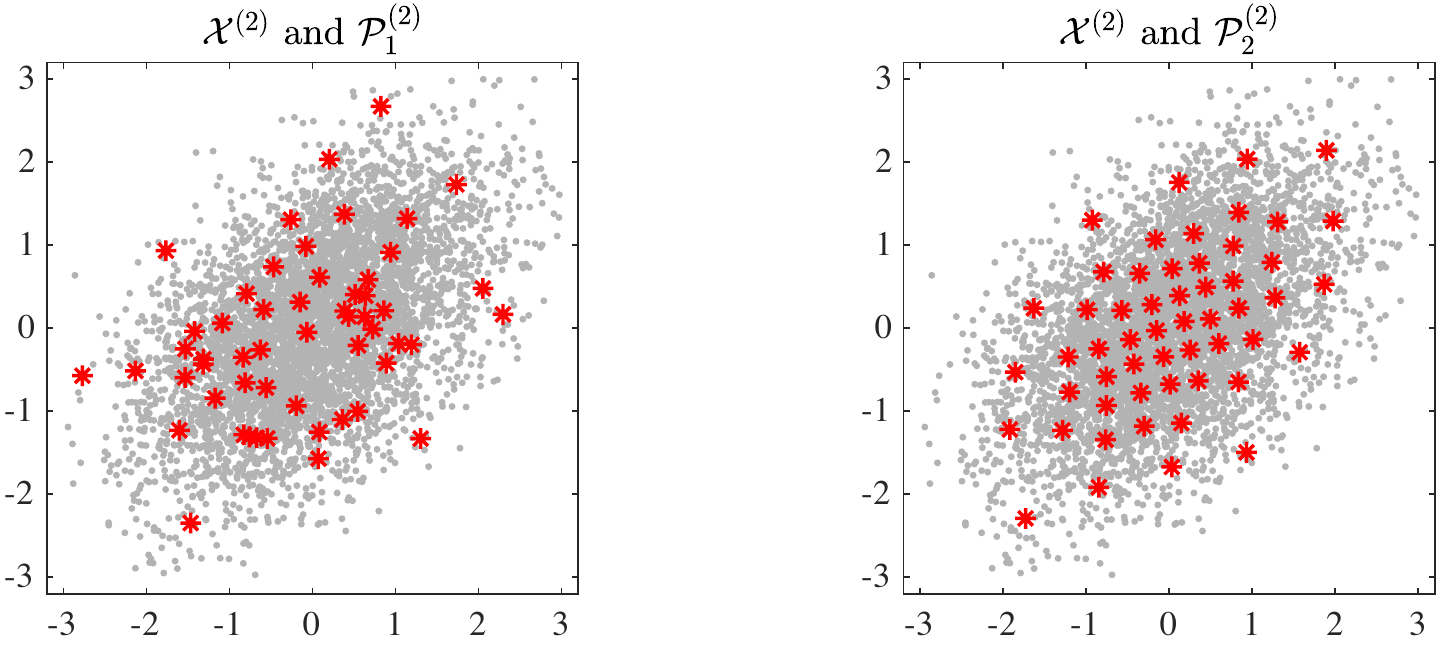}
\caption{The scatter plots of the original binormal data $\cx$ (grey points), the subsampled $\cp_1$ and $\cp_2$ (red asterisks). Upper: independent components; Lower: correlated components.}\label{def}
\end{figure}

Let us give a toy example to illustrate that the proposed GEFD is reasonable to measure
the representation of a small data with respect to the original big data.
Suppose the original dataset $\cx^{(1)} = \{\x_i^{(1)}, i = 1, \dots, N\}$ are generated from a binormal distribution with independent components, shown as the background grey points in Figure~\ref{def} (upper panel). We obtain $\cp_1^{(1)} = \{\bm{\xi}_k^{(1,1)}, k = 1, \dots, n\} \subset \cx^{(1)}$ by the URS method, and $\cp_2^{(1)} = \{\bm{\xi}_k^{(1,2)}, k = 1, \dots, n\} \subset \cx^{(1)}$ by the subsampling method in Section~\ref{Sect:DDS}.
Let us choose the kernel function $\bk^{\text M}$ of mixture discrepancy in (\ref{M}). According to the analytical expression (\ref{simple_GEFD}),
%the squared GEFD values of $\cp_j^{(1)} (j = 1\text{ and } 2)$ %and $\cp_2^{(1)}$  with respect to $\cx^{(1)}$ could calculated as follows,
%\begin{align} \label{eg_GEFD}
%D^2(\cp_j^{(1)};\cx^{(1)},\bk^{\text M})
%& =  \frac{1}{N^2} \sum_{i,k=1}^N \bk^{\text M}(T_{\cx^{(1)}}(\x_i^{(1)}),T_{\cx^{(1)}}(\x_k^{(1)})) - \frac{2}{Nn} \sum_{i=1}^N \sum_{k=1}^n \bk^{\text M}(T_{\cx^{(1)}}(\x_i^{(1)}),T_{\cx^{(1)}}(\bm{\xi}_k^{(1,j)}))\nonumber\\
%&\quad + \frac{1}{n^2} \sum_{i,k=1}^n \bk^{\text M}(T_{\cx^{(1)}}(\bm{\xi}_i^{(1,j)}),T_{\cx^{(1)}}(\bm{\xi}_k^{(1,j)})),
%\end{align}
%where the transformation $T_{\cx^{(1)}}$ is defined in (\ref{F_tra}).
it can be obtained that
$$
D^2(\cp_1^{(1)}; \cx^{(1)} ,\bk^{\text M}) = 1.1212\times 10^{-2} > 4.0424 \times 10^{-4} = D^2(\cp_2^{(1)}; \cx^{(1)},\bk^{\text M}).
$$
Thus $\cp_2^{(1)}$ is better than $\cp_1^{(1)}$ by the criterion of GEFD. This comparison is in accordance with the observed fact in Figure~\ref{def} (upper panel), where $\cp_2^{(1)}$ represents the full data $\cx$ better than $\cp_1^{(1)}$ does. % reasonable and efficient for assessing the representation of a small data with respect to the original data.
Meanwhile, it is worthy mentioning that the GEFD by Definition~\ref{GEFD} is also well-defined when the components of $\cx$ are correlated, although the transformation $T_{\cx}$ in (\ref{F_tra}) translates each coordinate of $\cx$ into $[0,1]$ independently. The right hand side of  (\ref{dpk2}) is a norm of the function $F_{\cx} - F_{\cp}$, where the ECDF $F_{\cx}$ is general enough to cover any joint distributions of $\cx$. For illustration, we give another example where the two components of $\cx^{(2)}$ are correlated, see Figure~\ref{def} (lower panel). We obtain two different subsamples $\cp_1^{(2)}$ and $\cp_2^{(2)}$ {respectively by the URS method and the subsampling method in Section~\ref{Sect:DDS},} and calculate their GEFD criteria,
$$
D^2(\cp_1^{(2)}; \cx^{(2)} ,\bk^{\text M}) = 7.7165\times 10^{-3} > 1.2544 \times 10^{-3} = D^2(\cp_2^{(2)}; \cx^{(2)},\bk^{\text M}).
$$
It is implied that $\cp_2^{(2)}$ is better than $\cp_1^{(2)}$ for representing the full data, which agrees with the intuitive fact in Figure~\ref{def} (lower panel).
Therefore, the GEFD is a reasonable goodness measure for a small subdata in representing the original full data.

% Given $\cx = \{\x_i, i = 1, \dots, N\}$ in Figure \ref{def} as the original full data set. $\cp_1$ is a subset of $\cx$ by the random method, and $\cp_2$ is another subset of $\cx$ obtained by the subsampling method in Section \ref{Sect:DDS}. Let the kernel function $\bk$ be $\bk^{\text M}$, the kernel function of mixture discrepancy in (\ref{M}). According to the computational expression of GEFD in (\ref{simple_GEFD}), the squared GEFD values of $\cp_1$ and $\cp_2$ with respect to $\cx$ are $D^2(\cp_1; \cx ,\bk^{\text M}) = 1.1212\times 10^{-2} > 4.0424 \times 10^{-4} = D^2(\cp_2; \cx,\bk^{\text M})$. Then $\cp_2$ is better than $\cp_1$ in the sense of GEFD. This result is in accordance with the intuitive fact reflected from Figure \ref{def}, which confirms that the GEFD is reasonable and efficient for assessing the representation of a small data with respect to the original data.

%\begin{figure}[h!]
%\centering
%\subfigure[The scatter plots of $\cx$ and $\cp_1$]{\label{def1}
%\includegraphics[width=0.4\textwidth]{eg1_2}}\hspace{1.5cm}
%\subfigure[The scatter plots of $\cx$ and $\cp_2$]{\label{def2}
%\includegraphics[width=0.4\textwidth]{eg1_1}}
%\caption{The scatter plots of $\cx$, $\cp_1$ and $\cp_2$.}\label{def}
%\end{figure}

\section{Properties of the GEFD}\label{Subsect:property}
When $N\rightarrow\infty$, the ECDF $F_{\cx}(\x)$ of $\cx$
converges to the CDF $F(\x)$ for $\x\in\br^s$,
%of some random vector underlying behind $\cx$,
and  the marginal ECDF $F_{\cx_{(j)}} (x_j)$
of the $j$th component  converges to the $j$th marginal CDF of $F(\x)$ for $j = 1, \dots, s$.
Under the joint independence condition, we can derive the asymptotic equivalence between the GEFD defined in (\ref{dpk2}) %of $\cp$ with respect to $F_{\cx}$ in
and the generalized $\ell_2$-discrepancy in (\ref{uniform_disc}) upon
the transformation $T_{\cx}$ in (\ref{F_tra}). % with respect to $F_{\mathrm{u}}$.

\begin{theorem}  \label{asymptotic_theorem}
Given a reference data $\cx \subset \br^s$ satisfying
the joint independence
\begin{equation} \label{ind} F_{\cx}(\x) = \prod_{j=1}^sF_{\cx_{(j)}}(x_j), \quad \forall~\x=(x_1, \dots, x_s) ^T\in \br^s, \end{equation}
suppose the number of repeated points %observations
within $\cx$ be upper
bounded by a constant $c_1$, and the kernel $\bk(u,v) = \prod_{j=1}^s
K(u_j,v_j)$ be Lipschitz continuous in the sense of
$|K(u_1,v) - K(u_2,v)| \leq c_2|u_1-u_2|,$ for any $ u_1,u_2,v \in [0,1]$,
and a constant $c_2$. Then for any small size data
$\cp \subset \br^s$, \Bea %\label{DDF2}
D^2(\cp; \cx,\bk)
= D^2(T_{\cx}(\cp); F_{\rm u}, \bk)+O(1/N^*),\Eea
where $N^* = min_{j\in\{1:s\}}N_j$ and $N_j$ is the number of
marginal distinct values $\cx_{(j)} = \{x_{ij}, i = 1, \dots, N\}$ along the
$j$th coordinate. \end{theorem}

%The proof of Theorem \ref{asymptotic_theorem} can be found in the Appendix.
For the commonly used kernel functions in (\ref{M}), the Lipschitz continuous condition is easily satisfied. Theorem~\ref{asymptotic_theorem}
%\sout{\revA{is important as it}}
translates the GEFD of  $\cp$ in $\br^s$ to
the generalized $\ell_2$-discrepancy of $T_{\cx}(\cp)$ in $C^s$,
subject to an approximation error of order $O(1/N^*)$.
For $\cx$ with large sample size $N$, the quantity $N^*$ is also large as each
continuous coordinate may entertain infinitely many distinct values.
% Here the points in $\cp$ can be chosen from $\cx$ or other points from $\br^s$.
Specifically for the one-dimensional case, we obtain the following asymptotic optimality result since  the equidistant point set $\{1/(2n), 3/(2n), \dots, (2n-1)/(2n)\}$
is known to have the lowest discrepancy for the kernels in (\ref{M}); see Zhou et al. \cite{Z13}.

\begin{coro}\label{1-dim}
For $\cx \subset \br$ with sample size $N$ , the point set $\cp$
given by $$ %\label{1-dim_eq}
\bm{\xi}_k = F_{\cx}^{-1} \left( \frac {2k-1}{2k} \right),
\quad k = 1, \dots, n,$$
is asymptotically optimal (as $N \rightarrow \infty$) with respect to
the GEFD
%generalized empirical $F$-discrepancy
using the kernel function $\bk^\mathrm{C}$, $\bk^\mathrm{W}$ or $\bk^\mathrm{M}$ defined in (\ref{M}). Here $F_{\cx}$ is the ECDF of $\cx$.
\end{coro}

%\subsection{Generalized empirical Koksma-Hlawka Inequality}

Hickernell \cite{hick1} pointed out an important property of the generalized $\ell_2$-discrepancy,
known as the famous Koksma-Hlawka Inequality.
%which gives a upper bound of the difference between the integration
%and the averaged value of a function in $C^s$.
Let $\cd = \{\bm{\zeta}_k, k = 1, \dots, n\}$ be a set of points on $C^s$,
and $f$ be a function on $C^s$, then
the upper bound of the difference between the integral
of $f$ over $C^s$ and the averaged value of $f$ among $\cd$
is given by % the Koksma-Hlawka Inequality as follows,
\begin{equation}\label{KH}
\left|\int_{C^s} f(\u) \mbox{d}\u - \frac 1n \sum_{k = 1}^n f(\bm{\zeta}_k)\right|
\leq D(\cd; F_{\mathrm{u}},\bk) V_2(f,\bk),
\end{equation}
where $V_2(f,\bk)$ is the generalized $\ell_2$-variation defined in \cite{hick1}.
Such Koksma-Hlawka inequality is an important result in numerical integration and
quasi-Monte Carlo methods.
Given the function $f$ and a reproducing kernel $\bk$, $V_2(f,\bk)$ is a fixed quantity.
When $V_2(f,\bk)$ is bounded in $C^s$, the lower the generalized $\ell_2$-discrepancy
$D(\cd; F_{\mathrm{u}},\bk)$, the smaller the upper bound in (\ref{KH}).

%To save time and cost, we %\revG{expect for}
%use a small data to replace the initial data
%\revG{to model and analysis, so we are interested in}
Now consider the difference between the averaged values of $f$ under two data sets. Given an $N$-size data set $\ce = \{\u_i, i = 1, \dots, N\} \subset C^s$, an $n$-size data set $\cd = \{\bzeta_k, k = 1, \dots, n\} \subset C^s$, and a reproducing kernel $\bk$ on $C^s\times C^s$, write
\begin{align} \nonumber
D_{\bk}^2 (\ce,\cd) = & \int_{C^{2s}} \bk(\u,\v)\mbox{d}
(F_{\ce}(\u)-F_{\cd}(\u))\mbox{d}(F_{\ce}(\v)-F_{\cd}(\v)) \\\label{DK}
= &  \frac{1}{N^2} \sum_{i,k = 1}^N \bk(\u_i,\u_k)
- \frac{2}{Nn} \sum_{i = 1}^N\sum_{k = 1}^n\bk(\u_i,\bm{\zeta}_k)
+ \frac{1}{n^2} \sum_{i,k = 1}^n \bk(\bm{\zeta}_i,\bm{\zeta}_k).
\end{align}
Here $D_{\bk}^2 (\ce,\cd)$ is indeed a squared norm $\| F_{\cd} - F_{\ce} \|_{\bk}^2$ induced by $\bk$. When the data set $\ce$ contains a large sample of points generated from the uniform distribution on $C^s$, $D_{\bk}^2 (\ce,\cd)$ becomes the empirical form of the squared generalized $\ell_2$-discrepancy. % $D(\cp; F_{\mathrm{u}},\bk)$, similar to (\ref{uniform_disc})
% with the uniform CDF $F_{\mathrm{u}}$ being replaced by the ECDF of $\ce$, $F_{\ce}$.
In this view, $D_{\bk}^2 (\ce,\cd)$ assesses the difference between
$\ce$ and $\cd$ in terms of their  empirical CDFs.  Then we obtain the following lemma
of the empirical form of (\ref{KH}) on the unit hypercube $C^s$.

\begin{lemma}[Empirical Koksma-Hlawka Inequality on $C^s$] \label{EKH}
Given two point sets $\ce$, $\cd \subset C^s$, a reproducing kernel $\bk$ on $C^s \times C^s$,
and the function $f$ with the bounded ${\ell}_2$-variation $V_2(f,\bk)$ on $C^s$, we have
%\begin{align*}
%\left| \frac 1N \sum_{\x \in \ce} f(\x) - \frac 1n
%\sum_{\z \in \cp} f(\z)\right| \leq
%D_{\bk}(\ce,\cp) V_2(f,\bk) ,\end{align*}
$$\left| \frac 1N
\sum_{\u \in \ce} f(\u) - \frac 1n\sum_{\bzeta \in \cd} f(\bzeta)\right| \leq
%\sum_{i = 1}^N f(\u_i) - \frac 1n\sum_{k = 1}^n f(\bm{\zeta}_k)\right| \leq
D_{\bk}(\ce,\cd) V_2(f,\bk) ,$$
where $D_{\bk}(\ce,\cd)$ is given by (\ref{DK}).
\end{lemma}

%The proof of Lemma \ref{EKH} can be find in the Appendix.
%Lemma \ref{EKH} indicates that the difference of averaged function values between different data sets is upper bounded by the product of the $L_2$-variation of such function and the difference of the two data sets measured by (\ref{DK}).

Note that for any two data sets $\cx=\{\x_i, i = 1, \dots, N\} \subset \br^s$,
and $\cp=\{\bm{\xi}_k, k = 1, \dots, n\} \subset \br^s$, through the
transformation $T_{\cx}$ in $(\ref{F_tra})$, it is easy to see that
$T_{\cx}(\cx) \subset C^s$ and $T_{\cx}(\cp) \subset C^s$.
According to the analytical expression of GEFD in (\ref{simple_GEFD}), we have
that $D^2(\cp;\cx,\bk)= D_{\bk}^2(T_{\cx}(\cx),T_{\cx}(\cp))$.
By combining Lemma~\ref{EKH} and the equivalence between
$D(\cp;\cx,\bk)$ and $D_{\bk}(T_{\cx}(\cx),T_{\cx}(\cp))$,
%yields Theorem \ref{EKH_theorem} directly.
we can deduce the empirical Koksma-Hlawka inequality in terms of GEFD defined on $\br^s$.

\begin{theorem}[Empirical Koksma-Hlawka Inequality on $\br^s$] \label{EKH_theorem}
Given a %\sout{\revA{general  reference}}
reference data $\cx \subset \br^s$,
$\bk(u,v) = \prod_{j=1}^s K(u_j,v_j)$ is a reproducing kernel function on $C^s \times C^s$ and $f$ has a bounded ${\ell}_2$-variation $V_2(f,\bk)$ on $C^s$,
%a reproducing kernel function with a production form
%$\bk(u,v) = \prod_{j=1}^s K(u_j,v_j)$ on $C^s \times C^s$ and
%\revA{a $L_2$-variation $V_2(f,\bk)$ bounded function $f$ on $C^s$,}
then for any point set $\cp \subset \br^s$,
$$ %\label{EKH_2}
\left| \frac 1N \sum_{\x \in \cx} f(T_{\cx}(\x))
- \frac 1n \sum_{\z \in \cp} f(T_{\cx}(\z)) \right|
\leq D(\cp;\cx,\bk) V_2(f,\bk),$$
where $T_{\cx}$ takes the form of (\ref{F_tra}), and $D(\cp;\cx,\bk)$ is given by Definition~\ref{GEFD}.
\end{theorem}

%Theorem \ref{EKH_theorem} is a direct result of Lemma \ref{EKH} and
%%the equation (\ref{D=DK}) which characterizes
%combined with the equivalence between
%$D(\cp;\cx,\bk)$ and $D_{\bk}(T_{\cx}(\cx),T_{\cx}(\cp))$.

It is worth noting that in Theorem~\ref{EKH_theorem} the reference data $\cx$ on $\br^s$ is not required to satisfy the joint independence condition as in Theorem~\ref{asymptotic_theorem}. %because of the generality of data sets in Lemma \ref{EKH}.
%\revG{Moreover, Theorem \ref{EKH_theorem} %only presents the importance of a minimal GEFD value, and draws none attention to how to make it.}
%presents the importance of a design with a minimal GEFD value. }Moreover,
Theorem \ref{EKH_theorem} provides another rationale for
the proposed GEFD criterion used for measuring the closeness of $\cp$ to $\cx$.
%generalized empirical $F$-discrepancy and indicates that for a given $\cx$, the point set $\cp$
% which minimizes the GEFD value $D(\cp;\cx,\bk)$ has good representation with respect to $\cx$.

\section{Data-driven Subsampling}\label{Sect:DDS}

%\revB{[[[give some explanation of the words ``data-driven" and ``data-driven space-filling" here]]]}

%\revA{ The adjective data-driven means that progress in an activity is compelled by data, rather than by intuition or by personal experience\footnote{https://en.wikipedia.org/wiki/Data-driven.}.

% In this section, we propose a data-driven subsampling method that the whole subsampling progress is compelled by a given data, rather than by the model.

Given a reference data $\cx \subset \br^s$, we want to find a small data $\cp \subset \br^s$ such that $\cp$ has a good representation of $\cx$. As discussed in the previous section, the goodness of representation can be measured by the GEFD criterion.
%Theorem~\ref{EKH_theorem} gives the practical significance of the data-driven subsampling.
%Theorem \ref{asymptotic_theorem} gives a guidance on the approach of finding a data-driven space-filling design for a constrained given data set. Note that
Theorem~\ref{asymptotic_theorem} translates {the} GEFD of $\cp \subset \br^s$ to the generalized $\ell_2$-discrepancy  of $T_{\cx}(\cp)$ in $C^s$, subject to an approximation error of order $1/N^*$.
Therefore, if $\cx$ satisfies the joint independence assumption (\ref{ind}), a good design $\cd$ in $C^s$ with low generalized $\ell_2$-discrepancy could lead to a point set $\cp$ with low GEFD by the inverse transformation of $T_{\cx}$, i.e. $\cp = T_{\cx}^{-1}(\cd)$.
Such inversion method can be based on an existing low-discrepancy design from the rich library of uniform experimental designs (Fang et al. \cite{F06}) or low discrepancy sequences or nets (Niederreiter \cite{N92}),
%, and construction methods such as good lattice point method (Fang et al. \cite{F18}),
whereas one may also use the R:UniDOE package (Zhang et al. \cite{Z18}) for real-time construction of nearly uniform designs. %we can obtain the data-driven space-filling design $\cp$ with low GEFD.

\begin{figure}[tp!]
\centering
\vspace{3mm}\subfigure[Original Space]{
\includegraphics[width=0.32\textwidth]{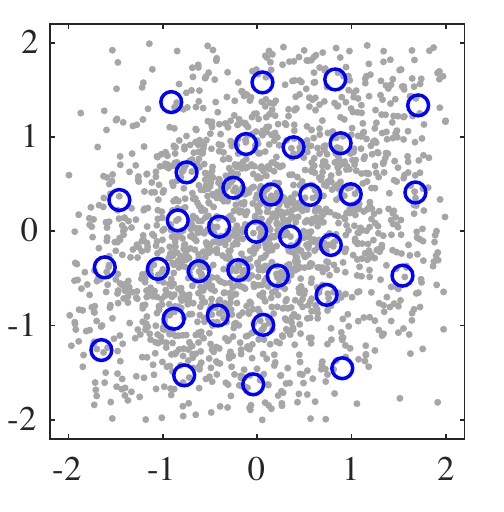}}\label{original}~~
\subfigure[Rotated Space]{
\includegraphics[width=0.32\textwidth]{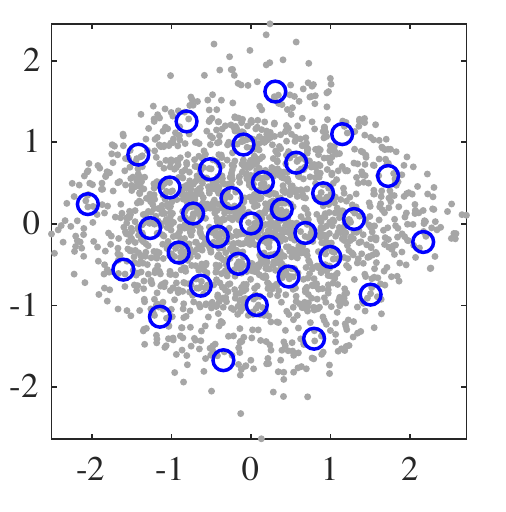}}\label{rotated}~~
\subfigure[Data-driven Subsampling]{
\includegraphics[width=0.32\textwidth]{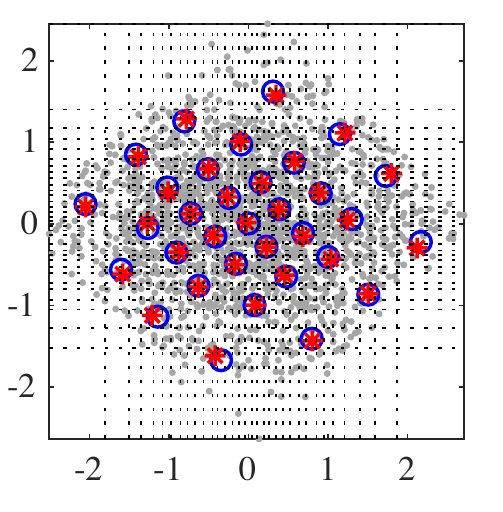}}\label{DDS}
\caption{ Data-driven space-filling design by the rotation-inversion
construction method and
data-driven subsampling by non-uniform stratification
and nearest neighbor on the rotated space.} \label{svd_fig}
\end{figure}

The joint independence assumption in (\ref{ind}) may be not satisfied for a real data set $\cx \subset \br^s$.  As a common practice, we can apply the statistical procedures such as the
principal component analysis (PCA) or independent component analysis (ICA)
to transform the reference data to a latent space, $\cz$.
For simplicity, we suggest to use the PCA approach in this paper, which is based on the singular value decomposition (SVD), to convert the data to have linearly uncorrelated coordinates. Based on the principal scores on the latent space, one can use the inversion method to find a small representation $\cq_{\cp}$ with low GEFD. Finally, such $\cq_{\cp}$ can be converted back to the original space as the desired small data representation of $\cx$ on $\br^s$.
The above procedure can be called the rotation-inversion construction, and it can be described more precisely in the following three steps: (i) performing SVD for $\cx$ to obtain the rotation $\V$, the singular-valued matrix $\Lambda$, and the rotated data $\cz$;
%(II) for each point $\bm{\zeta}_k \in \cd$, performing the $T_{\cz}^{-1}$ transformation through $$\bm{\eta}_{k} = T_{\cz}^{-1}(\bm{\zeta}_k) = \big(F_{\cz_{(1)}}^{-1}(\zeta_{k1}), \dots, F_{\cz_{(s)}}^{-1}(\zeta_{ks})\big),$$ to obtain the data-driven space-filling design $\cq_{\cp} = \{\bm{\eta}_k, k = 1, \dots, n\}$ in the space of $\cz$ ;
(ii) constructing the data-driven space-filling design $\cq_{\cp} = \{\bm{\eta}_k, k = 1, \dots, n\}$ in the space of $\cz$ by performing the $T_{\cz}^{-1}$ transformation on each $\bm{\zeta}_k \in \cd$ as follows,
\begin{equation}\label{T_inverse}
\bm{\eta}_{k} = T_{\cz}^{-1}(\bm{\zeta}_k) = \big(F_{\cz_{(1)}}^{-1}(\zeta_{k1}), \dots, F_{\cz_{(s)}}^{-1}(\zeta_{ks})\big)^T,\quad k = 1, \dots, n;
\end{equation}
(iii) generating the point set $\cp$ by $\bm{\xi}_k = \V \Lambda \bm{\eta}_k$ for each $k = 1, \dots, n$.
Such a procedure is computationally efficient for a large scale data in a low dimensional space. For an illustration, we apply the rotation-inversion construction to a two-dimensional reference data $\cx$ and obtain the new sampled points $\cp$, shown in Figure~\ref{svd_fig} (a) and (b). The original data $\cx$ is simulated from a truncated binormal distribution with correlation, same as that in Figure~\ref{def}. It can be found that such a small data $\cp$ (plotted as circles) represents the original data $\cx$ (plotted as dots) quite well.

%\revG{Algorithm \ref{Alg:SVD} provides the construction of data-driven
%space-filling designs under GEFD.
%\revB{It is simply rotating $\cx$ to $\cz$,
%then converting $\cp$ points by $T_{\cz}^{-1}$ transform,
%finally rotating back to the original space of $\cx$.[[[re-organize]]]}
%This is computationally efficient for big data in low dimensional space.
%For an illustration,
%we apply Algorithm \ref{Alg:SVD} to a two-dimensional data set and
%obtain the new sampling points, respectively shown as dot points
%and circle points in Figure \ref{svd_fig}.
%Clearly the data-driven space-filling design
%represents the underlying data \revA{very well}.}

%\begin{algorithm}[h!]
%\caption{Rotation-Inversion Construction}
%\label{Alg:SVD}
%\begin{algorithmic}[1]
%\REQUIRE
%Original data $\cx \subset \br^s$, a uniform design $\cd \subset C^s$.
%\STATE Perform SVD for $\cx$ to obtain the rotation $\V$, the
%singular-valued matrix $\Lambda$, and the rotated data $\cz$;
%\STATE For each point $\bm{\zeta}_k \in \cd$,
%perform the $T_{\cz}^{-1}$ transformion
%$$\eta_{kj} = F_{\cz_{(j)}}^{-1}(\zeta_{kj}), j = 1, \dots, s;$$
%\STATE Generate the point set $\cp$ by $\bm{\xi}_k = \V \Lambda \bm{\eta}_k$
%for each $k$.
%\ENSURE
%Data-driven space-filling design $\cp$.
%\end{algorithmic}\end{algorithm}

For the subsampling purpose, it is required that the obtained small data $\cp$ should be a subset of the original data $\cx$.
%Note that throughout the notation-inversion construction, the first step is simple but significant. Since it rotates $\cx$ to $\cz$ with the goal that, $\cz$ satisfies the joint independence in (\ref{ind}) to meets the demand of Theorem \ref{asymptotic_theorem}.
%However the coordinates of $\cz$ are only linearly uncorrelated after the first step, therefore $\cz$ could not attain the joint independence unless it is generated from the normal distribution of the correlation coefficient matrix as a diagonal matrix.
%With the randomness and complexity of the practical data set, $\cz$ may only be nearly independence. In this situation,
Since the rotated data $\cz$ is only nearly independent,  there exist some points
in $\cq_{\cp}$ not belonging to $\cz$. %i.e., elements of $\cq_{\cp} \backslash \cz$.
%It leads to that
%Then the obtained data-driven space-filling design $\cp$ through the procedure may not be a subset of $\cx$.
%To guarantee the subdata acquirement,
For each of such data point, we suggest to find its nearest neighbor in $\cz$ as a replacement. In Figure \ref{svd_fig} (c), the points of $\cq_{\cp}$ are shown as {the} circle points, and their nearest points from $\cz$ are shown as {the} star points.
%However, it is precisely b
%Because of the huge amount of $\cz$, the computation complexity becomes a crucial issue in the nearest searching process.
{To find the nearest neighbor of each point in $\cq_{\cp} \backslash \cz$, KD-tree is the conventional methods. However, the time complexity of constructing a KD-tree is $\mathcal{O}\left(PN\log N\right)$, the time complexity of searching the nearest neighbor for a design point is $\mathcal{O}\left(N^{\left(p-1\right)\slash p}\right)$. This can be very time consuming when the sample size N is very large, which is common in the big data era. Therefore, KD-tree cannot be applied to this case directly, it is critical to develop an effective subdata selection algorithm.}

%To address this issue,
%\revA{Zhang et al. \cite{Z19} suggested employing
%a non-unfiorm stratification on the whole search space to
%narrow the search scope to some small subsets determined by a radius.}
%To address this issue,
One way is to limit the search space for each point in $\cq_{\cp} \backslash \cz$.
We introduce a non-uniform stratification into the search step. Given the size of subdata $n$, by the following transformation for every coordinate $j$,
\begin{equation}\label{cut}
\kappa_{kj} = F_{\cz_{(j)}}^{-1}(k/n), ~k = 1, \dots, n-1,
\end{equation}
so each coordinate of the rotated space is divided into $n$ partitions.
It is approximately balanced when each partition has $\lfloor N/n \rfloor$ or $\lceil N/n \rceil$ points.
%The $n$ partitions are marked with $1, \dots, n$ along the increasing direction to indicate their position.
In fact, in (\ref{cut}), the obtained $\kappa_{1j}, \dots, \kappa_{n-1,j}$ are the quantiles of $\cz_{(j)}$. Denote $\kappa_{0j} = \min  \cz_{(j)}$, $\kappa_{nj} = \max  \cz_{(j)}$, and the $k$th partition as $\big(\kappa_{k-1,j},\kappa_{kj}\big]$ for $k = 1, \dots, n$. Then any value $z \in [\kappa_{0j}, \kappa_{nj}]$ locates in the partition marked with $\lceil nF_{\cz_{(j)}}(z) \rceil$. Conversely,  any $\zeta \in [0,1]$ can be converted into a value belonging to the $(\lceil n\zeta \rceil)$th partition of $ [\kappa_{0j}, \kappa_{nj}]$. These cuts in the $s$ coordinates altogether define a non-uniform stratification grid of the rotated space. See Figure \ref{svd_fig} (c) for such non-uniform grid (formed by the dotted lines). The whole rotated space is divided into the $n^s$ different cells within the grids. Label each cell with an $s$-dimensional index vector to distinguish different cells.
%Here the $j$th element of the cell index vector is determined by the partition mark along the $j$th coordinate of the points in this cell. For example, the index vector of the bottom left cell in Figure \ref{svd_fig} (c) is $(1,1)^T$.
%Each observation $\z_{i} = (z_{i1}, \dots, z_{is})^T \in \cz$ locates in the cell with index vector $$I_i =  (\lceil nF_{\cz_{(1)}}(z_{i1}) \rceil, \cdots, \lceil nF_{\cz_{(s)}}(z_{is}) \rceil)^T, ~~ i = 1, \dots, N.$$
%
%For a $n$-level uniform design $\cd = \{\bm{\zeta}_k, k = 1, \dots, n\}$ in $C^s$, the $n$ values of its each factor are a permutation of $1/(2n), 3/(2n), \dots, (2n-1)/(2n)$. By using the $n$-level uniform design $\cd$ in constructing $\cq_{\cp}$, each $\bm{\eta}_k = T_{\cz}^{-1}(\bm{\zeta}_k) = \big(F_{\cz_{(1)}}^{-1}(\zeta_{k1}), \dots, F_{\cz_{(s)}}^{-1}(\zeta_{ks})\big)$ is located in one and only one of the $n^s$ cells.
%Denote this cell by Cell$_k$, and its index vector by $J_k$.
%Note that $\zeta_{kj} \in \{1/(2n), 3/(2n), \dots, (2n-1)/(2n)\}$ and $F_{\cz_{(j)}}^{-1}(\zeta_{kj})$ belongs to the partition marked with $(2n\zeta_{kj}+1)/2 = \lceil n\zeta_{kj} \rceil$ along the $j$th coordinate, so $J_k = (\lceil n\zeta_{k1} \rceil, \dots, \lceil n\zeta_{ks} \rceil$).
%From the process of labeling the index vector for a cell,
It is easy to justify that any point $\z = (z_{1}, \dots, z_{s})^T$ in the rotated space  is located in the cell indexed by $(\lceil nF_{\cz_{(1)}}(z_{1}) \rceil, \cdots, \lceil nF_{\cz_{(s)}}(z_{s}) \rceil)^T$, which helps to locate $\z$ in this non-uniform stratification.
% and denote it as  the cell index vector of $\z$.
% In addition, by performing transformation (\ref{T_inverse}), any point $\bm{\zeta} = (\zeta_1, \dots, \zeta_s)^T \in C^s$ can be converted to a point having the cell index vector $(\lceil n\zeta_1 \rceil, \cdots, \lceil n\zeta_s \rceil)^T.$
To this end, for each $\z_i$ in $\cz$ we find its cell index vector $I_i =  (\lceil nF_{\cz_{(1)}}(z_{i1}) \rceil, \cdots, \lceil nF_{\cz_{(s)}}(z_{is}) \rceil)^T$.

\begin{algorithm}[t]
	\caption{Data-driven Subsampling}
	\label{Alg:DDS}
	\begin{algorithmic}[1]
		\REQUIRE
		The original data $\cx  = \{\x_1, \dots,\x_N\} \subset \br^s$, the neighboring radius $\tau$ and a uniform design $\cd = \{\bm{\zeta}_1, \dots, \bm{\zeta}_n\} \subset C^s$.
		\STATE \label{step:svd}
		%Perform SVD $\cx = \cz \Lambda \V^T$ for $\cx$ to obtain the rotation $\V$, the singular-valued matrix $\Lambda$, and the rotated data $\cz$;
		Perform SVD $\cx = \cz \Lambda \V^T$ for $\cx$ to obtain the rotated data $\cz  = \{\z_1, \dots,\z_N\}$ whose components are linearly uncorrelated;
		\STATE \label{step:stratify}
		Implement non-uniform stratification to the rotated space, and label the $n^s$ cells with the corresponding index vectors. Then for each $i  = 1, \dots, N$, the point $\z_i  = (z_{i1}, \dots, z_{is})^T \in \cz$ falls into the cell with index vector $I_i =  (\lceil nF_{\cz_{(1)}}(z_{i1}) \rceil, \cdots, \lceil nF_{\cz_{(s)}}(z_{is}) \rceil)^T;$
		\STATE \label{step:inverse}
		Apply the inverse transformation (\ref{T_inverse}) on the uniform design $\cd$ to obtain the data-driven space-filling design $\cq_{\cp} = \{\bm{\eta}_1, \dots, \bm{\eta}_n\}$ in the rotated space. Then for each $k = 1, \dots, n$, the design point $\bm{\eta}_k = T_{\cz}^{-1}(\bm{\zeta}_k) = \big(F_{\cz_{(1)}}^{-1}(\zeta_{k1}), \dots, F_{\cz_{(s)}}^{-1}(\zeta_{ks})\big)^T$ is located in the cell with index vector $J_k =  (\lceil n\zeta_{k1} \rceil, \cdots, \lceil n\zeta_{ks} \rceil)^T;$
		%For each $k = 1, \dots, n$, the data-driven space-filling design point $\bm{\eta}_k$ obtained by performing the inverse transformation (\ref{T_inverse}) on $\bm{\zeta}_k$ locates in the cell with index vector $$J_k =  (\lceil n\zeta_{k1} \rceil, \cdots, \lceil n\zeta_{ks} \rceil)^T;$$
		\STATE \label{step:search}
		For each $\bm{\eta}_k$, identify the points located in $\bm{\eta}_k$'s neighboring cells with their index vectors in the range of $\tau$ from $J_k$ in the version of ${\ell}_{\infty}$-norm, and find out the nearest sample index $i_k^*$, i.e. $i_k^* = \min_{ \left\{i:~\parallel I_i - J_k \parallel_{\infty} \leq \tau \right\} } \parallel \z_i - \bm{\eta}_k \parallel_2, \quad k = 1, \dots, n.$
		\ENSURE
		Data-driven subsample $\cp^{\dagger} = \{\x_{i_1^*}, \dots, \x_{i_n^*}\}$.
\end{algorithmic}\end{algorithm}
For an $n$-level uniform design $\cd = \{\bm{\zeta}_k, k = 1, \dots, n\}$ in $C^s$,
all its entries locate at the center of the bins of $(0,1/n],\dots,((n-1)/n,1]$.
%the $n$ values of its each factor are a permutation of $1/(2n), 3/(2n), \dots, (2n-1)/(2n)$.
By using $\cd$ to construct $\cq_{\cp}$, each $\bm{\eta}_k = T_{\cz}^{-1}(\bm{\zeta}_k) = \big(F_{\cz_{(1)}}^{-1}(\zeta_{k1}), \dots, F_{\cz_{(s)}}^{-1}(\zeta_{ks})\big)^T$ is located in one and only one of the $n^s$ cells.
Denote this cell by Cell$_k$, and its index vector by $J_k$. Obviously $J_k = (\lceil n\zeta_{k1} \rceil, \dots, \lceil n\zeta_{ks} \rceil)^T$ according to the process of labeling the index vector for a cell. Meanwhile, $\bm{\eta}_k$ is also labeled by $J_k$.
Then we can find a sampling point indexed by $i_k^*$ from
$\{i : \z_i \in \mathrm{Cell}_k \cap \cz\}$  that is nearest to $\bm{\eta}_k$.
Note that when the condition of the joint independence (\ref{ind}) is not satisfied,
there could exist cells consisting none of  points from $\cz$.
%observations.
Then search region may be increased to the neighboring cells
around $\bm{\eta}_k$.
%With regard to the determination method of neighbor cells,
%Here each neighboring cell is determined based on its index vector
%such that the
Since the non-uniform stratification in (\ref{cut}) is carried out  component by component independently,
each neighboring cell is determined by a radius $\tau$ in the meaning of ${\ell}_{\infty}$-norm of the difference between its index vector $J$ and $J_k$.
It is equivalent to that the $J$ is satisfied with $||J_k - J||_{\infty} \leq \tau$.
Then the points to be searched are locating in the cells
with index vectors belonging to the $s$-dimensional rectangular
$\big[\lceil n\zeta_{k1} \rceil - \tau, \lceil n\zeta_{k1} \rceil +\tau\big]
\otimes \dots \otimes \big[\lceil n\zeta_{ks} \rceil - \tau,
\lceil n\zeta_{ks} \rceil +\tau\big]$.
%\revC{here each neighboring cell is determined through a radius $\tau$ such that the its index vector $J$ is in the meaning of $\infty$-norm of the difference between its index vector $J$ and $J_k$. }
Therefore, there is always a nearest neighboring sample from $\cz$
that can be found for each $\bm{\eta}_k$.
%\revA{Finally, the searched data set of nearest points corresponds to a unique subdata set $\cp^{\dagger}$ in $\cx$.}
%(or equivalently nearest neighboring sample from $\cx$ for each design point $\bm{\xi}_k$).

We summarize the above procedure by Algorithm \ref{Alg:DDS}.  As an example, it can output the subsampled points shown in  Figure~\ref{svd_fig} (c).
For each design point (circle points) in the rotated space, a corresponding sample (star annotation) is found from $\cz$, which is closest the ideal design point. It is anticipated that
such kind of subdata $\cp^{\dagger}$ is similar to the designed $\cp$ and has low GEFD value.

\section{Accelerated Subsampling Procedure}
\label{Sect:ADDS}

In this section, we present an approximate approach to speed up the subsampling algorithm, which is useful for the implementation in high-dimensional cases.
%Except that, a
A parallel scheme based on sliced uniform designs is provided to deal with the complex situation where the storage of the original full data is decentralized by multiple servers.
% and impractical to be combined together.

Note that the essence of Algorithm \ref{Alg:DDS} is to partition the rotation space into $n^s$ intensive cells firstly, and to conduct the nearest searching process in the neighboring cells
dynamically determined by each point in $T_{\cz}^{-1}(\cd)$ with a certain radius $\tau$  in the meaning of ${\ell}_\infty$-norm.
For the high dimensional case, %and a moderate subsample size $n$,
the whole space is divided into the $n^s$ cells, and a relatively large radius $\tau$ is needed to guarantee the union of neighboring cells is nonempty.
However if $\tau$ is too large, it is still very time consuming to identify the points that belong to the target neighboring cells and further to search for the nearest sample index from the original data. Moreover,  it is not easy to find an appropriate radius to determine the neighboring cells.
% as well as identify the points belonging to the neighboring cells.

An alternative stratification strategy is to use a gross grid in the starting step.
For each coordinate $j$, we divide the $j$th coordinate of the rotated space into the $m$ partitions with the nodes $\{F_{\cz_{(j)}}^{-1}(1/m), \dots ,F_{\cz_{(j)}}^{-1}((m-1)/m)\}$
such that each partition has the
$\lfloor N/m \rfloor$ or $\lceil N/m \rceil$ points.
Then the whole space is divided into the $m^s$ blocks.
%Different from distinguishing different cells by their $s$-dimensional index vectors in Algorithm \ref{Alg:DDS}, here
Each block is labeled with a unique code in lexicographic order. Then each  point in $\cz$  can be labeled by the code of the block which the point belongs to. In this way, all the points are divided into at most the $m^s$ groups by their block codes.
%Label each block with a unique code in lexicographic order.
For each $\bm{\eta}_k\in T_{\cz}^{-1}(\cd)$, there is one and only one of the blocks with block code $C_k\in\{1,\dots,m^s\}$ containing $\bm{\eta}_k$ and the search space is narrowed to one of the $m^s$ blocks which consists of all the points with the same block code $C_k$.
%Since the stratification strategy uses no information about $T_{\cz}^{-1}(\revC{\cd})$, identifying the points which share the same block code is easier than identifying the points belonging to the neighboring cells determined by a radius in the high dimensional case.
The computational complexity of searching for a required code number among $N$ codes is much lower than that of searching for several different satisfied $s$-dimensional index vectors from $N$ cell index vectors, especially when $N$ is huge and $s$ is large.
It implies that the process of identifying the points labeled by their block codes accelerates Step \ref{step:search} in Algorithm \ref{Alg:DDS} in the %high dimensional.
high-dimensional cases.
We call this procedure the accelerated data-driven subsampling (ADDS) method. %\revB{[[[Divide it into two paragraphs]]]}

Compared with the DDS algorithm, ADDS adapts to a more gross non-uniform stratification. This stratification strategy uses no information about $T_{\cz}^{-1}(\cd)$. Moreover, since the labeling method is based a block code rather than an $s$-dimensional cell index vector,  it accelerates the speed of nearest search process.
See Algorithm \ref{Alg:ADDS} for the concrete procedure of ADDS.
%\revB{accelerated data-driven subsampling (ADDS) [[[give more explanation of the words ``ADDS", why call it as an ADDS?]]]} method.

\begin{algorithm}
\caption{Accelerated Data-driven Subsampling}
\label{Alg:ADDS}
\begin{algorithmic}[1]
\REQUIRE
The original data $\cx \subset \br^s$, a block parameter $m$ and a uniform design $\cd = \{\bm{\zeta}_k, k = 1, \dots, n\} \subset C^s$.
\STATE Perform SVD $\cx = \cz \Lambda \V^T$;
\STATE Parallel for each point $\z_i  = (z_{i1}, \dots, z_{is})^T\in \cz$,
label its block code in lexicographic order as
$$S_i =  (\lceil mF_{\cz_{(1)}}(z_{i1}) \rceil - 1, \cdots,
\lceil mF_{\cz_{(s)}}(z_{is}) \rceil -1)
\cdot(m^{s-1}, \dots, m^0)^T+1, ~~ i = 1, \dots, N.$$

\STATE Parallel for each $\bm{\eta}_k = T_{\cz}^{-1}(\bm{\zeta}_k)$ in
$T_{\cz}^{-1}(\cd)$, label its block code in lexicographic order as
$$C_k =  (\lceil m\zeta_{k1} \rceil - 1, \cdots,
\lceil m\zeta_{ks} \rceil - 1)\cdot
(m^{s-1}, \dots, m^0)^T+1, ~~ k = 1, \dots, n.$$
\STATE Parallel for each $\bm{\eta}_k \in T_{\cz}^{-1}(\cd)$,
find the nearest sample index $i_k^{**}$ from those sharing the same
block code, i.e. $\{ i: S_i = C_k\}$.
\ENSURE
Data-driven subsample
$\cp^{\dagger\dagger} = \{\x_{i^{**}}, k = 1, \dots, n\}$.
\end{algorithmic}\end{algorithm}

\begin{remark}\label{modify}
For the block parameter $m$, a small integer is enough
especially when the dimension of $\cz$ is high
($m = 2$ or $3$ is enough for the high dimensional case).
Thus for each $\bm{\eta}_k$, its subordinate block is usually nonempty.
If unfortunately, a slightly bigger $m$ causes that
some $\bm{\eta}_{k_0}$ locates in an empty block,
we could conduct the nearest searching step among
at most $2m$ closest blocks with the block codes
$C_{k_0} \pm m^{s-1}, \dots, C_{k_0} \pm m^0$.
Then the nearest searching step for $\bm{\eta}_{k_0}$
in Algorithm \ref{Alg:ADDS} can be modified as
``find the nearest sample index $i_{k_0}^{**}$ from
$\{ i: S_i \in \{C_{k_0} \pm m^{s-1}, \dots, C_{k_0} \pm m^0 \}\}$".
\end{remark}

\begin{figure}[t!]
\centering
\vspace{3mm}\subfigure[DDS with $\tau = 1$]{
\includegraphics[width=0.32\textwidth]{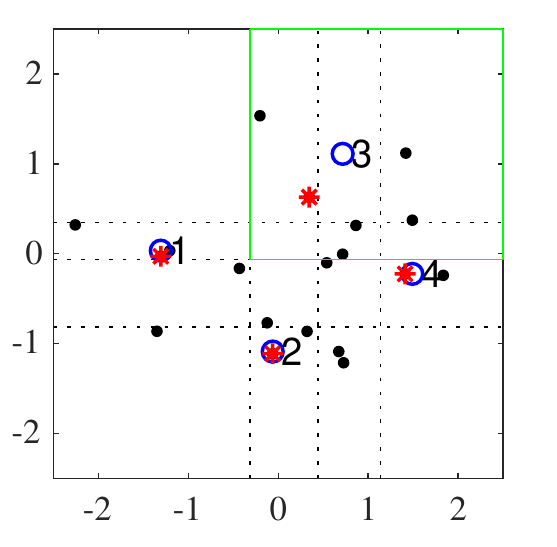}}\label{eg3_1}~~
\subfigure[ADDS with $m = 3$]{
\includegraphics[width=0.32\textwidth]{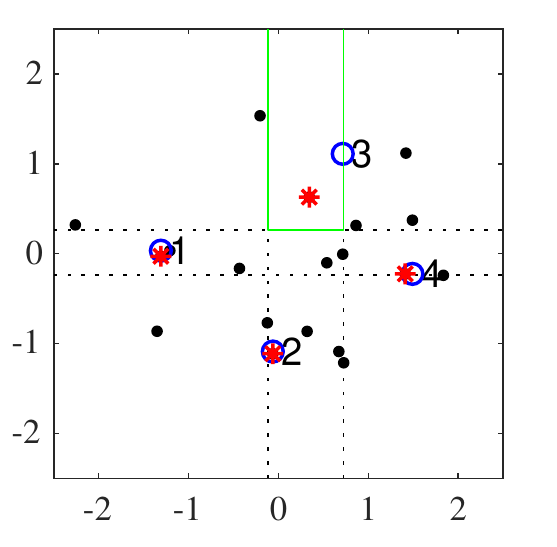}}\label{eg3_2}~~
\subfigure[ADDS with $m = 4$]{
\includegraphics[width=0.32\textwidth]{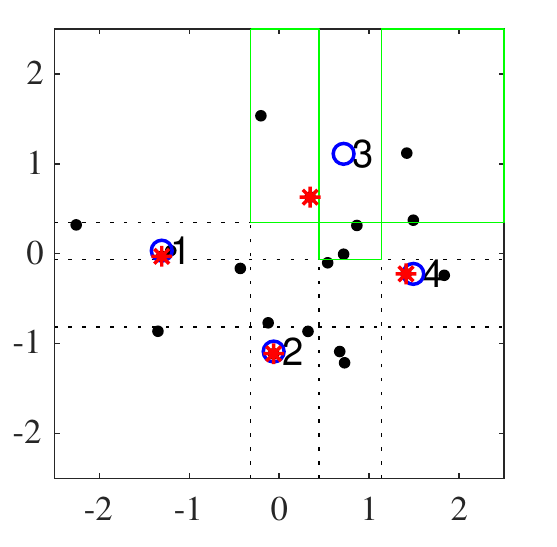}}\label{eg3_3}
\caption{An simple example to illustrate the difference between
DDS and ADDS.
The dotted lines give the non-uniform stratification for each coordinate;
The rectangles bounded by solid lines in (a), (b) and (c)
are respectively the nearest search scopes for the point ``3"
in DDS with the radius $\tau = 1$, and
ADDS with the block parameter $m = 3$ and $4$, respectively;
The star points in each subfigure are the corresponding
obtained subdata.}
\label{eg_Compare}
\end{figure}

Now we present the difference between the DDS and ADDS on the
nearest searching step through a simple case of $s = 2$.
$\cd = \{(1/8, 5/8), (3/8,1/8),(5/8,7/8),(7/8,3/8)\}$
is a $4$-run uniform design in $C^2$.
Assume the whole space $\cz$ consists of
the black points in Figure \ref{eg_Compare}.
Then the data-driven space-filling design
$T_{\cz}^{-1}(\cd)$ based on $\cd$
is the set of circle points in Figure \ref{eg_Compare}.
Consider the point with notation ``3" in $T_{\cz}^{-1}(\cd)$
as an example to show the different searching scopes for the DDS and ADDS.
For the DDS, each coordinate of $\cz$ is divided into $4$ partitions by
the dotted lines, see Figure \ref{eg_Compare} (a).
%Then the point ``3" is located in an empty cell in the non-uniform stratification. So a radius $\tau = 1$ is set and the neighboring cells of the point ``3" consist of the black points in the rectangle bounded by solid lines.
The point ``3" is located in the cell with index vector $(3,4)^T$. But this cell does not contain any point of $\cz$. Then a radius $\tau = 1$ is set and the neighboring cells to be searched have the index vector $(id_1, id_2)^T$ satisfying $\max\left\{\left|id_1 - 3\right|, \left|id_2 - 4\right|\right\} \leq 1$. Thus the nearest search process is implemented among the black points in the rectangle bounded by solid lines.
For the ADDS with $m = 3$,
the corresponding nearest search scope is the block that
contains the point ``3", i.e. the rectangle bounded by solid lines
in Figure \ref{eg_Compare} (b).
If a bigger $m = 4$ is implied,
i.e. the same non-uniform stratification with that in DDS,
the block containing the point ``3" contains none of points of $\cz$,
according to the modification recommended in Remark \ref{modify},
the nearest search scope for this point is the union of the points
contained in the three rectangles bounded by solid lines
in Figure \ref{eg_Compare} (c).
The final searched subdata sets are the star points in each
subfigure of Figure \ref{eg_Compare}.
In these example, the three searched subdata sets contain
the same points of $\cz$. {Note that the KD-tree algorithm can be applied to search the nearest sample index sharing the same block code. We can constructed a KD-tree for each block generated by ADDS with time complexity to be $\mathcal{O}\left(P\left(N\slash m^p\right)\log\left(N\slash m^p\right)\right)$ and search a nearest neighbour from the design with time complexity to be $\mathcal{O}\left(P\left(N\slash m^p\right)^{\left(p-1\right)\slash p}\right)$.}

%\begin{figure}[h!]
%\centering
%\vspace{3mm}
%\includegraphics[width=0.9\textwidth]{eg_Compare2}
%\caption{An simple example to illustrate the difference between
%DDS and ADDS.
%The dotted lines give the non-uniform stratification for each coordinate;
%The rectangles bounded by solid lines in the left, center
%and right panel are respectively the
%nearest search scopes for the point ``3"
%in DDS with the radius $\tau = 1$, and
%ADDS with the block parameter $m = 3$ and $4$, respectively;
%The star points in each subfigure are the corresponding
%obtained subdata.}
%\label{eg_Compare}
%\end{figure}

To further certify the performance of ADDS intuitively,
we consider the case of $s = 10$,
where the full data set $\cx$ is randomly generated from
the $s$-dimensional uniform distribution
with independent components each on [0,1].
Let the data size
$N = 10^4, 10^5$,$10^6$ and $10^7$, respectively.
Consider the data-driven subsampling procedure with
$\tau = 750, 550, 400$ and $300$ respectively
for the different $N$
according to Algorithm \ref{Alg:DDS},
and the %accelerated data-driven subsampling
{ADDS} procedure
with block parameters $m = 2$,
to obtain subsamples with size $n = 2000$.
Table \ref{time} shows the computing time for each $N$
carried out on a {server} Intel(R) Xeon(R) with CPU E5-2650 v4 and 2.20 GHz.
It can be seen that the accelerated approach significantly
speeds up the subsampling procedure.

\begin{table}[h!]
\centering
\caption{CPU seconds with $n = 2000$, $s=10$ and
different full data size $N$.}
\label{time}
\setlength{\tabcolsep}{15pt}
\begin{tabular}{ccrrr}\hline%\noalign{\smallskip}
\multirow{2}{*}{Method} & \multicolumn{4}{c}{$N$}
\\\cline{2-5} %\noalign{\smallskip}
& $10^4$ & \multicolumn{1}{c}{$10^5$} & \multicolumn{1}{c}{$10^6$}
& \multicolumn{1}{c}{$10^7$}\\\hline%\noalign{\smallskip}
DDS & \multicolumn{1}{r}{0.2668} & 1.2329 & 11.4039  & 143.5468
\\\hline%\noalign{\smallskip}
ADDS & \multicolumn{1}{r}{0.0998}  &  0.2930 & 2.6088 & 27.6287
\\\hline
\end{tabular}\end{table}

In practical application, we may encounter more complex situation where the storage of the initial data decentralizes in different servers because of memory constraint. In this case, it is difficult to obtain a subdata having good presentation with respect to this type of decentralized initial data. It seems impractical to combine the decentralized parts of the full data together to apply the DDS or ADDS based on a general uniform design directly. To deal with this problem, we use the idea of divide-and-conquer. Assume the initial data contains ${L}$ decentralized parts with the size of $N_{(1)}, \dots, N_{({L})}$, and the size of subdata sets to be subsampled from ${L}$ parts are $n_{(1)}, \dots, n_{({L})}$ according to the proportion of $N_{(1)}, \dots, N_{({L})}$. The ${L}$ decentralized parts can be seen as the ${L}$ resulting parts of the objectively dividing of the whole initial data. Then treat every part as the full data in Algorithm \ref{Alg:DDS} or \ref{Alg:ADDS} to perform the DDS or ADDS. Finally combining the ${L}$ obtained subdata sets from the ${L}$ parts acquires the subsample of the initial data.
%\revG{It should be noted that the $\ell$ parts are not conducted following Algorithm \ref{Alg:DDS} and \ref{Alg:ADDS} independently. Since what we want to study is the initial whole data, and the decentralized mode has been established rather than obtained after analyzing the total data. Here the relationship between the initial data and the decentralized parts is reflected through the uniform designs used for each parts.}
The $n_{(1)}$-, $ \dots, n_{({L})}$-run small uniform designs used as the input terms in DDS or ADDS
%\sout{\revG{are no longer the general uniform designs. They}}
have the following inherent connections: \vspace{-2mm}
 \begin{description}
   \item[(i)] The combination of the ${L}$ designs is an  $n$-level sliced uniform design $\cd$  on $C^s$; \vspace{-2mm}
   \item[(ii)] for each $l = 1, \dots, {L}$, the $\left(\sum_{k = 1}^{l}n_{(k-1)}+1\right)$th to $\left(\sum_{k = 1}^{l}n_{(k)}\right)$th run of $\cd$ corresponds to the $n_{(l)}$-run small uniform design $\cd_{(l)}$ used for performing DDS or ADDS for the $l$th parts;   \vspace{-2mm}
   \item[(iii)] for each $j = 1, \dots, s$ and $l = 1, \dots, {L}$, the $n_l$ entries of the $j$th factor in $\cd_{(l)}$ have exactly one element in each of the $n_{(l)}$ bins of $\left(0,1/n_{(l)}\right],\dots,\left((n_{(l)}-1)/n_{(l)},1\right]$.  \vspace{-2mm}
 \end{description}

\noindent Here, $n=n_1+\cdots+n_{{L}}$, and a sliced uniform design is the uniform design $\cd$ that can be partitioned into some slices $\cd_{(l)}, l = 1, \dots, {L},$  and each slice has good uniformity.
%The uniformity concerned here is required not only for the whole design $\cd$ but also for each slice $\cd_{(l)}$. Since there exist some relationships among different slices, the good combined uniformity of the $\cd$ and $\cd_{(l)}, l = 1, \dots, \ell,$ is needed.
%Therefore a combined uniformity measure should be adopt.
For measuring the uniformity of such a design, Chen et al. \cite{C16} used the combined centered ${\ell}_2$-discrepancy
%of $\cd$ and $\cd_{(l)}, l = 1, \dots, \ell,$ i.e. ,
to measure the uniformity of the sliced Latin hypercube designs where each slice has the same size. %Except that, in order to
Yuan et al. \cite{Y19} used a weighted average centered ${\ell}_2$-discrepancy to obtain uniform flexible sliced Latin hypercube designs of which the slices may have different run sizes.
%took a weighted average of the centered $L_2$-discrepancy of the whole design and that of each slice as the combined uniformity measure in their optimization algorithm.
Since the  mixture discrepancy is better than the centered ${\ell}_2$-discrepancy, we use the following
%Here replace the centered $L_2$-discrepancy with mixture discrepancy considering of the better properties.  The
weighted average mixture discrepancy
$${\rm WAMD}(\cd)  = \frac{1}{2}D(\cd;F_{\rm u}, \bk^M) + \sum_{l = 1}^{\ell} \frac{n_{(l)}}{2n} D(\cd_{(l)};F_{\rm u}, \bk^M)$$
to measure the uniformity of the sliced uniform designs.
%could be used as the combined uniformity measure.
Both the  flexible sliced Latin hypercube designs by Yuan et al. \cite{Y19} and the sliced Latin hypercube designs with arbitrary run sizes by Xu et al. \cite{X19} satisfy the above connection (iii).
The connections (i) and (ii) are reflected from the combined uniformity of the $\cd$ and $\cd_{(l)}, l = 1, \dots, \ell.$ Then, the sliced uniform designs can be obtained   by the optimization algorithm under the weighted average mixture discrepancy criterion.
%In fact, when $\cd$ satisfies
%The sliced Latin hypercube designs with arbitrary run sizes proposed by Xu et al. \cite{X19} could provide the $\ell$ small designs satisfying above relationships. However neither the full designs nor their slices can guarantee a good uniformity. At this time we
Given a sliced uniform design,
%\uwave{sliced uniform design}\revB{[[[give the detailed structure]]]},
we assign each slice to the corresponding part of the initial data, and conduct the DDS or ADDS in parallel to obtain the corresponding subdata of each part.
The final subsample obtained through the DDS or ADDS based on the sliced uniform design is the union of these subdata sets.

%
%For the more complex situation where the storage of
%the initial data decentralizes in different servers
%because of memory constraint,
%similar to the divide-and-conquer idea,
%we suggest the corresponding data-driven
%subsampling method based on the sliced uniform designs on $C^s$,
%where the whole design is a uniform design
%consisting of the same number of slices with that of servers.
%Each slice is also a small uniform design,
%of which the run size is determined by
%the volume distribution of the decentralized initial full data.
%After provided the sliced uniform designs,
%assign each slice to one storage of the initial full data, and
%conduct the data-driven subsampling method based on
%the corresponding slice to obtained a subsample of this
%part of the initial data in parallel.
%The final subdata obtained through the
%extended DDS method based on the sliced uniform designs
%is the union of these subsamples.

\section{Numerical Examples}\label{eg}

In this section, we show the model-free property of the subsampling method through both simulation data and a real case study.
When using Algorithm \ref{Alg:DDS}, we only select the  leading components that give no less than 85\% of variance explained in the rotated space.
This operation yields a lower dimensional rotated pace, and reduces the computational burden due to non-important components.

For the given subsample size $n$, and the determined dimension of the rotated pace $s$,
the $n$-run $s$-factor uniform design in Algorithm \ref{Alg:DDS} is constructed by the {leave-one-out}  good lattice point method with power generator.
%(see \cite{F18} for a comprehensive introduction).
%\revB{[[[give the brief introduction of the procedure of method]]]}
In this construction method, the generator vector is given by a positive integer $\alpha$ which satisfies that the great common divisor of $n+1$ and $\alpha$ is one and $\alpha, \alpha^2, \dots, \alpha^s$ are distinct.
Then for each $j = 1, \dots, s-1$, the remainders after dividing $\alpha^{j}, 2\alpha^{j}, \dots, n\alpha^{j}$ by $n+1$ are $n$ distinct integers.
Denote this generator vector by $\gamma_\alpha$, and it has a power form $(\alpha^{0},\alpha^{1},\dots,\alpha^{s-1})$.
%What's more, for each $j = 1, \dots, s-1$, the remainders after dividing $a^{j}, \dots, n*a^{j}$ by $n+1$ are $n$ distinct integers.
Then the corresponding design generated by $\gamma_\alpha$ is $\cd^{(\alpha)} = \{\bm{\zeta}_i^{(\alpha)}, i = 1,\dots,n\}$, where $\bm{\zeta}_i^{(\alpha)} = \mathrm{mod}\left(i\gamma_\alpha,n+1\right)/n-1/(2n)$ with $\mathrm{mod}\left(\cdot,\cdot\right)$ denoted as the modulus operation.
%Denote the corresponding design generated through $\gamma_\alpha$ by $\cd^{(\alpha)} = \{\bm{\zeta}_i^{(\alpha)}, i = 1,\dots,n\}$ with $\bm{\zeta}_i^{(\alpha)} = \mathrm{mod}\left(i\gamma_\alpha,n+1\right)/n-1/(2n)$.
Different values of $\alpha$ lead to different designs. %Here we use the mixture discrepancy because of its better properties as the uniformity criterion to compare the uniformity of different designs.
We use the mixture discrepancy as the uniformity criterion.
Finally, the uniform design $\cd$ constructed by the {leave-one-out} good lattice point method with power generator is the one which owes the smallest mixture discrepancy value among all possible $\cd^{(\alpha)}$.
This construction procedure provides a fast and effective method for selecting the $n$ design points from $n^s$ lattice points.
The designs constructed by
the {leave-one-out} good lattice point method with power generator
are indeed approximate uniform designs.
For comparison, another simplest model-free subsampling method
URS is also implemented.
To make the different subsampling methods comparable,
for each given subsample size $n$,
the implementation of URS is executed for $R$ repetitions
to obtain the mean, median and bounds of the results of
the URS's subsample.
%Note that, the designs constructed by
%the  leave one out  good lattice point method with power generator
%are indeed approximate uniform designs.
%In order to make the subdata obtained through DDS could represents the initial data well, the points of $\cd$ should scatter on $C^s$ as uniformly as possible. In addition, there exists an approximation error between the GEFD of the obtained subdata with respect to the initial data and the mixture discrepancy of $\cd$. To reduce the influence resulted from these mentioned aspects, here the random shift in the sense of modulo 1 is adopted to $\cd$. By this way, after
Moreover we use a random shift of $\varepsilon \in C^s$ for all design points in $\cd = \{\bm{\zeta}_k, k = 1, \dots, n\}$, to obtain another approximate uniform design $\cd_{\varepsilon} = \{\mathrm{mod}(\bm{\zeta}_k+\varepsilon,1), k = 1, \dots, n\}$. Performing this random shift for $R$ repetitions results $R$ approximate uniform designs. Then we also obtain the mean, median and bounds of the results corresponding to DDS.
%is adopted here}
%the random shift in the sense of modulo 1
%is adopted here for $I$ repetitions
%which results $I$ approximate uniform designs, to
%obtain the mean, median and bounds of the results
%corresponding to DDS.
%\revB{[[[give more explanation for this procedure]]]}

\subsection{Simulation Studies}
In this subsection, we utilize the
data-driven subsampling method
for both classification and regression problems.
For each kind of model, we compare the prediction property of
the trained models based on the subdata sets by different subsampling strategies.
It will be shown that the proposed DDS method possesses the model-free property,
and outperforms URS method in various settings.

In each simulation,% with the size of the full data being $N$,
we generate two $N$-size data sets respectively as the full data
$\cx_{\text{Full}}$ and the test data $\cx_{\text{Test}}$
through the same generation way
as well as the corresponding binary class label
(and response in regression problem)
$\cy_{\text{Full}}$ and $\cy_{\text{Test}}$.
To measure the prediction performance of
a specified type of model trained upon
the $n$-size subdata $\cp$ %obtained by a certain subsampling method
from $\cx_{\text{Full}}$,
we fit a corresponding model using $\cp$ to predict
the class label (or response) for each point in $\cx_{\text{Test}}$.
Denote the fitted model by $\hat{h}(\cdot)$.
For each $\x \in \cx_{\text{Test}}$,
the corresponding predicted class label (or response) is $\hat{h}(\x)$.
%i.e. $f(\x)$ for each $\x \in \cx_{\text{Test}}$.
Then a prediction error could be computed by $\epsilon
= \frac 1N \left\| \hat{h}(\cx_{\text{Test}}) - \cy_{\text{Test}}\right\|_2^2$
which represents the misclassification rate for the classification problem
with class labels being $0$ or $1$,
and mean squared prediction error in the regression case,
upon the test data $\cx_{\text{Test}}$.
Denote this criterion by Err and MSPE respectively for the
classification and regression problem.
The lower value of this criterion, the better of the
prediction performance.

%\subsubsection*{(A) The assumed model is the true model.
%%True model is a logistic model
%}

\bigskip
\noindent {\bf (A) Working model is true.}\\
For the classification simulation, let $N = 10^4$ and both the data sets
$\cx_{\text{Full}}$ and $\cx_{\text{Test}}$
be generated from the multinormial distribution $\cn(\mathbf{0}, \bm{\Sigma})$,
where $\bm{\Sigma}_{ij} = 0.5^{I(i\neq j)}$, $i, j = 1, \dots, 7$.
The kind of data set is also used in Wang et al. \cite{W18}.

Logistic regression model is a classical and widely used
classification method for its simplicity and effectiveness.
Consider the logistic regression as a classification model.
Fortunately, if the underlying model is the logistic model, %i.e.
for example,
the probability of the class label being $1$ for a point $\x$ is
$h(\x,\bm{\beta}) = 1/\big(1+\exp(-\x^T\bm{\beta})\big),$
where $\bm{\beta}$ is a $7\times 1$ vector of $0.5$,
the optimal subsampling methods (mMSE and mVc) in Wang et al. \cite{W18} could be adopted.
Consider the four different kinds of subsampling strategies:
URS, mVc, mMSE
%(\uwave{the subsample size $n = r_0+r$ with
%$r_0 = 200$ as the number of points in the first step of
%mVc and mMSE suggested in Wang et al. \cite{W18})}
%\revB{[[[use another sentence to describe it]]]}
and DDS, to %subsample for size
obtain the subdata with
$n = 400, 550, 750, 950, 1200$ and $1500$, respectively.
The {implementations} of mVc and mMSE are represented by a two-step algorithm. The first step obtains $r_0$ points randomly, and the other $n-r_0$ points are obtained in the second step based on their optimal subsampling probabilities. For this type of data sets, Wang et al. \cite{W18} suggested $r_0 = 200$ to well present the good performance of mVc and mMSE. For each $n$, every subsample strategy is executed
$1000$ times because of the randomness.
Recall that, the randomness of DDS is reflected from the
random shift for a initial constructed approximate uniform design.

\begin{figure}[t!]
\centering
\subfigure[Logistic Regression Classification]{
\includegraphics[width=2.7in]{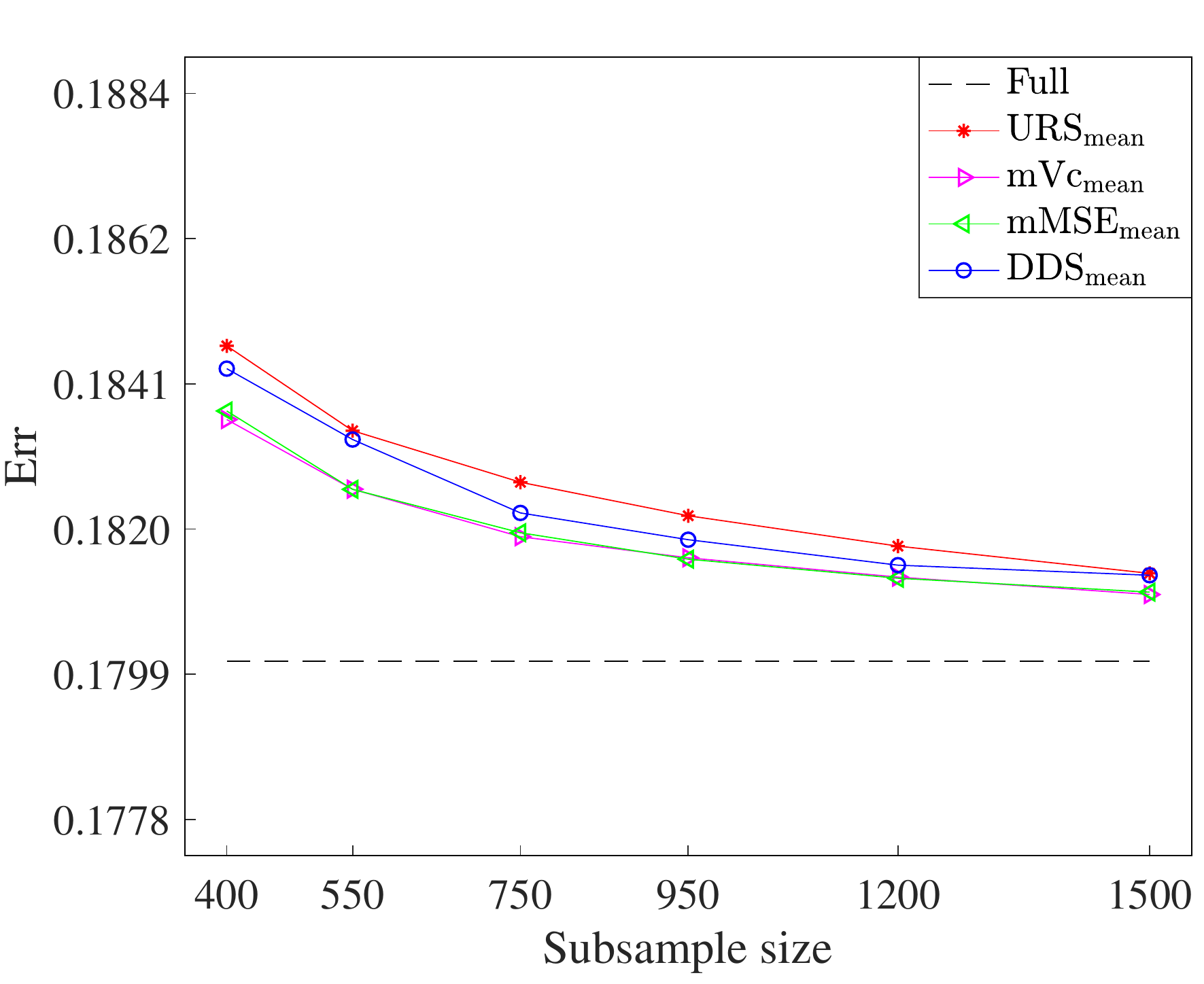}}
\hspace{3mm}
\subfigure[Logistic Regression Classification]{
\includegraphics[width=2.7in]{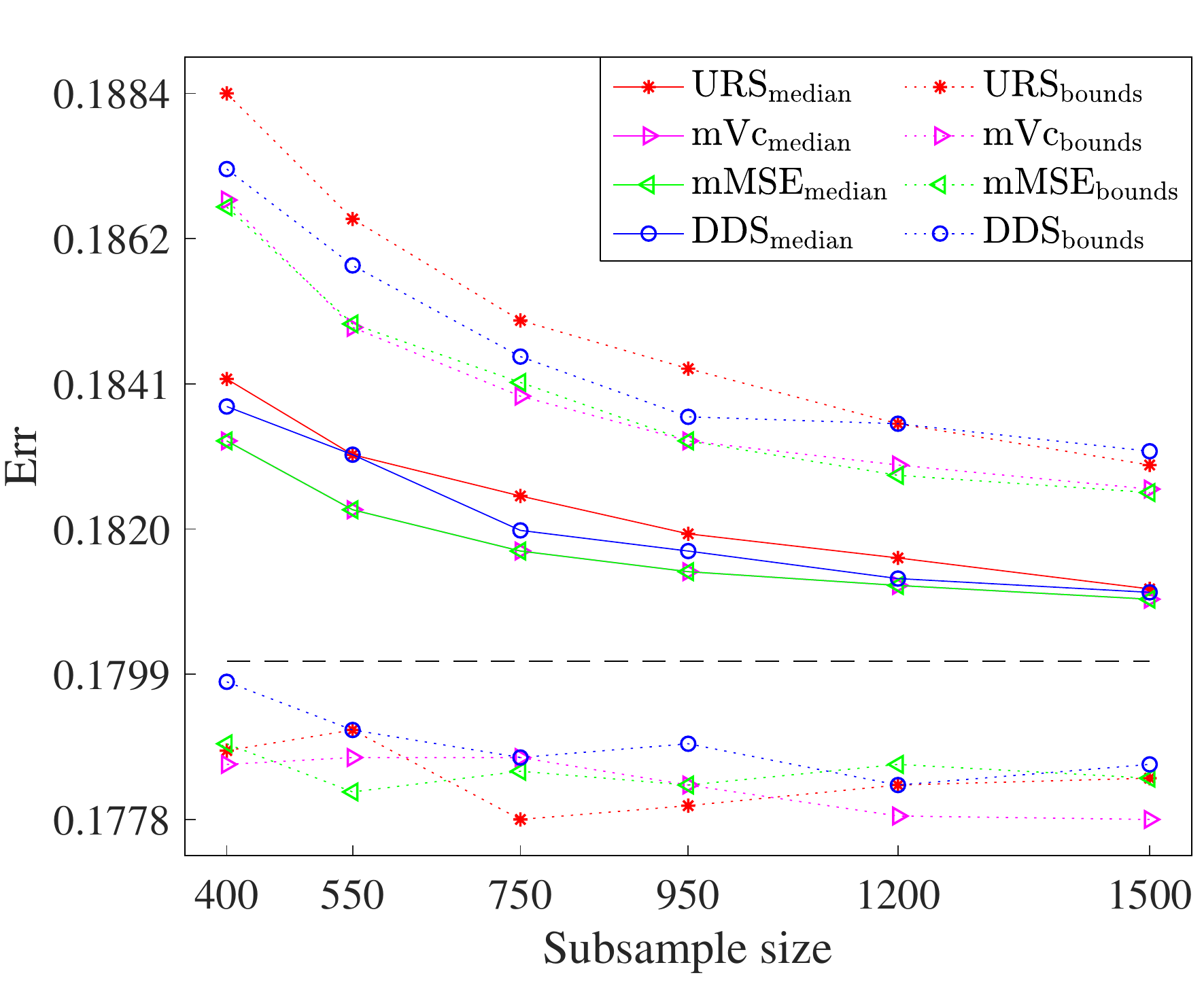}}
\caption{The classification errors of the fitted logistic regression
model based on the subdata sets obtained by
each subsampling strategy.
(a): the mean values of 1000 results for each methods;
(b): the median values and bounds of 1000 results for
each methods.}
\label{eg:LRC}
\end{figure}

Figure \ref{eg:LRC} contains the misclassification rates
of the fitted logistic regression models
based on the subdata sets
obtained by the four methods.
For comparison, the fitted logistic regression model based on
$\cx_{\text{Full}}$ is also considered.
For each subsample size $n$,
Figure \ref{eg:LRC} (a) presents
the mean values of the $1000$ results for the four subsampling methods
which are denoted by URS$_{\rm mean}$, mVc$_{\rm mean}$,
mMSE$_{\rm mean}$, and DDS$_{\rm mean}$;
Figure \ref{eg:LRC} (b) presents
the median values and lower and upper bounds of the $1000$ results
corresponding to the four methods
which are denoted by URS$_{\rm median}$, mVc$_{\rm median}$,
mMSE$_{\rm median}$, DDS$_{\rm median}$, and
URS$_{\rm bounds}$, mVc$_{\rm bounds}$,
mMSE$_{\rm bounds}$, DDS$_{\rm bounds}$, respectively.
In the version of mean and median, Figure \ref{eg:LRC}
illustrates that the mVc, mMSE and DDS all outperform URS.
The subsampling methods mVc and mMSE are based on
the logistic model, they have better performance than URS
when the model is correct, as shown in Wang et al. \cite{W18}.
The subdata sets obtained by DDS have better representation with respect to the original data than that by URS. %In this case, the better performance of DDS than URS means that
Therefore, the better similarity of the original data helps to better fit the model.
%\revB{The better performance of DDS than URS means that the better similarity of the original data is a better choice.[[[reoriganize]]]}
Figure \ref{eg:LRC} (b) also
shows a narrower interval of DDS than URS.
It means that the subdata from DDS is more robust than other methods. Moreover, the performance of DDS is as similar as that of
mVc and mMSE.

\bigskip
\noindent {\bf (B) Working model is misspecified.}\\
{We evaluate the performance of different subsampling strategies while working model is misspecified.}
We simulate the real life borehole example of the flow rate of water through a borehole from an upper aquifer to
a lower aquifer separated by an impermeable rock layer.
This example was investigated by many authors such as
Worley \cite{W87}, %An and Owen \cite{AO01},
Morris et al. \cite{M93},
Ho and Xu \cite{HX00}, %Fang and Lin \cite{FL03}
and Fang et al. \cite{F06}. %and Osuolale et al. \cite{O17}.
The response variable $\cy$, the flow rate through the borehole in
$m^3/yr$, is determined by a complex nonlinear function as follows,
\begin{equation}\label{borehole_fun}
\cy = \frac{2\pi T_{\rm u}(H_{\rm u} - H_{\rm l})}{\ln(r/r_{\rm w})\left[1+\dfrac{2LT_{\rm u}}
{\ln(r/r_{\rm w})r_{\rm w}^2K_{\rm w}^2}+\dfrac{T_{\rm u}}{T_{\rm l}}\right]}, \end{equation}
where the $8$ input variables with their usual input ranges are
listed as follows:
 $r_{\rm w} \in [0.05, 0.15]$ means the radius of borehole ($m$);
$r \in [100, 50000]$ means the radius of influence ($m$);
$T_{\rm u} \in [63070, 115600]$ means the transmissivity of upper aquifer
($m^2/yr$);
$T_{\rm l} \in [63.1, 116]$ means the
transmissivity of lower aquifer ($m^2/yr$);
$H_{\rm u} \in [990, 1110]$ means the
potentiometric head of upper aquifer ($m$);
$H_{\rm l} \in [700, 820]$ means the
potentiometric head of lower aquifer ($m$);
$L \in [1120, 1680]$ means the length of borehole ($m$);
$K_{\rm w} \in [9855, 12045]$ means the
hydraulic conductivity of borehole ($m/yr$).
The distribution of $r_{\rm w}$
is the normal distribution $\cn(0.10, 0.0161812^2)$,
the distribution of $r$ is the lognormal distribution
Lognormal$(7.71, 1.0056^2)$,
and the distributions of other
variables are all continuous uniform distribution on
their corresponding domains.

To compare the performances of different subsampling methods
for the large scale data sets, let
the size of the full data
$N = 10^6$, and the size of subdata
$n = 50, 80, 150, 250$ and $400$, respectively.
We generate the $N$-size 8-factor
$\cx_{\text{Full}}$ and $\cx_{\text{Test}}$
for training and testing, respectively, and the corresponding responses $\cy_{\text{Full}}$ and $\cy_{\text{Test}}$ through (\ref{borehole_fun}).
For the data $(\cx_{\text{Full}},\cy_{\text{Full}})$, we use different subsampling methods to obtain the subdata sets and use different models to fit each subdata set. The test data $(\cx_{\text{Test}},\cy_{\text{Test}})$ can be used to compare the performance of the different subsampling methods.

{First, we compare URS, DDS, IBOSS, kernel herding (KH) and support points (SP) under the simple linear model. IBOSS proposed by Wang et al. \cite{W19} is a kind of optimal subsampling for linear regression model. KH proposed by Chen et al.\cite{C12} and SP proposed by Mak and Joseph\cite{M18} are two popular data-based subsampling methods.
	According to the generation method of $\cx_{\text{Full}}$,
	the components of the data are mutually independent,
	so there do not need the rotation step in the process of DDS any more.
	For each subsample size $n$, URS, DDS and SP are all executed
	$1000 $ times because of the randomness. IBOSS and KH are both executed one time.
	For comparison, we also
	fit the same model for the full data.
}
\par
\begin{figure}[t!]
\centering
\subfigure[Mean MSPE of Linear Regression upon the Original Data]{
\includegraphics[width=2.7in]{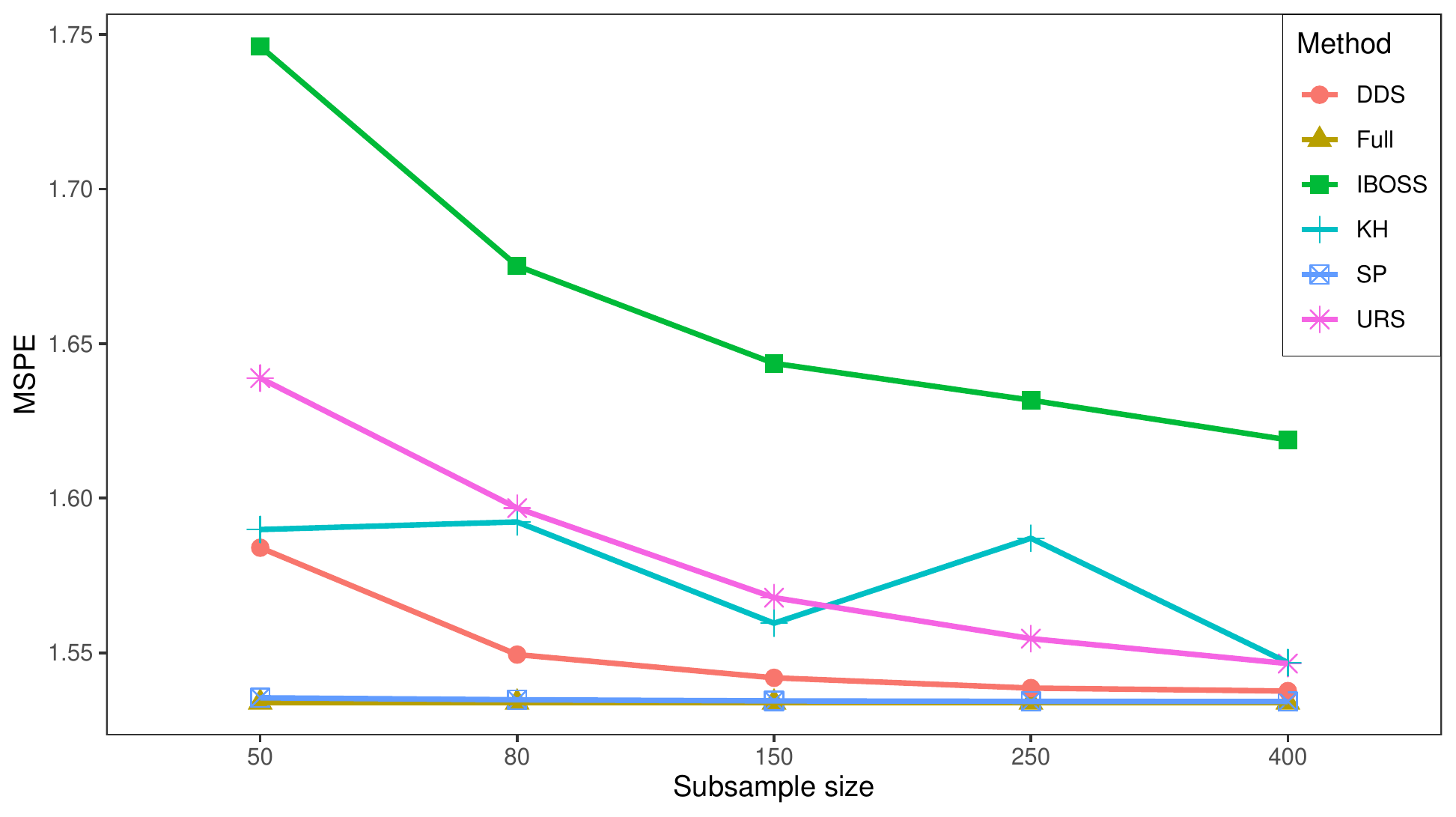}}\hspace{3mm}
\subfigure[Median MSPE of Linear Regression upon the Original Data]{
	\includegraphics[width=2.7in]{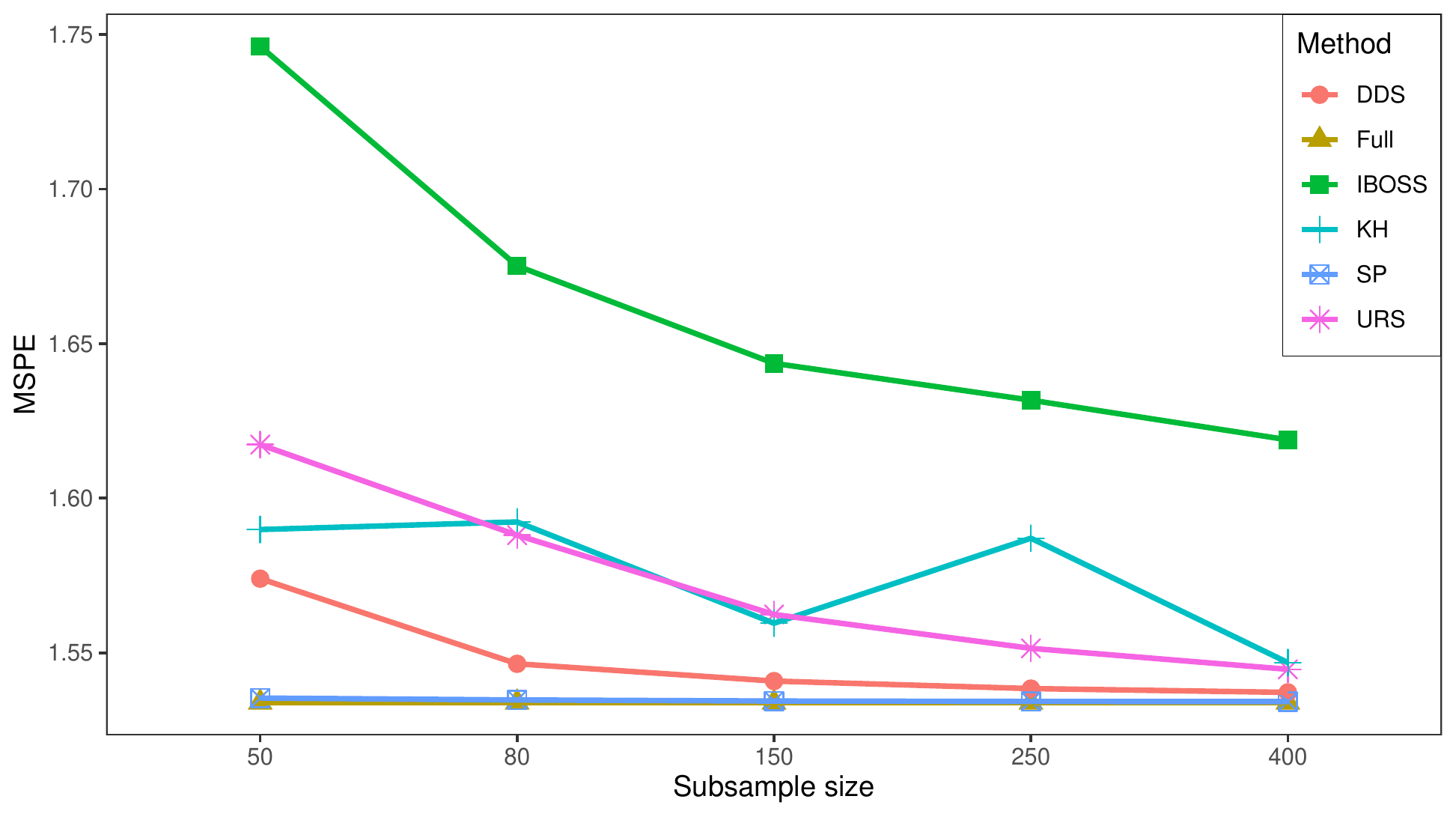}}\hspace{3mm}\\
\subfigure[Mean MSPE of Generalized Linear Regression upon the Transformed Data]{
\includegraphics[width=2.7in]{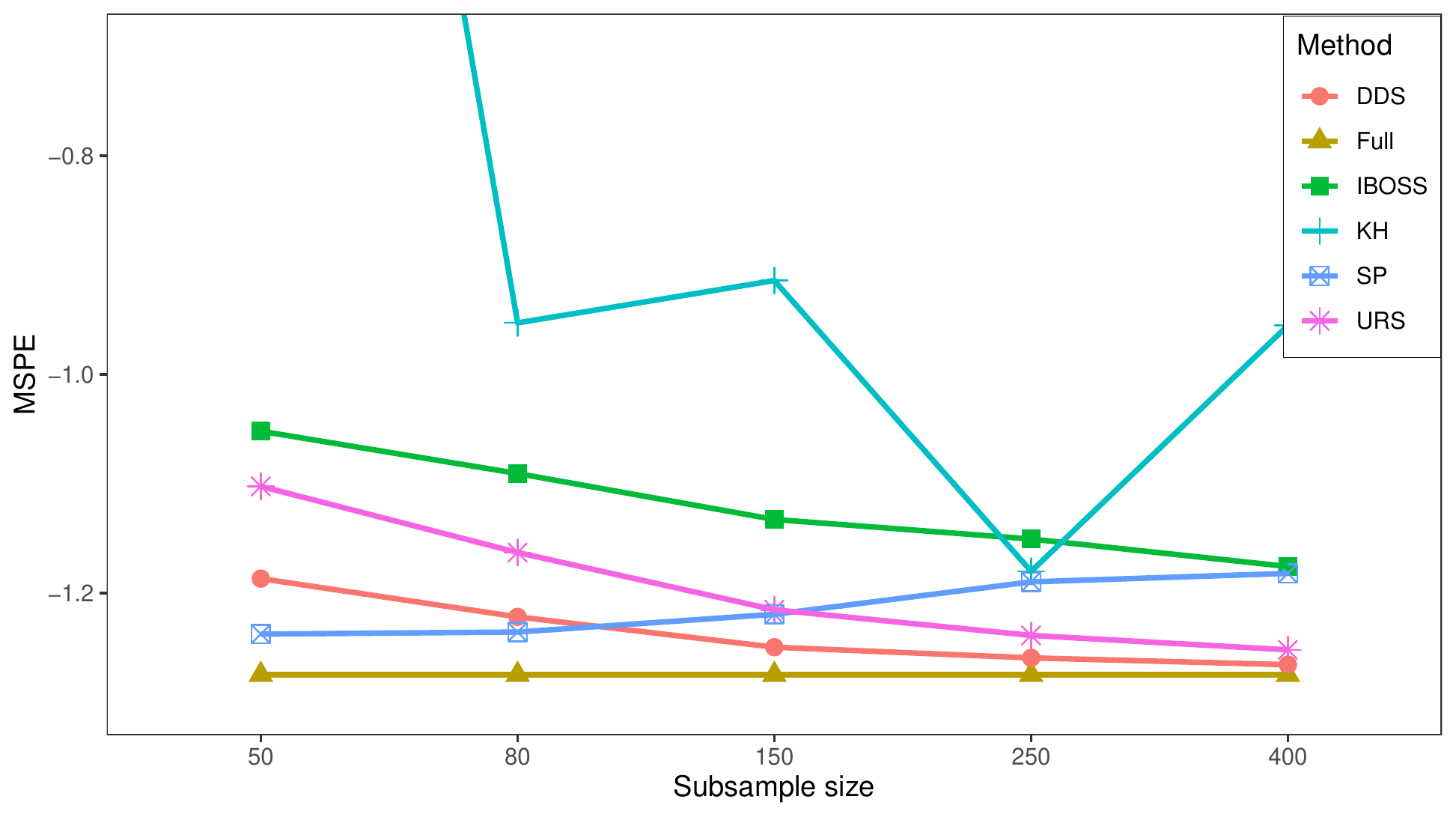}}
\subfigure[Median MSPE of Generalized Linear Regression upon the Transformed Data]{
	\includegraphics[width=2.7in]{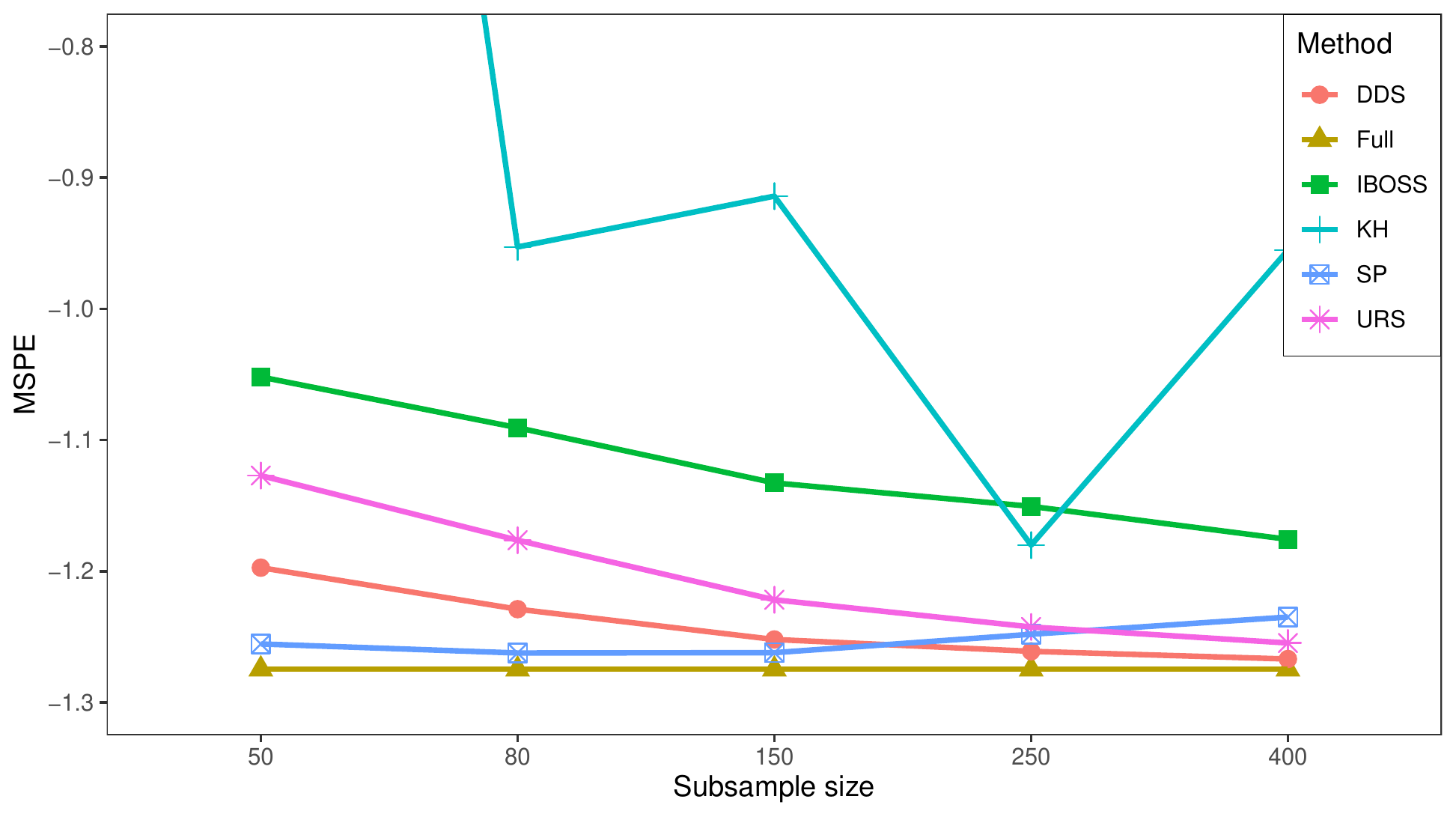}}
\caption{The MSPE values of the fitted linear regression model based on the subdata sets obtained through different subsample strategies from the full data sets with the original form and the transformed form for the borehole experiments.} \label{Fig:Borehole}
\end{figure}

 {
	Figures \ref{Fig:Borehole} (a) and \ref{Fig:Borehole} (b) show
	the MSPE values of the fitted linear regression model
	based on the full data and the
	subdata sets with different subsample size obtained by the
	five subsampling methods. To present the results more explicitly, the vertical axises in \ref{Fig:Borehole} are logarithmically transformed.
	For all subsampling methods except KH,
	the MSPE values of the fitted models based on the subdata sets
	are more close to that based on full data sets
	as the subsampling size arises.
	As shown in Figure \ref{Fig:Borehole},
	the MSPE values of the DDS are much lower than that of URS
	especially when the subsample size $n$ is relatively small,
	and are close to that of the full data
	no matter from the version of mean or median. Moreover, DDS performs better than KH and IBOSS. In addition,
	Table \ref{tblr} provide in Appendix B shows for most subsample size $n$, the upper bounds of the MSPE values of DDS are even lower than the MSPE values of IBOSS, KH, the mean and median MSPE values of URS. Thus the prediction ability of the fitted linear models based on the subsamples by DDS significantly outperforms that of URS, KH and IBOSS.
 	Moreover, under the MSPE criterion, the model-based IBOSS performs worst among the three subsampling methods, which may be caused by that the simple linear model is not enough to capture the relationship between the output $\cy$ and the $8$ input variables. The MSPE values of SP is the lowest in this case because our original variables are generated by some simple distribution. SP is adept in mining such distribution from the data and using them to make prediction. If some nonlinear transformation are applied to the parameters, it will be difficult for SP to handle the data, which can be demonstrated by the following model.}

{To compare the robustness of the sampling methods, we consider generalized regression models.
	A careful study by Ho and Xu \cite{HX00} suggested fitting
	$\log(\cy)$ with $10$ terms:
	$\log(r_{\rm w})$, $\log(r)$, $H_{\rm u}$, $H_{\rm l}$, $L$, $K_{\rm w}$,
	$H_{\rm u}*H_{\rm l}$, $H_{\rm u}^2$, $H_{\rm l}^2$ and $L^2$.
	In the generalized linear regression model,
	there are only $6$ significant variables:
	$r_{\rm w}$, $r$, $H_{\rm u}$, $H_{\rm l}$, $L$ and $K_{\rm w}$. Then
	the DDS is executed on these $6$ components
	of the original full data $\cx_{\text{Full}}$.
	We transform the subsamples obtained by URS and DDS and
	conduct IBOSS, KH and SP based on the transformed full data sets $\widetilde{\cx}_{\text{Full}}$ with above $10$ components in models to obtain the corresponding subdata sets with $10$ components.
}	

{	
	%The right panel in
	Figure  \ref{Fig:Borehole} (c) and \ref{Fig:Borehole} (d)
	show  the values of MSPE for the fitted regression model
	based on the transformed full data and
	subdata sets with different subsample size obtained by the
	five subsampling methods.
	It is obviously that the prediction performance
	of the regression model upon the transformed data set
	achieves the significant improvement  compared  with that in Figure \ref{Fig:Borehole} (a).
	DDS performs better than URS, IBOSS, KH for each subsample size $n$. The MSPE values of DDS is the lowest while the subsample size $n$ is relatively big. SP makes more accurate prediction than DDS while $n$ is relatively small. However, the MSPE value of SP arises as the subsampling size arises and it becomes lower than that of URS when $n\geq 150$, which shows SP failing to capture more information from the transformed data while the subsampling size arises. In contrast, the performance of DDS is consistently well.
}

{	
	From the performances of the DDS, URS, IBOSS, KH and SP under
	the linear regression
	and the generalized linear regression model, it is known that
	the model-based IBOSS perform worse than
	the three model-free subsampling methods URS, DDS and SP
	when the model does not fit the data very well. 	Therefore, model-free subsampling methods make a good presentation of the full data, which derives the benefit for the modeling procedure, especially for the cases that the true model is unknown. DDS is the most robust model-free subsampling method which performs well for both two models and different subsample size $n$.
}

\subsection{Real Case Study}
\begin{figure}[t!]
	\centering
	  \subfigure[Mean MSPE of Linear Regression]{
		\includegraphics[width=2.7in]{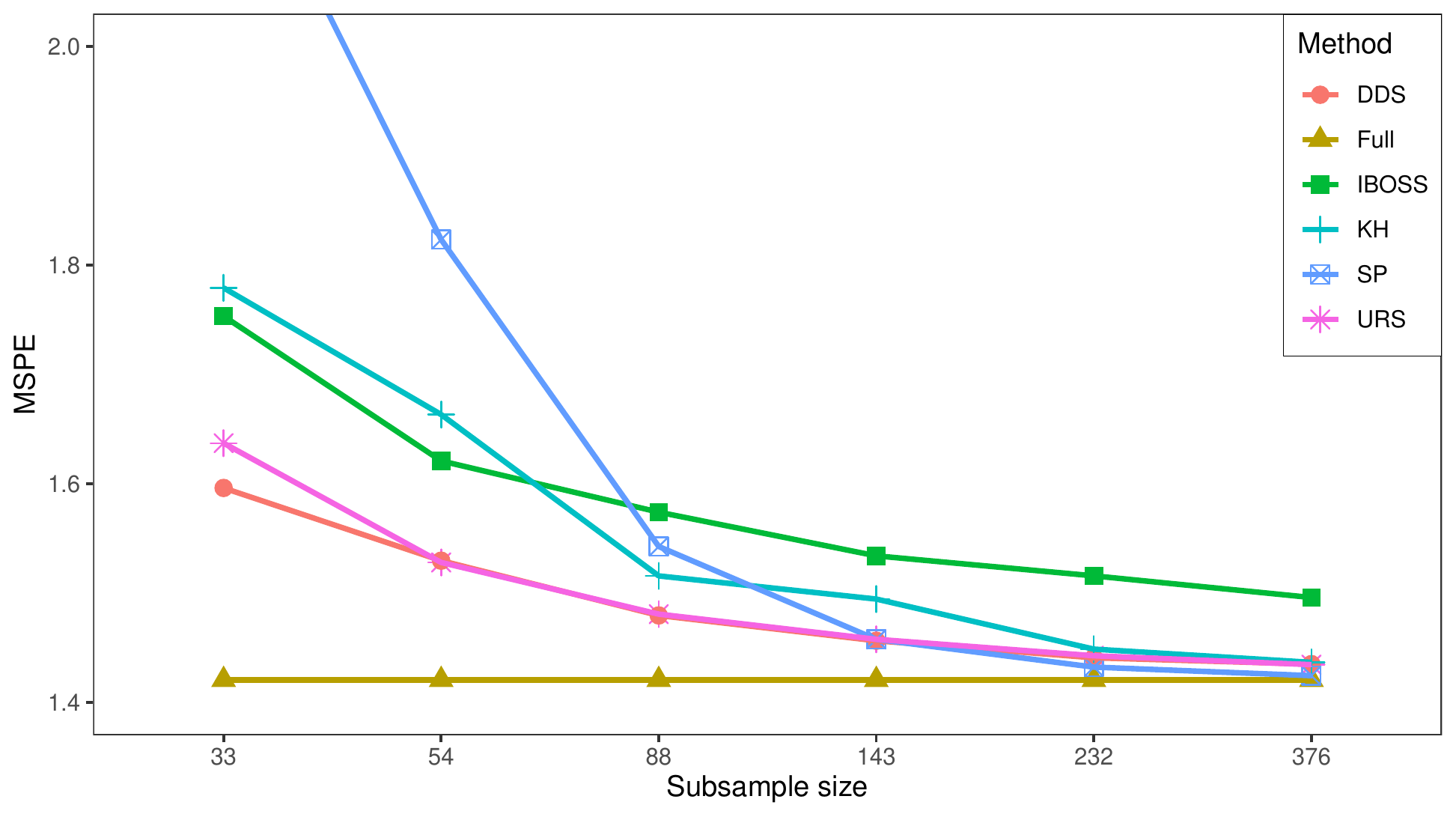}}\hspace{3mm}
	\subfigure[Median MSPE of Linear Regression]{
		\includegraphics[width=2.7in]{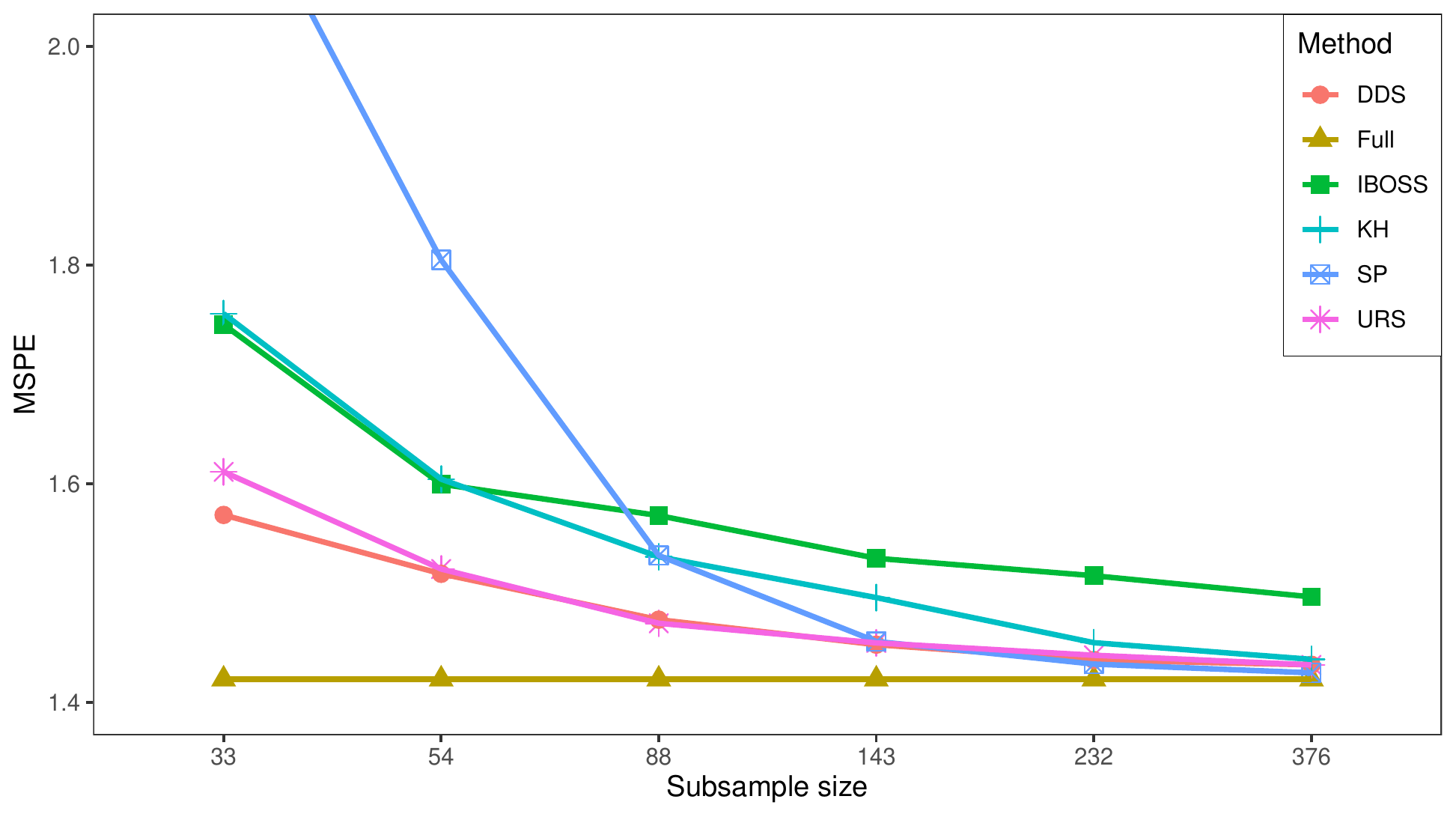}}\hspace{3mm}\\
	\subfigure[Mean MSPE of Gaussion Process Regression]{
		\includegraphics[width=2.7in]{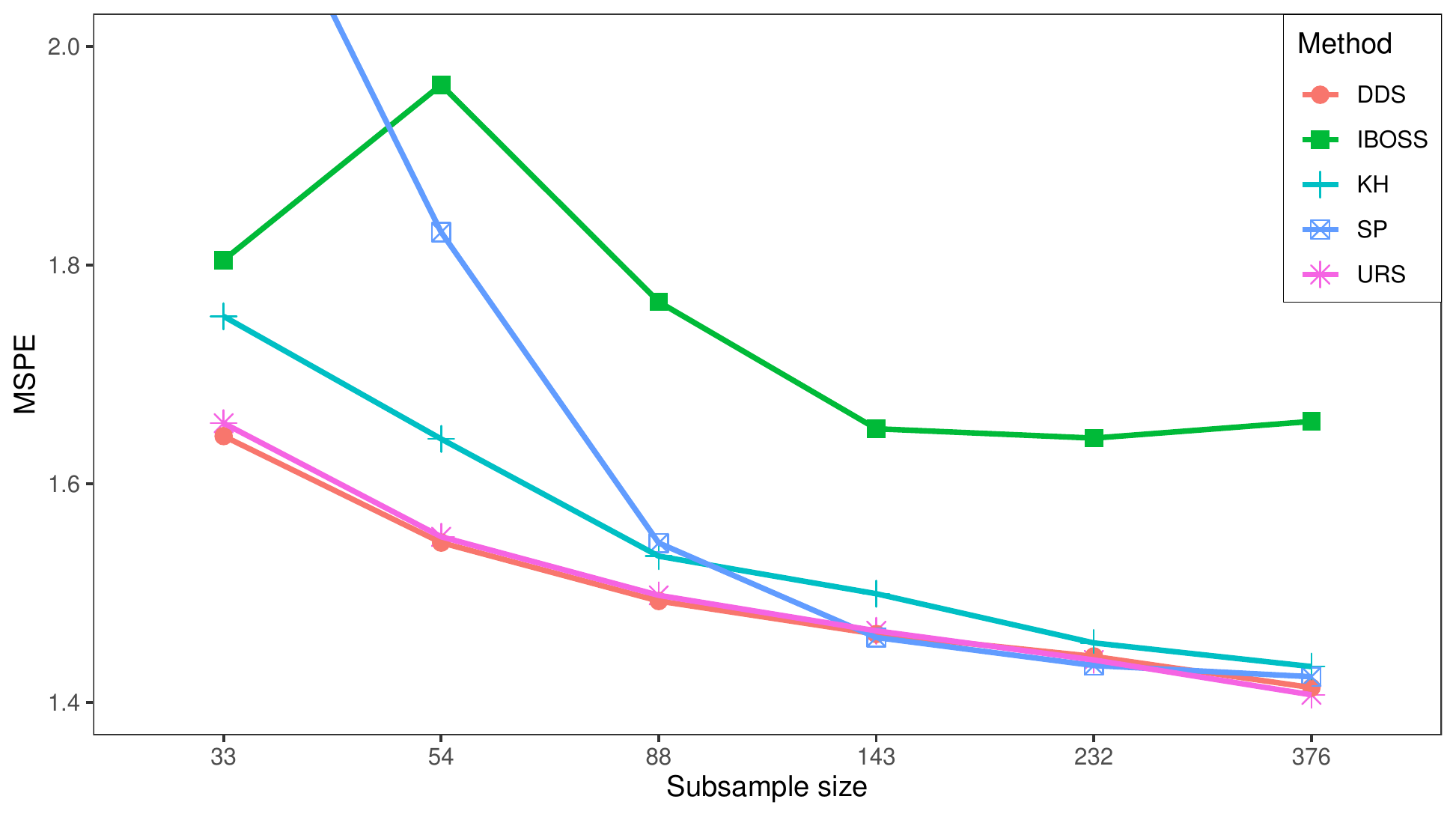}}
	\subfigure[Median MSPE of Gaussion Process Regression]{
		\includegraphics[width=2.7in]{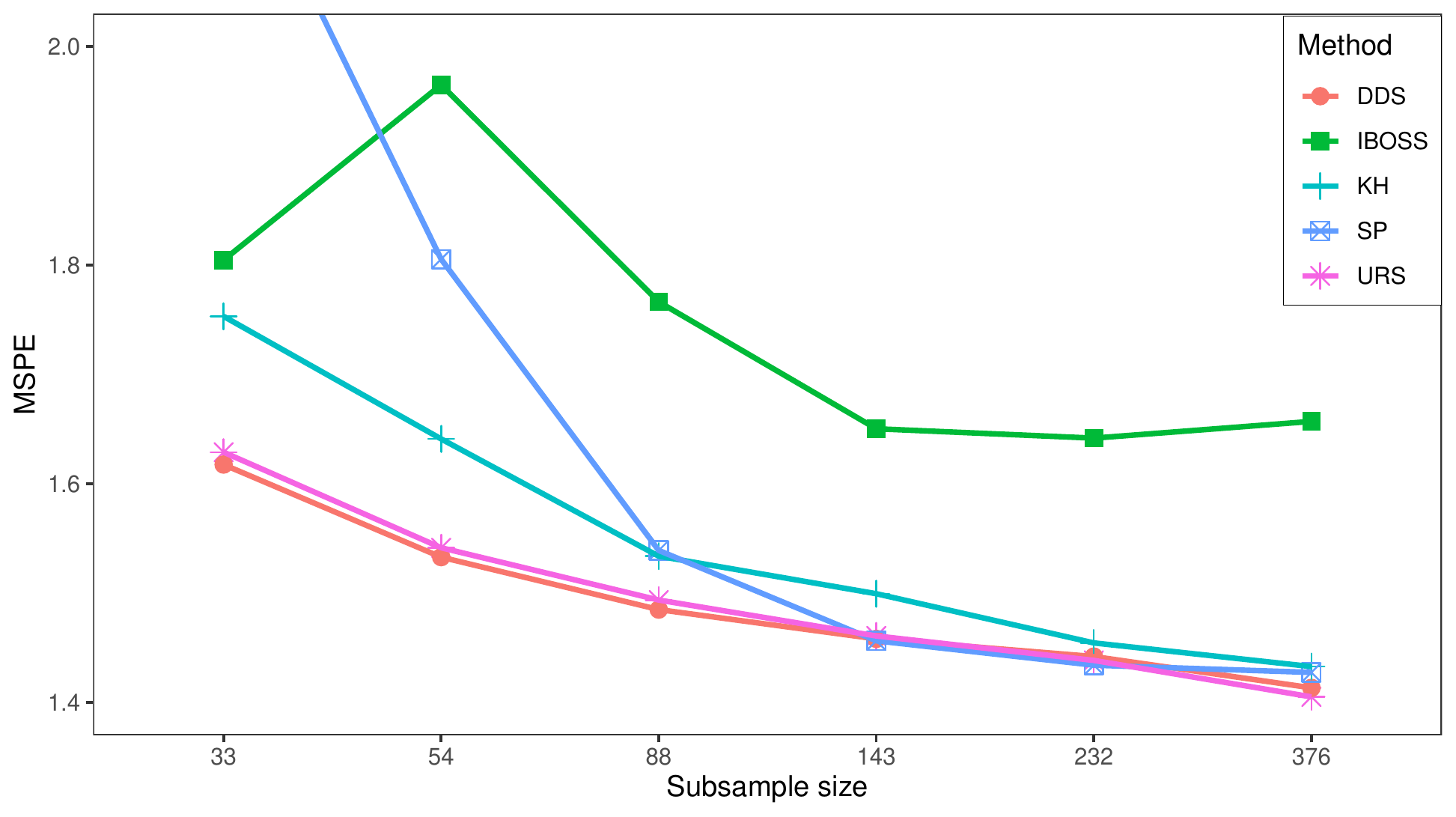}}
	\caption{The MSPE values of the fitted linear regression model
		and Gaussian process regression model
		established on different subsample strategies
		for the protein tertiary structure data.}
	\label{Protein}
\end{figure}
In this subsection, we use the DDS for a real case study of
%the protein tertiary structure data
the physicochemical properties of protein tertiary structure data set from the UCI machine learning repository (Dheeru and Taniskidou \cite{D17}). It contains $45730$ samples with 9 continuous attributes and 1 response variable.

{
First, we get rid of some outliers which results
$45253$ samples and then rescale the data
as the original data to use. Different from the simulation example of the borehole experiment, there is no guidance on the model for this real data. We
consider the two models, linear regression model and Gaussian process regression model for the subdata by DDS, URS, IBOSS, KH and SP.}
The MSPE values of the $5$-fold cross-validation
are calculated to compare the
performance of different subsampling strategies.
In the $5$-fold cross-validation, there are $5$
full data sets $\cx_{\text{Full},1}, \dots, \cx_{\text{Full},5}$ and
$5$ test data sets $\cx_{\text{Test},1}, \dots, \cx_{\text{Test},5}$.
For each $\cx_{\text{Full},l}$, in the process of DDS,
we use the first two dominated components which
may explain more than $85\%$ of total variation in $\cx_{\text{Full},l}$
by the principal component analysis.
So the dimension of the rotated space is reduced to $2$.
Here we select some numbers from the leave one out Fibonacci sequence as the subsample size $n$ for the good uniformity of the
corresponding $2$-dimensional designs on $C^2$.
The detail of the Fibonacci sequence can be seen in Fang et al. \cite{F18}.

{
For each fitted model and subsample size $n$,
the subsampling of URS, DDS and SP are repeated $100$ times
but no repetition for deterministic subsampling method IBOSS and KH in each fold.
Therefore, the mean, median and bounds of the MSPE results
are determined by $500$, $500$, $500$, $5$ and $5$ values respectively for URS, DDS, SP, IBOSS and KH.
}
{
Because of the simplicity of linear regression model,
fitting such the model based on the full data is also considered.
Figure \ref{Protein} (a) and \ref{Protein} (b) present the MSPE values of the fitted linear regression models
for the subdata obtained by different subsampling methods.  To present the results more explicitly, the vertical axises in  Figure \ref{Protein} are logarithmically transformed.
The MSPE values of DDS are lower than that of URS from the
versions of both mean and median.
What's more, the performance of DDS are better than IBOSS and KH. By Figure \ref{add2} (a) provided in Appendix B, the upper bound of the MSPE values of DDS are lower than that of IBOSS and KH for relatively large subsample size $n$. SP leads to lower MSPE values than DDS for $n\geq 232$, however the performance of SP is the worst while $n$ is relatively small, which is not robust for the subsample size.}

{
For the Gaussian process regression model,
training a model based on the full data in each fold is omitted
because of its high complexity.
For the subsampling methods,
we set the same basis function, kernel function and
hyper-parameter optimization method for each subdata
obtained by each subsampling method.
The corresponding MSPE values are shown in Figure  \ref{Protein} (c) and \ref{Protein} (d).
Compared to the result of the simple linear regression model,
the performance of IBOSS in this high complexity model
even become worse. URS, DDS and SP have similar performance when the subsample size $n$ is relatively large. However, DDS performs slightly better than URS when $n$ is relatively small while other methods are significantly worse than URS.}

{
These results of the real case study confirm that,
if there is no a priori knowledge about the true model,
model-free subsampling methods are more appropriate. Moreover, DDS performs more robust than other model-free subsampling method for different model specificaiton and subsample sizes.}

\section{Conclusion}
In this paper we provide a data-driven subsampling method that is model-free.
%to obtain subsamples, which
%are as much similar as the original data, whatever
%the distribution of the original data is.
This similarity of the subsamples with respect to the original data is measured by the generalized empirical $F$-discrepancy, which is asymptotic equivalent
to the classical generalized ${\ell}_2$-discrepancy in the theory of
uniform designs.
% and controls the upper bound of averaged function values of two data sets.
%These properties of the generalized empirical $F$-discrepancy yield the subsampling construction method and guarantee the model-free property.
The numerical examples show {the effectiveness and robustness} of the proposed DDS method, as well as its accelerated version (ADDS) in the high-dimensional cases.
%We further provide an accelerated algorithm  to speed up the data-driven subsampling process for high-dimensional cases.
In addition, both DDS and ADDS have the capacity of parallel computation for the cases of decentralized data storage according to the use of sliced uniform designs.

%The main results of this paper are established on the independent assumption of the variables or the approximate independence through PCA.
%For more complexity data such as manifold data with nonlinearly correlated attributions in which the PCA is invalid, the divide-and-conquer strategy could be applied, that is to divide the original data into multiple sub-regions of nearly independent or linearly correlated samples, and to apply the proposed method  for each sub-region.
%We will explore more effective approaches to solve the nonlinearly correlation in the future work.

\section*{Acknowledgements}
This work was supported by Hong Kong GRF (17306519), National Natural Science Foundation of China (11871288) and   Natural Science Foundation of Tianjin (19JCZDJC31100). The authors
would like to thank Hongquan Xu for his valuable comments. The first two authors contributed equally to this work.

\section*{Appendix A}

{\bf \textit{Proof of Theorem \ref{asymptotic_theorem}}}.
For %\revG{ease of presentation}
simplicity,
denote $\widetilde{\x} = T_{\cx}(\x)$ for each $\x \in \cx \cup \cp$,
and $\widetilde{\cx} = T_{\cx}(\cx)$,
$\widetilde{\cp} = T_{\cx}(\cp)$,
then $\widetilde{\cx}, \widetilde{\cp} \subset C^s$, and
the expression of $D^2(\cp;\cx,\bk)$ in (\ref{simple_GEFD})
can be rewritten as,
\begin{align}  \nonumber
D^2(\cp;\cx,\bk)
& = \frac{1}{N^2}\sum_{i,k=1}^N \bk(\widetilde{\x_i},\widetilde{\x_k})
- \frac{2}{Nn}\sum_{i=1}^N \sum_{k=1}^n
\bk(\widetilde{\x_i},\widetilde{\bm{\xi}_k})
+ \frac{1}{n^2}\sum_{i,k=1}^n \bk(\widetilde{\bm{\xi}_i},\widetilde{\bm{\xi}_k})\\
\label{GEFDs}
& = \int_{C^{2s}}\bk(\u,\v)\mbox{d}F_{\widetilde{\cx}}(\u)
\mbox{d}F_{\widetilde{\cx}}(\v) - \frac{2}{n} \sum_{k = 1}^n
\int_{C^{2s}}\bk(\u,\widetilde{\bm{\xi}_k})\mbox{d}F_{\widetilde{\cx}}(\u)
+ \frac{1}{n^2}\sum_{i,k=1}^n \bk(\widetilde{\bm{\xi}_i},\widetilde{\bm{\xi}_k}),
\end{align} where $F_{\widetilde{\cx}}$ and $F_{\widetilde{\cp}}$
are the ECDFs of $\widetilde{\cx}$ and $\widetilde{\cp}$, respectively.
Because of the joint independence assumption for $\cx$ in (\ref{ind}),
$\widetilde{\cx}$ also satisfies the joint independence as follows,
$$
F_{\widetilde{\cx}}(\u) = \prod_{j=1}^s F_{\widetilde{\cx}_{(j)}}(u_j),
~~~\forall~\u = (u_1, \dots, u_s)^T \in C^s, $$ where
$\widetilde{\cx}_{(j)} = \{F_{\cx_{(j)}}(x_{ij}), i = 1, \dots, N\}$
is the set of values along the $j$th component of $\widetilde{\cx}$,
then by using the product form of $\bk$,
the integrations over $C^{2s}$ and $C^s$ in (\ref{GEFDs}) could be
derived into the product form as follows,
\begin{align}  \label{prod_2_int}
\int_{C^{2s}} \bk(\u,\v) \mbox{d}F_{\widetilde{\cx}}(\u)
\mbox{d}F_{\widetilde{\cx}}(\v)
= & \prod_{j=1}^s \int_0^1 \int_0^1 K(u_j,v_j) \mbox{d}
F_{\widetilde{\cx}_{(j)}}(u_j) \mbox{d}F_{\widetilde{\cx}_{(j)}}(v_j),\\
\label{prod_1_int}
\int_{C^s} \bk(\u,\widetilde{\bm{\xi}_k})\mbox{d}F_{\widetilde{\cx}}(\u)
= &\prod_{j=1}^s \int_0^1 K(u_j,F_{\cx_{(j)}}(\xi_{kj}))
\mbox{d}F_{\widetilde{\cx}_{(j)}}(u_j).
\end{align}

For each $j = 1, \dots, s$,
there are also $N_j$ different values $\hat{x}_{1j}<\dots<
\hat{x}_{N_jj}=1$ in ${\widetilde{\cx}_{(j)}}$. Denote
$\hat{x}_{0j}=0$, and the replication of $\hat{x}_{ij}$ by $n_{ij}$,
then $\hat{x}_{ij}-\hat{x}_{i-1,j}=n_{ij}/N$, $i=1,\dots,N_j$.
Based on the Mean Value Theorem,
for any $v \in [0,1]$,
\begin{equation*}   \int_0^1K(u,v)\mbox{d}u = \sum_{i=1}^{N_j}
\int_{\hat{x}_{i-1,j}}^{\hat{x}_{ij}} K(u,v) \mbox{d}u
= \sum_{i=1}^{N_j} K(u_{ij}^*,v) \cdot \frac{n_{ij}}{N},\end{equation*}
where $u_{ij}^*\in\left[\hat{x}_{i-1,j}, \hat{x}_{ij}\right]$.
Recall that the reproducing kernel $K$ satisfies the
Lipschitz continuous with  a constant $c_2$.
It can be easily obtained that
\begin{align}\nonumber
\left|\int_0^1K(u,v) \mbox{d}F_{\widetilde{\cx}_{(j)}}(u) -
\int_0^1K(u,v)\mbox{d}u\right|
& = \left|\frac 1N\sum_{i=1}^{N_j}n_{ij}\left(K(\hat{x}_{ij},
v)-K(u_{ij}^*,v)\right)\right|\\\label{upper}
& \leq \frac{c_2}{N}\sum_{i=1}^{N_j}n_{ij}\left|\hat{x}_{ij}-u_{ij}^*\right|
\leq \frac{c_2}{N^2}\sum_{i=1}^{N_j}n_{ij}^2.
\end{align}
%\begin{align}\nonumber
%\left|\int_0^1K(u,v) \mbox{d}F_{\widetilde{\cx}_{(j)}}(u) -
%\int_0^1K(u,v)\mbox{d}u\right|
%& = \left|\frac 1N\sum_{i=1}^{N_j}n_{ij}\left(K(\hat{x}_{ij},
%v)-K(u_{ij}^*,v)\right)\right|\\\nonumber
%& \leq \frac 1N\sum_{i=1}^{N_j}n_{ij}
%\left|K(\hat{x}_{ij},v)-K(u_{ij}^*,v)\right|\\\nonumber
%& \leq \frac{c_2}{N}\sum_{i=1}^{N_j}n_{ij}(\hat{x}_{ij}-u_{ij}^*)\\
%\label{upper} &\leq \frac{c_2}{N^2}\sum_{i=1}^{N_j}n_{ij}^2.
%\end{align}
Note that, the independent condition (\ref{ind}) also yields
the following constraints about $N, N_j$ and $n_{ij}$,
%According to the independent condition (\ref{ind}),
$N\geq \prod\limits_{l=1}^sN_l$ and
$n_{ij} \leq c_1\prod\limits_{l\neq j}^sN_l$ for
$i=1, \dots,N_j, j=1, \dots, s,$
%\begin{align*} \left\{
%\begin{array} {l} n_{ij} \leq c_1\prod\limits_{l\neq j}^sN_l, \\
%N\geq \prod\limits_{l=1}^sN_l,\end{array} \right. \ \ \
%~~~i=1, \dots,N_j,~~ j=1, \dots, s,\end{align*}
then the upper bound in (\ref{upper}) could be shrunk as follows,
\begin{equation*}
\frac{c_2}{N^2}\sum_{i=1}^{N_j}n_{ij}^2\leq
\frac{c_2}{\left(\prod_{l=1}^sN_l\right)^2}\sum_{i=1}^{N_j}
\left[c_1^2\cdot\left(\frac{\prod_{l=1}^sN_l}{N_j}\right)^2\right]
=\frac{c_2c_1^2}{N_j}=\O(1/N_j),
\end{equation*}
which leads to
\begin{equation} \label{1_int}
\int_0^1K(u,v)\mbox{d}F_{\widetilde{\cx}_{(j)}}(u)=
\int_0^1K(u,v)\mbox{d}u+\O(1/N_j).\end{equation}
Integrating (\ref{1_int}) from 0 to 1 with the reference distribution
$F_{\widetilde{\cx}_{(j)}}$ yields
\begin{align}\nonumber  \int_0^1\int_0^1K(u,v)\mbox{d}
F_{\widetilde{\cx}_{(j)}}(u) \mbox{d}F_{\widetilde{\cx}_{(j)}}(v)
& =  \int_0^1 \left[ \int_0^1 K(v,u) \mbox{d}F_{\widetilde{\cx}_{(j)}}(v)
\right] \mbox{d}u + \O(1/N_j) \\\nonumber
& =  \int_0^1 \left[ \int_0^1 K(v,u) \mbox{d}v + \O(1/N_j) \right]
\mbox{d}u + \O(1/N_j)\\ \label{2_int}
& =  \int_0^1 \int_0^1 K(u,v) \mbox{d}u \mbox{d}v + \O(1/N_j).
\end{align}
%\begin{align}\nonumber  \int_0^1\int_0^1K(u,v)\mbox{d}
%F_{\widetilde{\cx}_{(j)}}(u) \mbox{d}F_{\widetilde{\cx}_{(j)}}(v)
%& =  \int_0^1\left[\int_0^1K(u,v) \mbox{d}u + \O(1/N_j)
%\right] \mbox{d}F_{\widetilde{\cx}_{(j)}}(v) \\\nonumber
%& =  \int_0^1 \left[ \int_0^1 K(v,u) \mbox{d}F_{\widetilde{\cx}_{(j)}}(v)
%\right] \mbox{d}u + \O(1/N_j) \\\nonumber
%& =  \int_0^1 \left[ \int_0^1 K(v,u) \mbox{d}v + \O(1/N_j) \right]
%\mbox{d}u + \O(1/N_j)\\ \label{2_int}
%& =  \int_0^1 \int_0^1 K(u,v) \mbox{d}u \mbox{d}v + \O(1/N_j).
%\end{align}
Substituting (\ref{2_int}) into (\ref{prod_2_int}), and
(\ref{1_int}) into (\ref{prod_1_int}) obtains
\begin{align} \nonumber
\int_{C^{2s}} \bk(\u,\v) \mbox{d}F_{\widetilde{\cx}}(\u)
\mbox{d}F_{\widetilde{\cx}}(\v)
& = \prod_{j=1}^s \left[\int_0^1 \int_0^1 K(u,v)
\mbox{d}u \mbox{d}v + \O(1/N_j) \right]\\
\label{prod_2_int2}
& =  \int_{C^{2s}} \bk(\u,\v) \mbox{d}\u \mbox{d}\v + \O(1/N^*),
\end{align} and \begin{align}
\nonumber \int_{C^s} \bk(\u,\widetilde{\bm{\xi}_k})
\mbox{d}F_{\widetilde{\cx}}(\u)
& = \prod_{j=1}^s\left[\int_0^1K(u,F_{\cx_{(j)}}(\xi_{kj}))
\mbox{d}u+\O(1/N_j)\right]\\
\label{prod_1_int2}
& = \int_{C^s}\bk(\u,\widetilde{\bm{\xi}_k})\mbox{d}\u+\O(1/N^*),
\end{align}
respectively. Finally, substituting (\ref{prod_2_int2}) and (\ref{prod_1_int2}) into (\ref{GEFDs}) completes the proof.

\bigskip
To prove Lemma 1, we give the following lemma. Its proof can be referred to  Lemma 3.3 in \cite{hick1}, and we omit it.

\begin{lemma}\label{components}
Define the kernel function on $C^s\times C^s$
with a product form $\bk(\u,\v) = \prod_{j = 1}^s K(u_j,y_j),$
where $K(u_j,v_j)$ is defined in (\ref{K_expression}). It follows that the components of $\bk$ are
\begin{equation} \label{K_u}
\bk_{\vartheta}(\u_{\vartheta},\v_{\vartheta}) = \left\{\ba{ll}
\cl_{\cs}(\bk(\u,\v)) = 1, & \quad \vartheta = \emptyset, \\
\prod_{j\in \vartheta} [K(u_j,v_j) - 1], & \quad \vartheta \neq \emptyset,
\ea \right. \end{equation}  where $\cl_j$ denotes the operator $\cl$ defined in (\ref{defi_L}) on the $j$th coordinate and $\cl_{\vartheta} = \prod_{j \in \vartheta} \cl_j$. Moreover, any $f \in \bm{X}_p(C^s)$ has the following properties,
\begin{equation}\label{f_u} f_{\vartheta}(\v_{\vartheta}) =
\beta^{-2|\vartheta|} \int_{C^{\vartheta}}
\frac{\partial^{|\vartheta|}f_{\vartheta}}{\partial \u_{\vartheta}}
\frac{\partial^{|\vartheta|}\bk_{\vartheta}(\cdot,\v_{\vartheta})}
{\partial \u_{\vartheta}} \text{d}\u_{\vartheta}, \nonumber
\end{equation} and
$ f(\v) = \langle f, \bk(\cdot, \v)\rangle. $
That indicates $\bk$ is a reproducing kernel for $\bm{X}_p(C^s)$.
\end{lemma}

\bigskip
\noindent {\bf \textit{Proof of Lemma \ref{EKH}}:}
We give the details of the derivation from the two aspects of
one-dimensional cases and multi-dimensional cases.

Define the Bernoulli polynomials $\{B_n(x)\}_{n = 0}^{\infty}$
by the generating function $\frac{te^{xt}}{e^t-1} = \sum_{n = 0}^{\infty}B_n(x)\frac{t^n}{n!}.$
The first few are $B_0(x) = 1, B_1(x) = x-1/2, B_3(x) = x^2 - x +1/6.$

\smallskip
{\bf (i) Consider the case of $\bm{s = 1}$.}\smallskip\\
For any $p$, $1 \leq p \leq \infty$, let
\bea \label{X_p}
X_p \equiv \left\{f: \frac {\text{d}f}{\text{d}u} \in L_p([0,1])
\right\}.\eea
Suppose $K(\cdot,\cdot)$ have the following form
\bea \label{K_expression}
K(u,v) = M + \beta ^2\left[ \lambda(u) + \lambda(v) +
\frac 12 B_2\left(\bm{\{}u-v\bm{\}}\right) + B_1(u)B_1(v)\right], \eea
where $\bm{\{}\cdot\bm{\}}$ denotes the fractional part of a real number,
$\lambda$ is a function in the space $X_{\infty}$
satisfying $\int_0^1 \lambda(x)\text{d}u = 0$,
$\beta$ is an arbitrary positive constant, and
$M = 1 + \beta^2 \int_0^1 \left(\frac{\text{d}\lambda}{\text{d}u}\right)^2\text{d}u$.
Then for any $v \in [0,1)$, the function %$H^{(y)} (\cdot) = K(\cdot,y) \in X_2$,
$K(\cdot,v) \in X_2$, and for any function $f \in X_2$,
the following results could be derived,
\begin{equation} \label{K_partial}
\frac{\partial K}{\partial u} = \beta^2 \left[ \frac{\text{d}\lambda}{\text{d}u}
+ B_1\left(\bm{\{}u-v\bm{\}}\right) + B_0(u)B_1(v)\right],\end{equation}
$$\beta^{-2} \int_0^1  \frac{\partial K}{\partial u}
\frac{\text{d}f}{\text{d}u} \text{d}u
= f(v) - \int_0^1 \left(f - \frac{\text{d}\lambda}{\text{d}u}
\frac{\text{d}f}{\text{d}u}\right)\text{d}u.$$
Define a linear function of $f$ as follows
\begin{equation} \label{defi_L}
\cl(f) \equiv \int_0^1\left( f - \frac{\text{d}\lambda}{\text{d}u}
\frac{\text{d}f}{\text{d}u}\right)\text{d}u.\end{equation}
Define  an  inner product and the induced norm as follows
\begin{equation}\label{defi_inner-norm}
\langle f, g \rangle = \cl(f)\cl(g) + \beta^{-2}\int_0^1
\frac{\text{d}f}{\text{d}u}\frac{\text{d}g}{\text{d}u}\text{d}u,\qquad
 ||| f |||_p  = \left\| \left(\cl(f),\beta^{-1}\frac{\text{d}f}{\text{d}u}\right)
\right\|_p.\end{equation}
Note that \begin{equation} \label{L(K)}
\cl(K(\cdot,v)) = \int_0^1\left(K - \frac{\text{d}\lambda}{\text{d}u}
\frac{\partial K}{\partial u}\right)\text{d}u =
M - \beta^2\int_0^1\left(\frac{\text{d}\lambda}{\text{d}u}\right)^2\text{d}u
= 1,\end{equation}
which leads to $$
\langle K(\cdot,v), f \rangle = \cl(f) + \beta^{-2}\int_0^1
\frac{\partial K}{\partial u}\frac{\text{d}f}{\text{d}u} \text{d}u = f(v).$$
This result indicates $K$ defined in (\ref{K_expression})
is a reproducing kernel.
For the linear function $\cl$,
by the Riesz Representation Theorem, there exists $\psi \in X_2$
such that $\cl(f) = \langle \psi, f \rangle$ for any $f \in X_2$.
As $K$ is a reproducing kernel, $\psi$ is the representer of $\cl$,
and from (\ref{L(K)}), we can obtain $$
\psi(v) = \langle K(\cdot, v), \psi \rangle = \cl(K(\cdot, v)) = 1,$$
which implys that
the constant 1 is the representer for the linear functional $\cl$, i.e.
\begin{equation} \label{L(f)} \cl(f) = \langle 1, f \rangle, \quad \forall f \in X_2. \end{equation}
Denote \begin{equation} \label{defi_T(f)} \ct(f) \equiv
\frac 1N \sum_{u \in \ce} f(u) - \frac 1n\sum_{\zeta \in \cd} f(\zeta),
%\frac 1N \sum_{i = 1}^N f(u_i) - \frac 1n\sum_{k = 1}^n f(\zeta_k),
\end{equation}
then $\ct$ is also a bounded linear function.
By the Riesz Representation Theorem, there exists $\phi \in X_2$
such that \begin{equation} \label{Riesz_T}
\ct(f) = \langle \phi, f \rangle,~~~~\forall f \in X_2. \end{equation}
Note that $K$ is a reproducing kernel, then
$\phi(v) = \langle K(\cdot, v), \phi \rangle$,
based on (\ref{Riesz_T}),
$\phi(v)$ is the results of the transformed $K(\cdot, v)$ under the
linear function $\ct$, i.e.
$$\phi(v) = \ct(K(\cdot, v)) = \frac 1N
\sum_{u \in \ce} K(u, v) - \frac 1n \sum_{\zeta \in \cd} K(\zeta, v).$$
%\sum_{i = 1}^N K(u_i, v) - \frac 1n \sum_{k = 1}^n K(\zeta_k, v).$$
From (\ref{L(f)}), (\ref{Riesz_T}) and (\ref{defi_T(f)}), it can be easily
obtained that $\cl(\phi) = \langle 1, \phi \rangle =  \ct(1) = 0$,
which leads to
\begin{equation}\label{phi_norm}
 ||| \phi |||_2^2  = \left\| \left(\cl(\phi), \beta^{-1}
\frac{\text{d}\phi}{\text{d}v} \right)\right\|_2^2
= \int_0^1 \left(\beta^{-1}\frac{\text{d}\phi}{\text{d}v}\right)^2 \text{d}v.
\end{equation}
For any $f \in X_p$,
define the generalized ${\ell}_p$-variation as
\bea \label{defi_V(f)} V_p(f) \equiv  |||f-\cl(f)|||_p  =
\left \| \beta^{-1}\frac{\text{d}f}{\text{d}v} \right\|_p,\eea
then we could obtain the following inequality by the
H$\ddot{\text{o}}$lder inequality,
\begin{equation}\label{res:s=1}
|\ct(f)| = |\langle \phi, f \rangle| = \beta^{-2} \left|
\int_0^1 \frac{\text{d}\phi}{\text{d}v}
\frac{\text{d}f}{\text{d}v}\text{d}v\right| \leq  ||| \phi |||_2  V_2(f).
\end{equation}
On the basis of the partial derivative of $K$ in (\ref{K_partial}),
we have
\begin{equation} \label{d_phi}
\frac{\text{d}\phi}{\text{d}v} = \frac{\beta^2}{N}
\sum_{u\in \ce}\left[B_1\left(\bm{\{}v-u\bm{\}}\right) + B_1(u)\right]
%\sum_{i=1}^N\left[B_1\left(\bm{\{}v-u_i\bm{\}}\right) + B_1(u_i)\right]
- \frac{\beta^2}{n}
\sum_{\zeta \in \cd}\left[B_1\left(\bm{\{}v-\zeta\bm{\}}\right) + B_1(\zeta)\right].
%\sum_{k = 1}^n\left[B_1\left(\bm{\{}v-\zeta_k\bm{\}}\right) + B_1(\zeta_k)\right].
\end{equation}
Note that for any $u,\zeta\in[0,1]$, according to the properties of $B_n$,
\begin{equation} \label{int_each}
\int_0^1 \left[ B_1\left(\bm{\{}v-u\bm{\}}\right)  + B_1(u) \right]
\left[ B_1\left(\bm{\{}v-\zeta\bm{\}}\right)  + B_1(\zeta)\right] \text{d}v
= \frac 12 B_2\left(\bm{\{}u-\zeta\bm{\}}\right) + B_1(u)B_1(\zeta),\end{equation}
then substituting (\ref{d_phi}) into (\ref{phi_norm}) and using
the integral result in (\ref{int_each}),
we can compute the squared norm of $\phi$ as follows
\begin{align*}  ||| \phi |||_2^2
&= \frac{\beta^2}{N^2}\sum_{u, \check{u}\in \ce} \left[\frac 12 B_2\left(\bm{\{}u-\check{u}\bm{\}}\right) + B_1(u)B_1(\check{u})\right] - \frac{2\beta^2}{Nn}\sum_{u \in \ce, \zeta\in \cd} \left[\frac 12 B_2\left(\bm{\{}u-\zeta\bm{\}}\right) + B_1(u)B_1(\zeta)\right]\\
&\quad + \frac{\beta^2}{n^2}\sum_{\zeta, \check{\zeta}\in \cd} \left[\frac 12 B_2\left(\bm{\{}\zeta-\check{\zeta}\bm{\}}\right) + B_1(\zeta)B_1(\check{\zeta})\right] \\
&= \frac{1}{N^2}\sum_{u, \check{u}\in \ce} K(u,\check{u}) - \frac{2}{Nn} \sum_{u \in \ce, \zeta\in \cd}K(u,\zeta) + \frac{1}{n^2}\sum_{\zeta, \check{\zeta}\in \cd} K(\zeta,\check{\zeta}) = D_K^2(\ce,\cd).\end{align*}
%\begin{align*}
%||| \phi |||_2^2 &= \frac{\beta^2}{N^2}\sum_{i,k = 1}^N \left[\frac 12
%B_2\left(\bm{\{}u_i-u_k\bm{\}}\right) + B_1(u_i)B_1(u_k)\right]
%- \frac{2\beta^2}{Nn}\sum_{i = 1}^N\sum_{k = 1}^n \left[\frac 12
%B_2\left(\bm{\{}u_i-\zeta_k\bm{\}}\right) + B_1(u_i)B_1(\zeta_k)\right]\\
%&\quad + \frac{\beta^2}{n^2}\sum_{i,k = 1}^n \left[\frac 12
%B_2\left(\bm{\{}\zeta_i-\zeta_k\bm{\}}\right) + B_1(\zeta_i)B_1(\zeta_k)\right] \\
%&= \frac{1}{N^2}\sum_{i,k = 1}^N K(u_i,u_k)
%- \frac{2}{Nn} \sum_{i = 1}^N\sum_{k = 1}^nK(u_i,\zeta_k)
%+ \frac{1}{n^2}\sum_{i,k = 1}^n K(\zeta_i,\zeta_k)
%= D_K^2(\ce,\cd).\end{align*}
Replacing  $||| \phi |||_2$  in (\ref{res:s=1}) with $D_K(\ce,\cp)$,
we achieve the inequality in Lemma \ref{EKH} for $s = 1$.

\smallskip
{\bf (ii) Consider the case of $\bm{s \geq 2}$.}\smallskip\\
Let $\cs = \{1, \dots, s\}$ be the set of coordinate indices. For any
$\vartheta \subseteq \cs$, let $|\vartheta|$ denote its cardinality and $C^\vartheta = [0,1]^\vartheta$
the $|\vartheta|$-dimensional unit cube involving the coordinates in $\vartheta$,
$\u_\vartheta$ the vector containing the components of $\u$ whose indices are in $\vartheta$,
and $\mathrm{d} \u_\vartheta = \prod_{j\in \vartheta} \mathrm{d}u_j$ the uniform measure on
$C^{\vartheta}$.
The multidimensional generalization of $X_p$,
the space of integrands defined in (\ref{X_p}),
is a space of functions whose mixed partial derivatives are all
integrable, \begin{equation*} %\label{multi_X_p}
\X_p \equiv \X_p(C^s) \equiv \left\{ f: \frac{\partial^{|\vartheta|}f}{\partial \u_{\vartheta}} \in L_p(C^{\vartheta}), ~ \forall \vartheta \subseteq \cs\right\}.\end{equation*}
Let $\cl_j$ denote the operator $\cl$ defined in (\ref{defi_L})
on the $j$th coordinate, $\cl_{\vartheta} = \prod_{j \in \vartheta} \cl_j$,
and $\cl_{\emptyset}$ be defined as the identity.
For any $f \in \X_p(C^s)$, iteratively define its components
$f_{\vartheta} = \cl_{\cs-\vartheta}f - \sum_{\omega \subset \vartheta} f_{\omega}, \forall \vartheta \subseteq \cs.$
These components possess the following properties:
$$ \cl_j(f_{\vartheta}) = \left\{\ba{ll} 0, & ~ j \in \vartheta,\\ f_{\vartheta}, & ~ j \notin \vartheta,
\ea \right.  \quad \text{and} \quad
f = \sum_{\vartheta \subseteq \cs} f_{\vartheta}. $$
Then define the inner product $\langle \cdot, \cdot \rangle$ on
$\X_2(C^s)$ and the norm  $|||\cdot|||_p$  on $\X_p(C^s)$ as the
generalizations of those in (\ref{defi_inner-norm}) as follows,$$
\langle f, g \rangle = \sum_{\vartheta \subseteq \cs} \beta^{-2|\vartheta|} \int_{C^{\vartheta}}
\frac{\partial^{|\vartheta|}f_{\vartheta}}{\partial \u_{\vartheta}}\frac{\partial^{|\vartheta|}g_{\vartheta}}{\partial \u_{\vartheta}}
\text{d}\u_{\vartheta}, \qquad  |||f|||_p  = \left \| \left(\beta^{-|\vartheta|}\frac{\partial^{|\vartheta|}f_{\vartheta}}
{\partial \u_{\vartheta}}\right)_{\vartheta \subseteq \cs}\right \|_p .$$
Define the kernel function on $C^s\times C^s$
with a product form $\bk(\u,\v) = \prod_{j = 1}^s K(u_j,v_j),$
where $K(u_j,v_j)$ is defined in (\ref{K_expression}).
 Then $\bk$ is the reproducing kernel for $\X_p(C^s)$ indicated from Lemma \ref{components}. % (at the end of the Appendix).

Now define the linear function $\ct$ with the same form as
the case of $s=1$ in (\ref{defi_T(f)}),
then its representer $\phi$ satisfies
$\ct(\bk(\cdot, \v)) = \langle \phi,  \bk(\cdot, \v)\rangle$.
%Lemma \ref{components} indicates that $\bk$ is the reproducing kernel for $\X_p(C^s)$, which implies
Note that $\bk$ is a reproducing kernel, and
$\phi(\v) = \langle \bk(\cdot, \v), \phi \rangle$,
so $\phi(\v)$ could be presented as the following linear combination
of the values of $\bk(\cdot, \v)$ upon the two sets $\ce$ and $\cd$,
$$\phi(\v) = \ct(\bk(\cdot, \v)) = \frac 1N \sum_{\u \in \ce} \bk(\u, \v)
- \frac 1n \sum_{\bzeta \in \cd} \bk(\bzeta, \v),$$
%$$\phi(\v) = \ct(\bk(\cdot, \v)) = \frac 1N \sum_{i = 1}^N \bk(\u_i, \v)
%- \frac 1n \sum_{k = 1}^n \bk(\bm{\zeta}_k, \v),$$
then its components are
$\phi_{\vartheta}(\v_{\vartheta}) = \frac{1}{N}\sum_{\u \in \ce} \prod_{j \in \vartheta} \left[K(u_{j},v_j) -1\right] - \frac{1}{n}\sum_{\bm{\zeta} \in \cd} \prod_{j \in \vartheta} \left[K(\zeta_{j},v_j) -1\right]$
%$\phi_{\vartheta}(\v_{\vartheta}) = \frac{1}{N}\sum_{i = 1}^N\prod_{j \in \vartheta}
%\left[K(u_{ij},v_j) -1\right] - \frac{1}{n}\sum_{k = 1}^n\prod_{j \in \vartheta}
%\left[K(\zeta_{ij},y_j) -1\right]$
by using the expression of $\bk$'s components in (\ref{K_u}).
On the basis of the partial derivative of $K$ in (\ref{K_partial}),
the partial derivative of $\phi$ could be computed as
\begin{align*}
\frac{\partial^{|\vartheta|}\phi_{\vartheta}}{\partial \v_{\vartheta}}
& = \frac{\beta^{2|\vartheta|}}{N}\sum_{\u \in \ce}\prod_{j \in \vartheta} \left[ \lambda'(v_j) + B_1\left(\bm{\{}v_j - u_j\bm{\}}\right)  + B_1(u_j) \right ]\\
& \quad - \frac{\beta^{2|\vartheta|}}{n}\sum_{\bzeta \in \cd}\prod_{j \in \vartheta} \left[ \lambda'(v_j) + B_1\left(\bm{\{}v_j - \zeta_j\bm{\}}\right)  + B_1(\zeta_j) \right], \end{align*}
%\begin{align*}
%\frac{\partial^{|\vartheta|}\phi_{\vartheta}}{\partial \v_{\vartheta}}
%& = \frac{\beta^{2|\vartheta|}}{N}\sum_{i=1}^N\prod_{j \in \vartheta} \left[\lambda'(v_j) + B_1\left(\bm{\{}v_j - u_{ij}\bm{\}}\right)  + B_1(u_{ij}) \right ]\\
%& \quad - \frac{\beta^{2|\vartheta|}}{n} \sum_{k = 1}^n\prod_{j \in \vartheta}\left[\lambda'(v_j) + B_1\left(\bm{\{}v_j - \zeta_{ij}\bm{\}}\right)  + B_1(\zeta_{ij}) \right ],
%\end{align*}
where $\lambda'$ is the derivative of $\lambda$.
For any $u,\zeta \in [0,1]$, according to the properties of $B_n$,
the constraint of $M$ and the form of $K$ in (\ref{K_expression}),
we have
\begin{align*}
& \int_0^1\left[\lambda'(v) + B_1\left(\bm{\{}v - u\bm{\}}\right) + B_1(u) \right]
\left[\lambda'(v) + B_1\left(\bm{\{}v - \zeta\bm{\}}\right) + B_1(\zeta) \right] \text{d}v\\
& = \int_0^1\left(\lambda'(v)\right)^2\text{d}v +  \lambda(u) + \lambda(\zeta) +
\frac 12 B_2\left(\bm{\{}u - \zeta\bm{\}}\right) + B_1(u)B_1(\zeta) \\
& = \beta^{-2}(M-1) + \lambda(u) + \lambda(\zeta) +
\frac 12 B_2\left(\bm{\{}u - \zeta\bm{\}}\right) + B_1(u)B_1(\zeta) \\
& = \beta^{-2}\left(K(u,\zeta) - 1\right).\end{align*}
%Then,
 Combining this result with the formula of the components of $\bk$ presented in Lemma \ref{components}, we have
\begin{align*}
\int_{C^{\vartheta}}\left(\frac{\partial^{|\vartheta|}\phi_{\vartheta}}{\partial \v_{\vartheta}}\right)^2 \text{d}\v_{\vartheta}
& = \frac{\beta^{2|\vartheta|}}{N^2}\sum_{\u, \check{\u} \in \ce}\prod_{j \in \vartheta} \left[K(u_j,\check{u}_j)-1\right] - \frac{2\beta^{2|\vartheta|}}{Nn}\sum_{\u \in \ce, \bzeta \in \cd}\prod_{j \in \vartheta} \left[K(u_j,\zeta_j)-1\right]\\
& \quad +\frac{\beta^{2|\vartheta|}}{n^2}\sum_{\bzeta, \check{\bzeta} \in \cd}\prod_{j \in \vartheta} \left[K(\zeta_j,\check{\zeta}_j)-1\right]\\
& = \frac{\beta^{2|\vartheta|}}{N^2}\sum_{\u, \check{\u} \in \ce} \bk_{\vartheta}(\u_{\vartheta},\check{\u}_{\vartheta})
- \frac{2\beta^{2|\vartheta|}}{Nn}\sum_{\u \in \ce, \bzeta \in \cd}\bk_{\vartheta}(\u_{\vartheta},\check{\bzeta}_{\vartheta}) + \frac{\beta^{2|\vartheta|}}{n^2}\sum_{\bzeta, \check{\bzeta} \in \cd}
\bk_{\vartheta}(\bzeta_{\vartheta},\check{\bzeta}_{\vartheta}). \end{align*}
%\begin{align*}
%\int_{C^{\vartheta}}\left(\frac{\partial^{|\vartheta|}\phi_{\vartheta}}{\partial \v_{\vartheta}}\right)^2 \text{d}\v_{\vartheta}
%& = \frac{\beta^{2|\vartheta|}}{N^2}\sum_{i,k=1}^N\prod_{j \in \vartheta}
%\left[K(u_{ij},u_{kj})-1\right] - \frac{2\beta^{2|\vartheta|}}{Nn}\sum_{i=1}^N\sum_{k=1}^n\prod_{j \in \vartheta}\left[K(u_{ij},\zeta_{kj})-1\right]\\
%& \quad +\frac{\beta^{2|\vartheta|}}{n^2}\sum_{i,k=1}^n\prod_{j \in \vartheta} \left[K(\zeta_{ij},\zeta_{kj})-1\right]\\
%& = \frac{\beta^{2|\vartheta|}}{N^2}\sum_{i,k=1}^N\bk_{\vartheta}(\u_i,\u_{kj})
%- \frac{2\beta^{2|\vartheta|}}{Nn}\sum_{i=1}^N\sum_{k=1}^n\bk_{\vartheta}(u_{ij},\zeta_{kj})\\
%& \quad + \frac{\beta^{2|\vartheta|}}{n^2}\sum_{i,k=1}^n\bk_{\vartheta}(\zeta_{ij},\zeta_{kj}). \end{align*}
For any $f \in \X_p(C^s)$, let
$V_{p,\vartheta} \equiv \left\| \frac{\partial^{|\vartheta|}f_{\vartheta}} {\partial \v_{\vartheta}} \right\|_p$
%\Bea %\label{Vpu}
%V_{p,\vartheta} \equiv \left\| \frac{\partial^{|\vartheta|}f_{\vartheta}} {\partial \y_{\vartheta}} \right\|_p.\Eea
and define the generalized ${\ell}_p$-variation as \begin{equation}  \label{Vp}
V_p(f) \equiv  |||f - \cl_{\cs}(f)|||_p  = \left\| \left(\beta^{-|\vartheta|}V_{p,\vartheta}(f)\right)_
{\vartheta \neq \emptyset} \right\|_p =  \left\| \left(\beta^{-|\vartheta|}
\frac{\partial^{|\vartheta|}f_{\vartheta}} {\partial \v_{\vartheta}} \right)_{\vartheta \neq \emptyset}
\right\|_p. \end{equation}
Recall that $\bk$ is the reproducing kernel,
combined with the Riesz Representation Theorem,
it can be inferred that $\cl_S(\phi) = 0$, and
\begin{align} \nonumber
|\ct(f)| &=  |\langle \phi, f \rangle| =
\left|\sum_{\vartheta \neq \emptyset} \beta^{-2|\vartheta|} \int_{C^{\vartheta}}
\frac{\partial^{|\vartheta|}\phi_{\vartheta}}{\partial \v_{\vartheta}}\frac{\partial^{|\vartheta|}f_{\vartheta}}{\partial \v_{\vartheta}}
\text{d}\v_{\vartheta}\right| \\ \label{res:s>1}
&\leq \sum_{\vartheta \neq \emptyset} \beta^{-2|\vartheta|}  \int_{C^{\vartheta}} \left|
\frac{\partial^{|\vartheta|}\phi_{\vartheta}}{\partial \v_{\vartheta}}\frac{\partial^{|\vartheta|}f_{\vartheta}}{\partial \v_{\vartheta}}
\right|\text{d}\v_{\vartheta}
%\\ \nonumber &
\leq \sum_{\vartheta \neq \emptyset}
\left\|\beta^{-|\vartheta|}\frac{\partial^{|\vartheta|}\phi_{\vartheta}}{\partial \v_{\vartheta}}\right\|_2
\left\|\beta^{-|\vartheta|}\frac{\partial^{|\vartheta|}f_{\vartheta}}{\partial \v_{\vartheta}}\right\|_2
%\\\label{res:s>1} &
\leq  |||\phi|||_2  V_2(f), \end{align}
by H$\ddot{\text{o}}$lder inequality and Cauchy--Schwarz inequality,
where \begin{align*}\label{norm_phi2}  |||\phi|||_2^2
&= \left \| \left( \beta^{-|\vartheta|} \frac{\partial^{|\vartheta|}\phi_{\vartheta}} {\partial \v_{\vartheta}} \right)_{\vartheta \subseteq \cs}\right \|_2^2 = \sum_{\vartheta \subseteq \cs}\left \| \beta^{-|\vartheta|}\frac{\partial^{|\vartheta|}\phi_{\vartheta}} {\partial \v_{\vartheta}}\right \|_2^2 = \sum_{\vartheta \subseteq \cs} \int_{C^{\vartheta}} \left(\beta^{-|\vartheta|} \frac{\partial^{|\vartheta|}\phi_{\vartheta}}{\partial \v_{\vartheta}}\right)^2 \text{d}\v_{\vartheta}\\
& = \frac{1}{N^2}\sum_{\u, \check{\u} \in \ce}\sum_{\vartheta \subseteq \cs} \bk_{\vartheta}(\u_{\vartheta}, \check{\u}_{\vartheta})
- \frac{2}{Nn}\sum_{\u \in \ce, \bzeta \in \cd}\sum_{\vartheta \subseteq \cs} \bk_{\vartheta}(\u_{\vartheta}, \bzeta_{\vartheta}) + \frac{1}{n^2}\sum_{\bzeta, \check{\bzeta} \in \cd}\sum_{\vartheta \subseteq \cs}
 \bk_{\vartheta}(\bzeta_{\vartheta}, \check{\bzeta}_{\vartheta})\\\nonumber
&= \frac{1}{N^2}\sum_{\u, \check{\u} \in \ce}\bk(\u, \check{\u})
- \frac{2}{Nn}\sum_{\u \in \ce, \bzeta \in \cd}\bk(\u, \bzeta)
+ \frac{1}{n^2}\sum_{\bzeta, \check{\bzeta}\in \cd}\bk(\bzeta, \check{\bzeta})
= D_{\bk}^2(\ce,\cd).
\end{align*}
Note that, the definition of the generalized ${\ell}_p$-variation
$V_p(f)$ both in (\ref{defi_V(f)}) for $s = 1$
and in (\ref{Vp}) for $s \geq 2$
are relative to $\beta$, a term in the kernel function $K$,
therefore $V_p(f)$ could be denoted by $V_p(f,K)$ for $s = 1$
and $V_p(f,\bk)$ for $s>1$.
Replace  $||| \phi |||_2$  in (\ref{res:s>1}) with $D_{\bk}(\ce,\cp)$
and we complete the proof.
%\\\revB{[[[All the notation of $||| \cdot |||_p$ have been colored by blue.
%This notation is referred to Hickernell \cite{hick1}. This norm is different from the normal norm $|| \cdot ||_p$ ( $|| \cdot ||_p$ of a function $f$ and a function vector $(f_1, \dots, f_s)$ are $\left(\int_{0}^1 |f|^p {\rm d}x\right)^{1/p}$ and $\left(\sum_{j = 1}^s\int_{0}^1 |f_j|^p {\rm d}x\right)^{1/p}$, respectively). While the $||| \cdot |||_p$ could represented by an expression of $|| \cdot ||_p$, see (\ref{defi_inner-norm}), (\ref{phi_norm}), (\ref{defi_V(f)}) and (\ref{Vp}). So to separate from the normal norm, the notation $||| \cdot |||_p$ is used here.]]]}
\section*{Appendix B}
\begin{figure}
	\centering
	\subfigure[The Upper Bound of MSPE of Linear Regression upon the Original Data]{
		\includegraphics[width=2.7in]{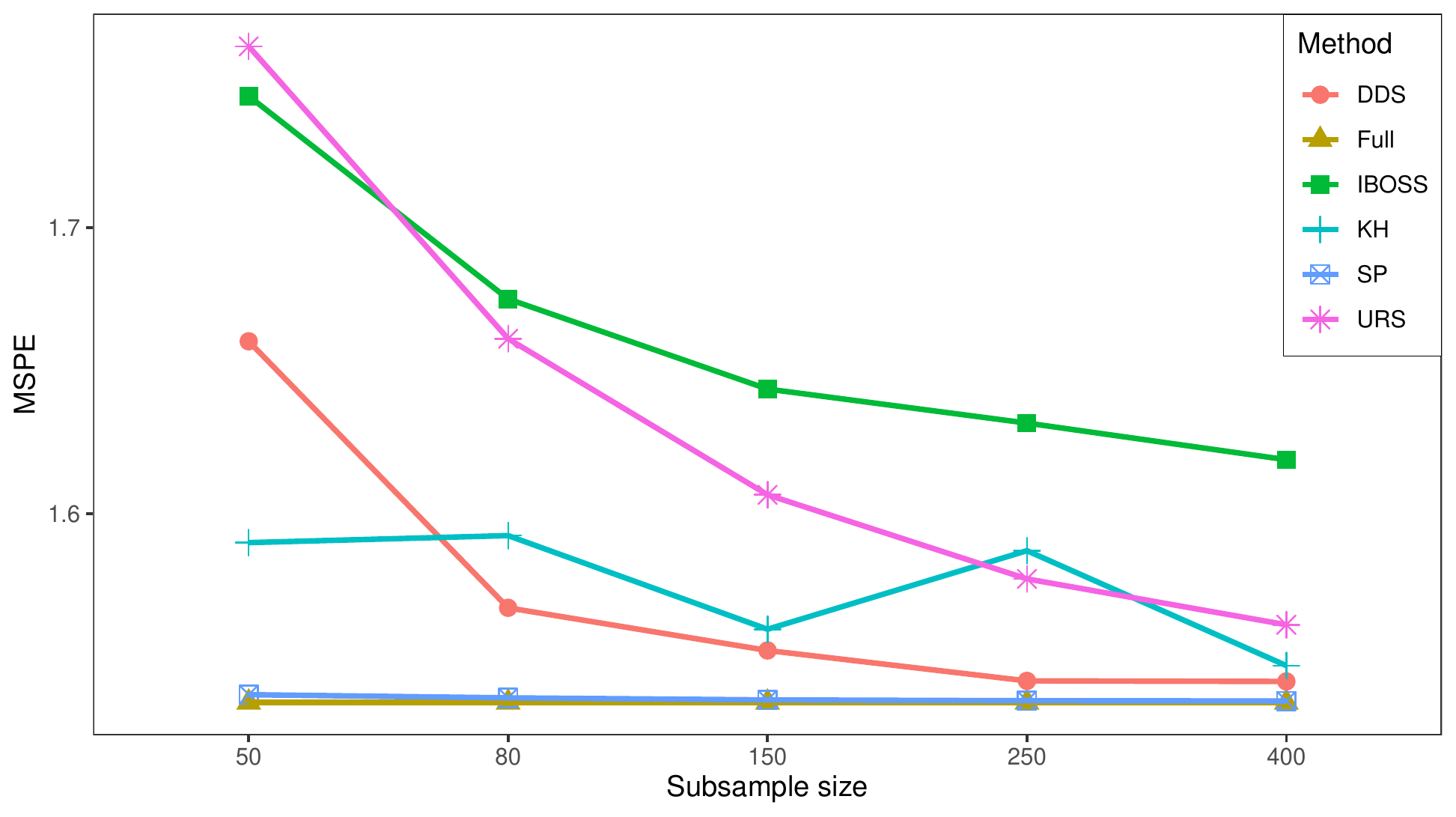}}\hspace{3mm}
	\subfigure[The Lower Bound  MSPE of Linear Regression upon the Original Data]{
		\includegraphics[width=2.7in]{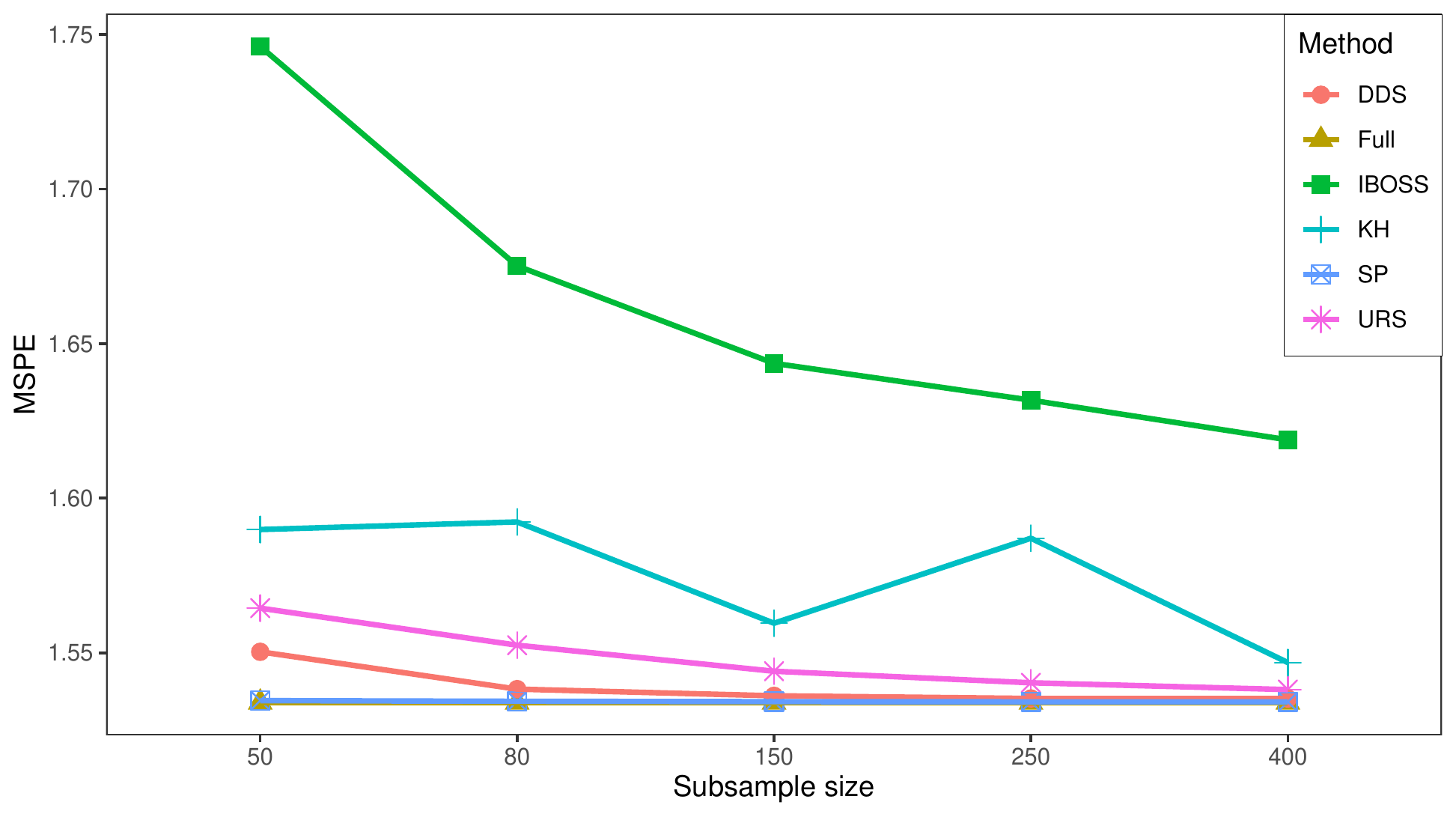}}\hspace{3mm}\\
	\subfigure[The Upper Bound of MSPE of Generalized Linear Regression upon the Transformed Data]{
		\includegraphics[width=2.7in]{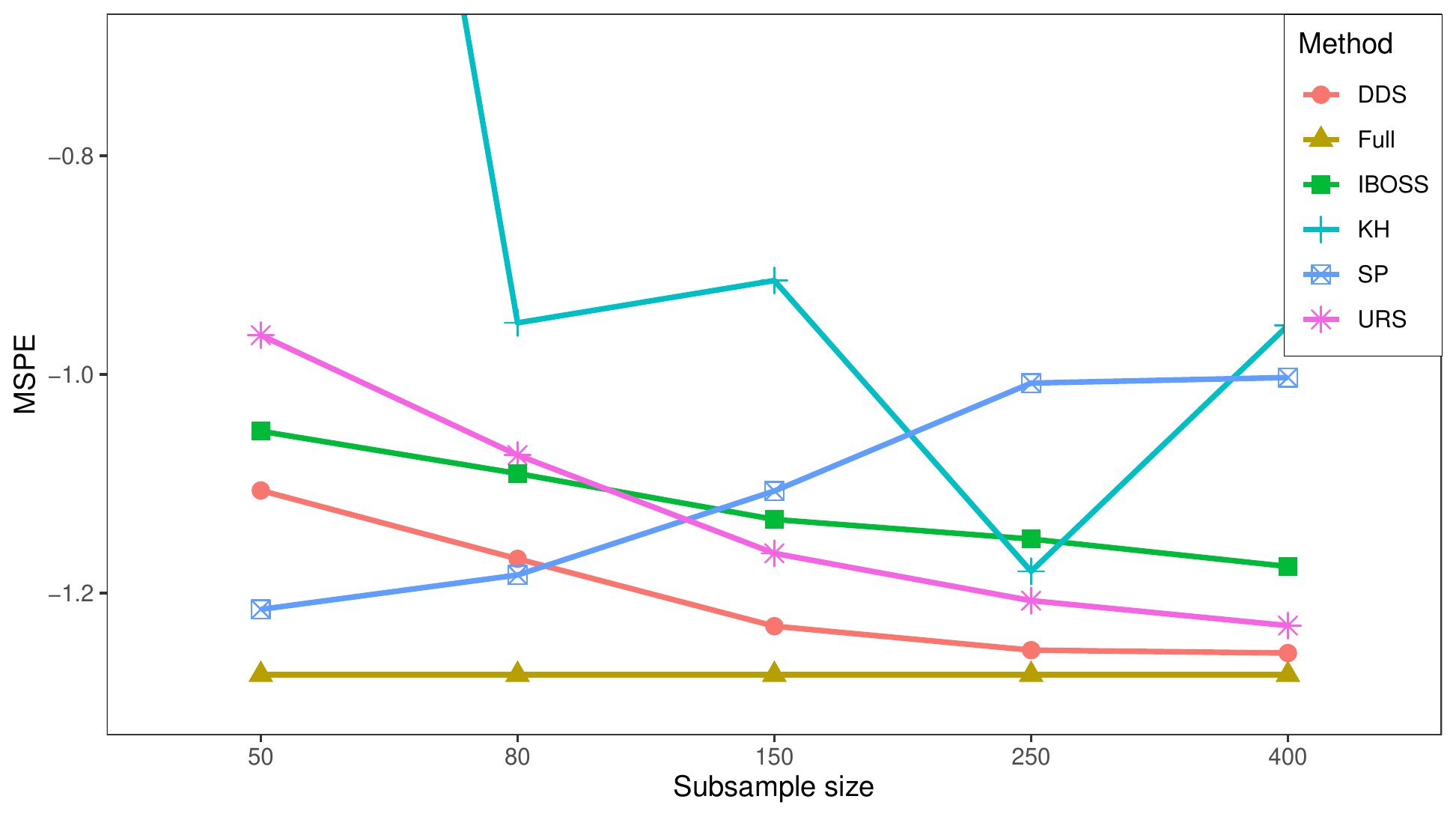}}
	\subfigure[The Lower Bound of MSPE of Generalized Linear Regression upon the Transformed Data]{
		\includegraphics[width=2.7in]{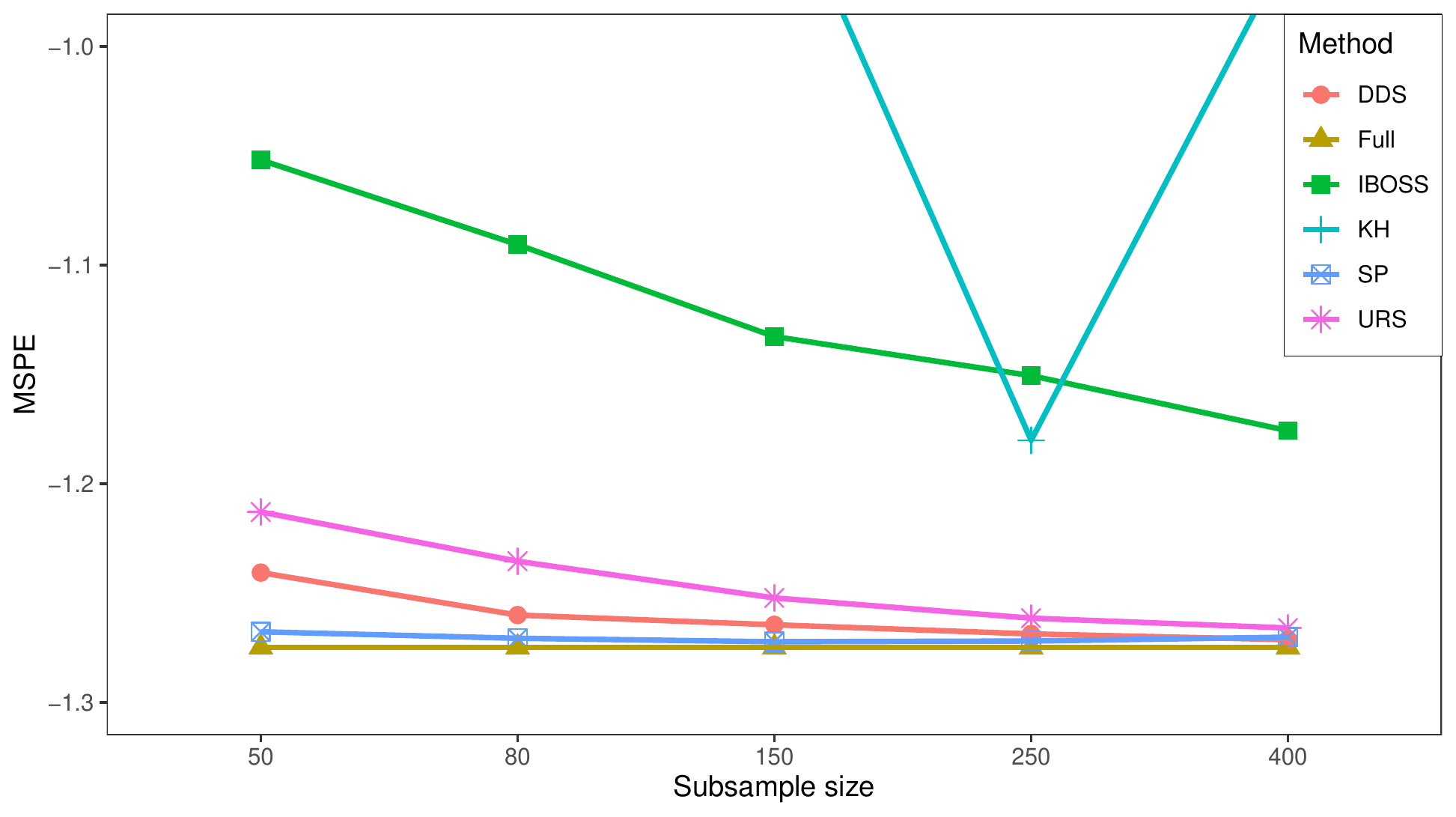}}
	\caption{The bound of MSPE values of the fitted linear regression model based on the subdata sets obtained through different subsample strategies from the full data sets with the original form and the transformed form for the borehole experiments.} \label{add1}
\end{figure}

\begin{figure}
	\centering
	\subfigure[The Upper Bound of MSPE of Linear Regression]{
		\includegraphics[width=2.7in]{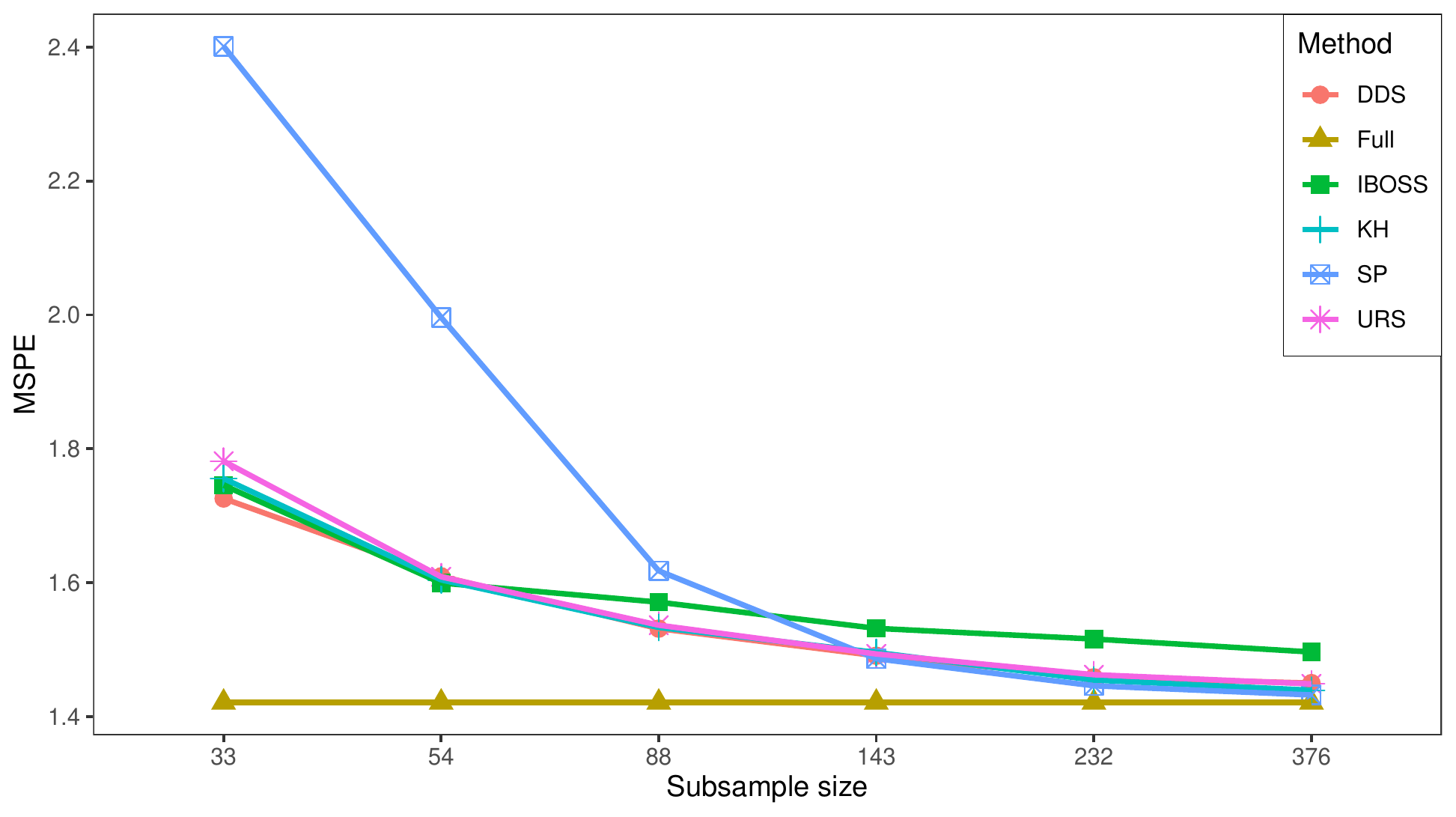}}\hspace{3mm}
	\subfigure[The Lower Bound of MSPE of Linear Regression]{
		\includegraphics[width=2.7in]{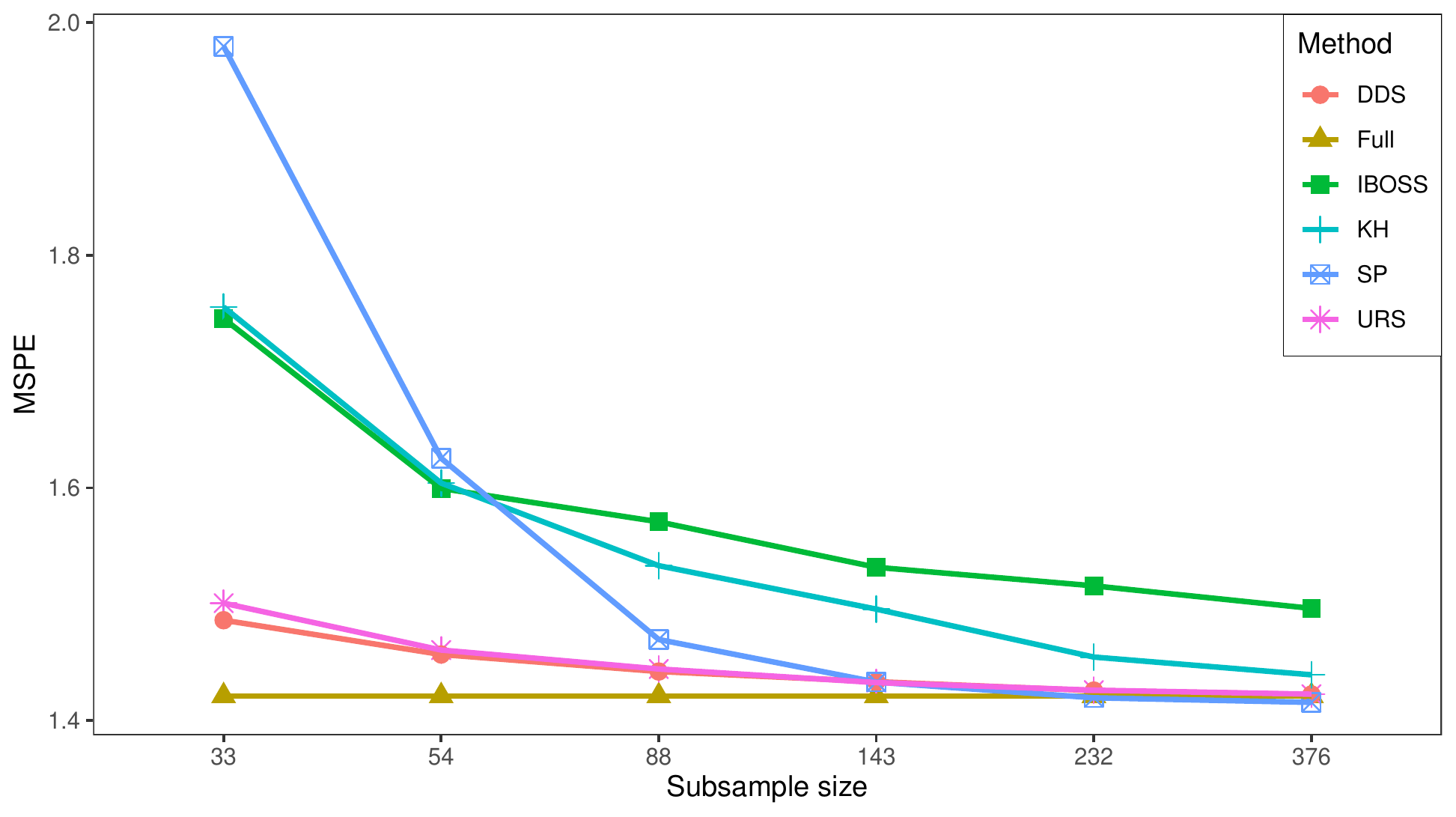}}\hspace{3mm}\\
	\subfigure[The Upper Bound of MSPE of Gaussion Process Regression]{
		\includegraphics[width=2.7in]{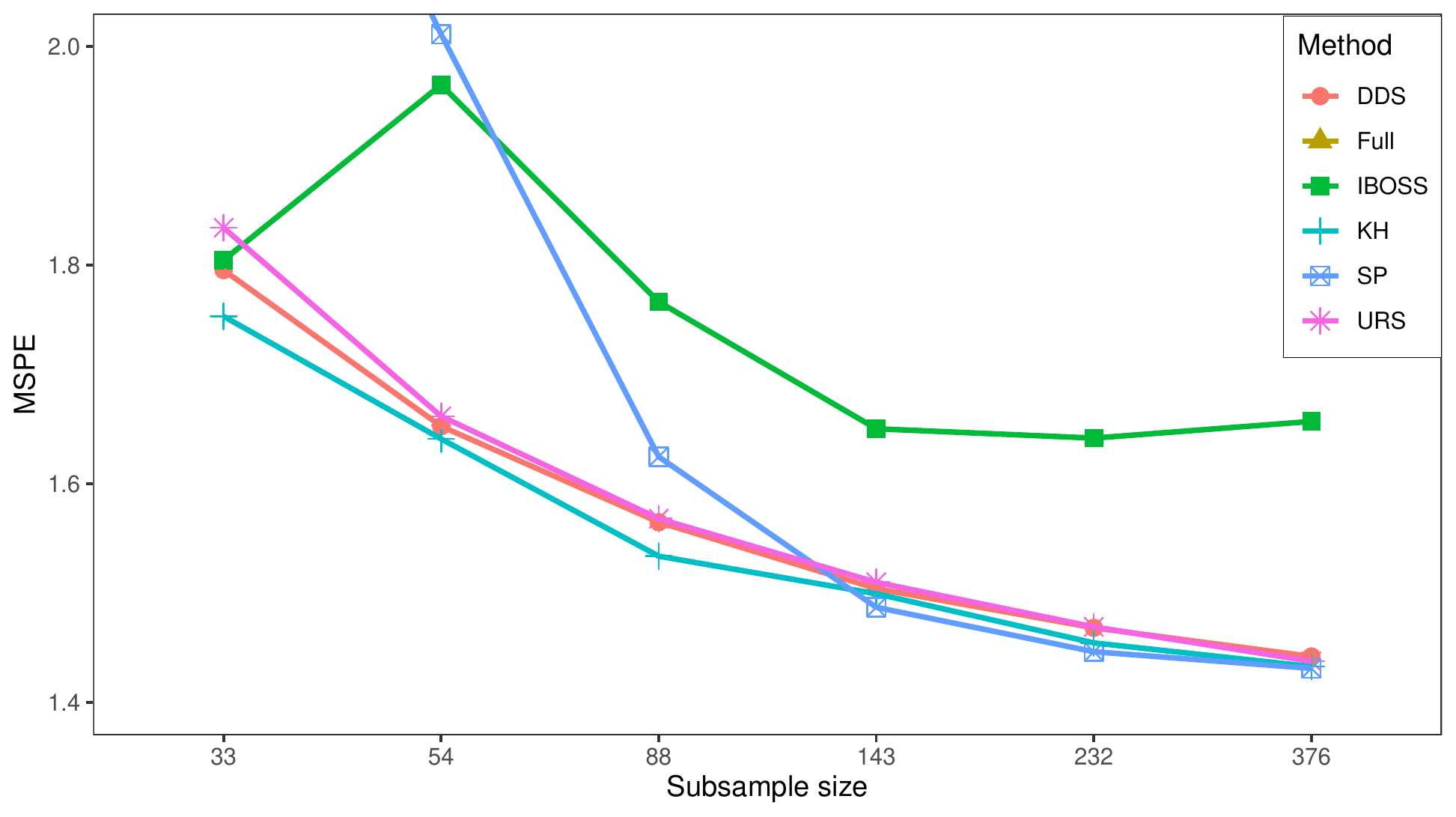}}
	\subfigure[The Lower Bound of MSPE of Gaussion Process Regression]{
		\includegraphics[width=2.7in]{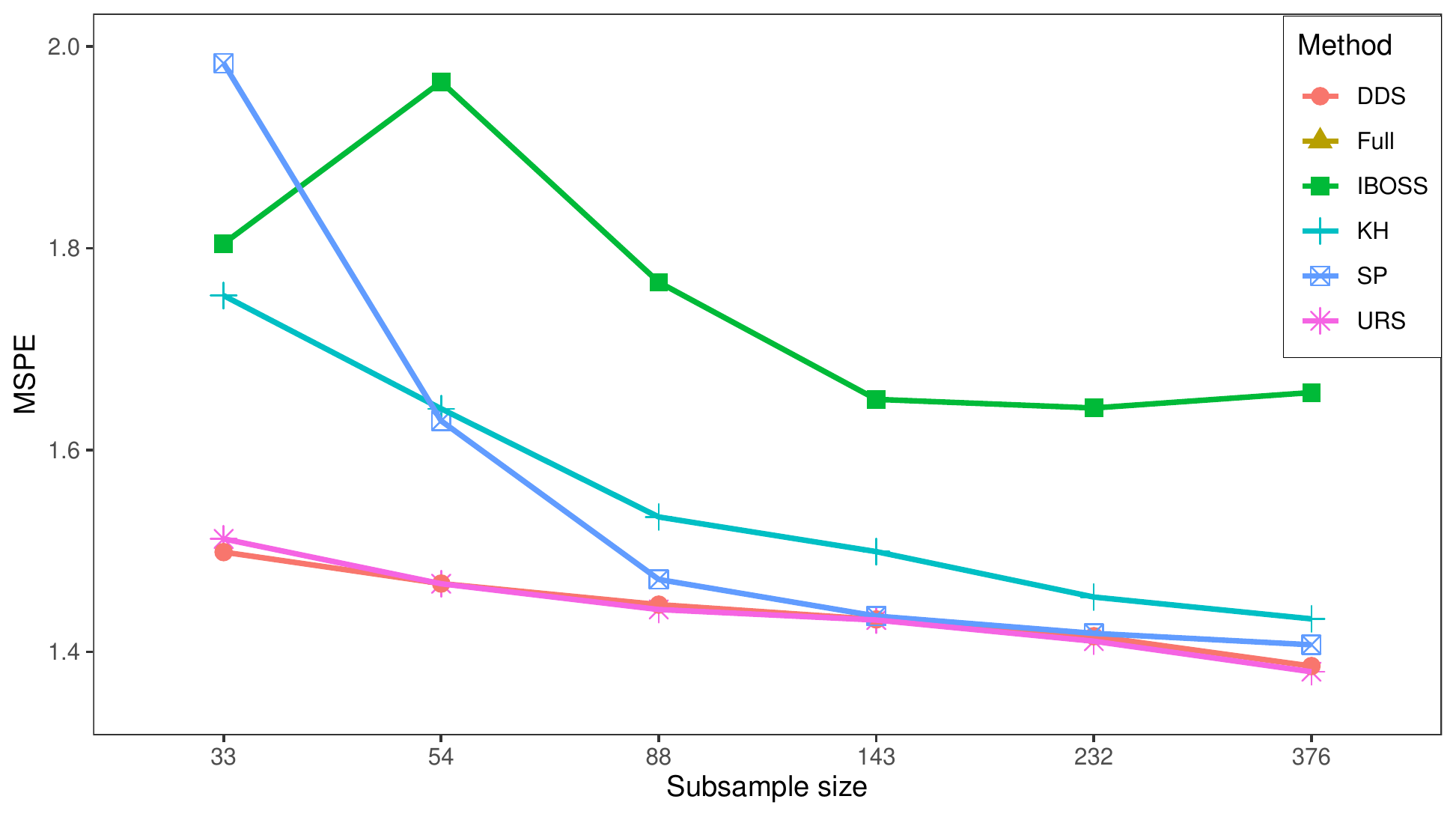}}
	\caption{The bound of MSPE values of the fitted linear regression model
		and Gaussian process regression model
		established on different subsample strategies
		for the protein tertiary structure data.}
	\label{add2}
\end{figure}

The MSPE values of numerical examples are given in Tables \ref{tblr}, \ref{tbglr}, \ref{tplr} and \ref{tpgpr}. For the deterministic method IBOSS and KH, the MSPE values for each subsample size $n$ are given. For URS, DDS and SP, the mean, median, upper bound and low bound of MSPE for each subsample size $n$ are given for these methods are executed multiple replications. Figures \ref{add1}, \ref{add2} are provide to compare the upper and lower bound of MSPE for different methods in each numerical example.
\begin{table*}[htbp]
	\caption{The MSPE values of different methods applied to the real life borehole example while working model is misspecified as the linear regression model. The MSPE value of prediction based on full data is 34.19551.}
	\label{tblr}
	\makeatletter\def\@captype{table}\makeatother
	\centering
 \begin{tabular}{lccccc}
 	\toprule
 	Method & n=50 & n=80 & n=150 & n=250 & n=400 \\
 	\midrule
 	URS\_Mean&43.54&39.52&36.97&35.86&35.20 \\
 	URS\_Median&41.43&38.73&36.51&35.60&35.05\\
 	URS\_Upper &58.01&45.84&40.42&37.77&36.40 \\
 	URS\_Lower &36.68&35.68&35.00&34.70&34.52 \\
 	DDS\_Mean&38.37&35.43&34.83&34.51&34.48 \\
 	DDS\_Median&37.50&35.19&34.74&34.55&34.45\\
 	DDS\_Upper &45.74&36.90&35.66&34.79&34.77 \\
 	DDS\_Lower &35.51&34.53&34.36&34.30&34.29 \\
 	SP\_Mean&34.31&34.26&34.23&34.22&34.22 \\
 	SP\_Median&34.31&34.26&34.23&34.22&34.21\\
 	SP\_Upper &34.41&34.32&34.26&34.24&34.23 \\
 	SP\_Lower &34.24&34.22&34.21&34.21&34.21\\
 	IBOSS&55.73&47.33&44.02&42.83&41.58 \\
 	KH&38.89&39.11&36.27&38.64&35.22 \\

 	\bottomrule
	\end{tabular}

\end{table*}
\begin{table*}[htbp]
	\caption{The MSPE values $\left(\times 10 ^{-2}\right)$ of different methods applied to the real life borehole example while working model is misspecified as the generlized linear regression model. The MSPE value of prediction based on full data is $5.312\times 10^{-2}$.}
	\label{tbglr}
	\makeatletter\def\@captype{table}\makeatother
	\centering
	\begin{tabular}{lccccc}
		\toprule
		Method & n=50 & n=80 & n=150 & n=250 & n=400 \\
		\midrule
		URS\_Mean&7.900&6.872&6.091&5.771&5.598 \\
		URS\_Media&7.463&6.662&6.001&5.721&5.562\\
		URS\_Upper &10.861&8.437&6.860&6.209&5.891\\
		URS\_Lower &6.125&5.815&5.595&5.477&5.420 \\
		DDS\_Mean&6.505&6.001&5.630&5.504&5.426\\
		DDS\_Median&6.349&5.903&5.597&5.482&5.410\\
		DDS\_Upper &7.831&6.781&5.884&5.595&5.560 \\
		DDS\_Lower &5.746&5.495&5.439&5.387&5.354 \\
		SP\_Mean&5.787&5.812&6.033&6.460&6.576 \\
		SP\_Median&5.553&5.466&5.469&5.648&5.822\\
		SP\_Upper &6.096&6.554&7.823&9.818&9.935 \\
		SP\_Lower &5.399&5.362&5.342&5.346&5.368\\
		IBOSS&8.873&8.116&7.367&7.070&6.672 \\
		KH&251.026&11.146&12.186&6.604&11.085 \\
		\bottomrule
	\end{tabular}
\end{table*}
\begin{table*}[htbp]
	\caption{The MSPE values of different methods applied to the real case example while model is specified as the linear regression model. The MSPE value of prediction based on full data is 26.34.}
	\label{tplr}
	\makeatletter\def\@captype{table}\makeatother
	\centering
	\begin{tabular}{lcccccc}
		\toprule
		Method & n=33 & n=54 & n=88 & n=143 & n=232 & n=376 \\
		\midrule
		URS\_Mean&43.36&33.73&30.25&28.69&27.71&27.20 \\
		URS\_Media &40.82&33.26&29.67&28.47&27.74&27.18\\
		URS\_Upper &60.46&40.62&34.40&31.16&29.01&28.11\\
		URS\_Lower &31.69&28.88&27.82&27.08&26.68&26.47 \\
		DDS\_Mean&39.46&33.85&30.16&28.60&27.59&27.22 \\
		DDS\_Median&37.28&32.93&29.90&28.36&27.46&27.19\\
		DDS\_Upper &53.18&40.69&33.99&30.96&28.74&28.18 \\
		DDS\_Lower &30.64&28.63&27.68&27.13&26.66&26.45 \\
		SP\_Mean&166.75&66.58&34.88&28.68&27.06&26.57 \\
		SP\_Median&151.97&63.78&34.22&28.56&27.23&26.73\\
		SP\_Upper &251.84&99.13&41.47&30.65&27.93&27.06 \\
		SP\_Lower &95.41&42.22&29.49&27.11&26.28&26.04\\
		IBOSS&56.66&41.77&37.50&34.20&32.79&31.33 \\
		KH&60.14&46.05&32.78&31.23&28.10&27.33 \\
		\bottomrule
	\end{tabular}
\end{table*}
\begin{table*}[htbp]
	\caption{The MSPE values of different methods applied to the real case example while model is specified as the Gaussian process regression model.}
	\label{tpgpr}
	\makeatletter\def\@captype{table}\makeatother
	\centering
	\begin{tabular}{lcccccc}
		\toprule
		Method & n=33 & n=54 & n=88 & n=143 & n=232 &n=376 \\
		\midrule
		URS\_Mean&45.23&35.60&31.47&29.20&27.47&25.52 \\
		URS\_Media&42.54&34.79&31.16&28.89&27.44&25.41\\
		URS\_Upper &68.25&45.88&36.98&32.35&29.45&27.39\\
		URS\_Lower &32.51&29.35&27.67&27.01&25.74&24.01 \\
		DDS\_Mean&44.01&35.17&31.10&29.01&27.67&25.91\\
		DDS\_Median&41.44&34.11&30.54&28.71&27.66&25.90\\
		DDS\_Upper &62.42&44.93&36.71&31.92&29.39&27.67 \\
		DDS\_Lower &31.54&29.36&27.98&27.05&26.04&24.32 \\
		SP\_Mean&170.16&67.62&35.14&28.80&27.15&26.51 \\
		SP\_Median&163.25&63.88&34.59&28.60&27.17&26.75\\
		SP\_Upper &265.56&102.58&42.13&30.69&27.95&26.98 \\
		SP\_Lower &96.22&42.55&29.63&27.26&26.21&25.53\\
		IBOSS&63.76&92.23&58.39&44.70&43.82&45.39 \\
		KH&56.64&43.75&34.18&31.58&28.47&27.09 \\
		\bottomrule
	\end{tabular}
\end{table*}

\newpage
\begin{thebibliography}{99}
\small {
%\bibitem{A50} Aronszajn, N. (1950), Theory of reproducing kernels,
%\textit{Transactions of the American Mathematical Society}
%{\bf 68}, 337--404.

%\bibitem{AO01}
%An, J. and Owen, A. (2001),
%Quasi-regression,
%\textit{Journal of Complexity} {\bf 17}(4), 588--607.

\bibitem{B18} Bottou, L., Curtis, F. E. and Nocedal, J. (2018), Optimization methods for large-scale machine learning, \textit{SIAM Review} {\bf 60}(2), 223--311.

\bibitem{C12} Chen, Y., Welling, M. and Smola, A.(2012),
Chen, Yutian and Welling, Max and Smola, Alex,
\textit{Proceedings of the Twenty-Sixth Conference on Uncertainty in Artiﬁcial Intelligence.}

\bibitem{C16} Chen, H., Huang, H. Z., Lin, D. K. J. and Liu, M. Q. (2016),
Uniform sliced Latin hypercube designs,
\textit{Applied Stochastic Models in Business and Industry} {\bf 32}(5), 574--584.

\bibitem{D17} Dheeru, D. and Karra Taniskidou, E. (2017),
\textit{UCI Machine Learning Repository},
URL: http://archive.ics.uci.edu/ml.

\bibitem{F06} Fang, K. T., Li, R. and Sudjianto, A. (2006),
\textit{Design and Modeling for Computer Experiments},
Chapman and Hall/CRC, London.

%\bibitem{FL03} Fang, K. T. and Lin, D. K. J. (2003),
%Uniform designs and their application in industry,
%\textit{in} R. Khattree and C. R. Rao,
%\textit{Handbook on Statistic in Industry} pp. 131--170,
%Elsevier, North--Holland, Amsterdam.

\bibitem {F18}
Fang, K. T., Liu, M. Q., Qin, H. and Zhou, Y. D. (2018),
Theory and Application of Uniform Experimental designs,
{Springer \& Science Press, Singapore \& Beijing.}

\bibitem{FW94}  Fang, K. T. and Wang, Y. (1994),
\textit{Number-theoretic Methods in Statistics},
Chapman and Hall, London.

\bibitem{hick1}  Hickernell, F. J. (1998a),
A generalized discrepancy and quadrature error bound,
\textit{Mathematics of computation} {\bf 67}(221), 299--322.

\bibitem{hick2}  Hickernell, F. J. (1998b), Lattice rules: how well do they measure up? In: P. Hellekalek and G. Larcher, Eds., \textit{Random and Quasi-Random Point Sets}, Lecture Notes in Statistics, vol. 138. Springer, New York, 109--166.

\bibitem{HX00} Ho, W. M. and Xu, Z. Q. (2000),
Applications of uniform design to computer experiments,
\textit{Journal of Chinese Statistical Association}
{\bf 38}, 395--410.


%\bibitem{O17}
%Osuolale, K. A., Yahya, W. B. and Adeleke, B. L. (2017),
%Statistical Modeling and Analysis of Borehole Computer Experiment,
%\textit{International Journal of Advanced Science and Technology}
%\textbf{107}, 21--32.

\bibitem{M11} Mahoney, M. W. (2011),
Randomized algorithms for matrices and data,
\textit{Foundations and Trend} $R$
\textit{in Machine Learning}
{\bf 3}(2), 123--224.

\bibitem{M18} Mak, S. and Joseph V. R. (2018),
Support points,
\textit{Annals of Statistics}, {\bf46}(6A), 2562--2592.

%\bibitem{M79}
%McKay, M. D., Beckman, R. J. and Conover, W. J. (1979),
%Comparison of three methods for selecting values of
%input variables in the analysis of output from a computer code,
%\textit{Technometrics} {\bf 21(2)}: 239--245.

\bibitem{M93}
Morris, M. D., Mitchell, T. J. and Ylvisaker, D. (1993),
Bayesian design and analysis of computer experiments:
use of derivatives in surface prediction,
\textit{Technometrics}, {\bf 35}(3), 243--255.

\bibitem{N92} Niederreiter, H. (1992),
Random Number Generation and Quasi-Monte Carlo Methods,
\textit{SIAM CBMS-NSF Regional Conference Series in
Applied Mathematics, Philadephia}.

%\bibitem{U06} Uzilov, A. V., Keegan, J.M. and Mathews, D.H. (2006),
%Detection of non-coding RNAs on the basis of predicted secondary
%structure formation free energy change,
%\textit{BMC Bioinformatics} {\bf 7}:173,
%DOI: 10.1186/1471-2015-7-173.

\bibitem{W18} Wang, H. Y., Zhu, R. and Ma, P. (2018),
Optimal Subsampling for Large Sample Logistic Regression,
\textit{Journal of the American Statistical Association},
{\bf 113}(522), 829--844.
%DOI: 10.1080/01621459.2017.1408468.

\bibitem{W19} Wang, H. Y., Yang, M. and Stufken, J. (2019),
Information-Based Optimal Subdata Selection for Big Data
Linear Regression,
\textit{Journal of the Amarican Statistical Association}
{\bf 114}(525), 393--405.
%DOI: 10.1080/01621459.2017.1408468.

%\bibitem{WF90} Wang, Y. and Fang, K.T., (1990),
%Number theoretic methods in applied statistics,
%\textit{Chinese Annals of Mathematics}, {\bf 11(1)}, 51--65.

%\bibitem{W72} Warnock, T.T. (1972), Computational
%investigations of low discrepancy point sets. In: S.K. Zaremba Eds.,
%\textit{Applications of Number Theory to Numerical Analysis},
%Academic Press, New York, 319--343.

\bibitem{W16} Weyl, H. (1916), $\ddot{u}$ber die gleichverteilung der zahlem mod eins,
\textit{Mathematische Annalen}, {\bf 77}(3), 313--352.

\bibitem{W14} Woodruff, D. P. (2014),
Sketching as a tool for numerical linear algebra,
\textit{Foundations and Trend} $\textcircled{R}$
\textit{in Theoretical Computer Science}, {\bf10}(1--2): 1--157.

\bibitem{W87}
Worley, B. A. (1987),
Deterministic uncertainty analysis (No. CONF-871101-30),
Oak Ridge National Lab., TN (USA).



\bibitem{X19} Xu, J., He, X., Duan, X. Y. and Wang, Z. M. (2018),
Sliced Latin Hypercube Designs for Computer
Experiments With Unequal Batch Sizes, IEEE access, {\bf 6}, 60396-60402.

\bibitem{Y19} Yuan, R., Guo, B. and Liu, M. Q. (2019),
Flexible sliced Latin hypercube designs with slices of different sizes.
\textit{Statistical Papers}, online. %https://doi.org/10.1007/s00362-019-01127-6.



\bibitem{Z18}
Zhang, A. J., Li, H. Y., Quan, S. J., and Yang, Z. B. (2018),
UniDOE: Uniform Design of Experiments,
\textit{R package version 1.0.2}.

%\revA{\bibitem{Z19}  Zhang, A. J., Zhang, M. and Zhou, Y. D., }

%\bibitem{Z11}  Zhou, Q., Qian, P. Z. and Zhou, S. (2011), A simple approach to emulation for computer models with qualitative and quantitative factors, \textit{Technometrics}, {\bf 53}(3), 266--273.

\bibitem{Z13}  Zhou, Y. D., Fang, K. T. and Ning, J. H. (2013), Mixture discrepancy for quasi-random point sets, \textit{Journal of Complexity}, {\bf 29}(3--4), 283--301.

}
\end{thebibliography}
\end{document}